\documentclass[12pt]{article}
    \usepackage[dvipsnames]{xcolor}
    \usepackage[a4paper, head=18pt]{geometry}
	\usepackage[english]{babel}
	\usepackage{amsmath}
	\usepackage{amssymb}
	\usepackage{etoolbox}
    \usepackage[T1]{fontenc}
    \usepackage{mathptmx}    
	\usepackage{graphicx}
	\usepackage{subcaption}
	\usepackage{enumitem}
	\usepackage{url}
	\usepackage{float}
    \usepackage[skip=2pt,font=small]{caption}
	\usepackage{booktabs}
	\usepackage{array}      
	\usepackage{makecell}   
	\usepackage{arydshln}   
	\usepackage{threeparttablex}
    \usepackage{pgfplots}
	\usepackage{flafter}
	\usepackage{multirow}
    \usepackage[hang, flushmargin]{footmisc}
    \usepackage[hyperfootnotes=false,colorlinks]{hyperref}
    \usepackage{footnotebackref}
	\usepackage{adjustbox}
	\usepackage{setspace}
	\usepackage{placeins}
	\usepackage[toc]{appendix}
    \usepackage{eurosym}
    \usepackage[nameinlink, noabbrev]{cleveref}
    \usepackage{titletoc} 

\renewcommand{\thetable}{\Roman{table}}
\renewcommand{\thefigure}{\Roman{figure}}
    
\hypersetup{
linkcolor=BrickRed
,citecolor=Blue
,filecolor=Mulberry
,urlcolor=NavyBlue
,menucolor=BrickRed
,runcolor=Mulberry
,linkbordercolor=BrickRed
,citebordercolor=NavyBlue
,filebordercolor=Mulberry
,urlbordercolor=NavyBlue
,menubordercolor=BrickRed
,runbordercolor=Mulberry
}

\AtBeginDocument{%
  \begingroup\doublespacing\normalsize
  \global\footnotesep=\dimexpr\baselineskip-\dp\strutbox\relax
  \endgroup}

\usepackage{titlesec}
\titleformat{\section}
  {\normalfont\fontsize{14}{16}\selectfont\bfseries}
  {\thesection}{1em}{}

\titleformat{\subsection}
  {\normalfont\fontsize{12}{14}\selectfont\bfseries}
  {\thesubsection}{1em}{}

    \usepackage[]{natbib}
    \setcitestyle{authoryear,open={(},close={)},comma} 
    
\usepackage{multibib}
\newcites{supp}{Supplementary References}

\makeatletter
\def\OA@dropnum#1#2\@nil#3{\def#3{#2}}
\renewcommand\makeatother

\makeatletter

\newcount\nat@length@list
\newcount\nat@tempcount

\renewcommand\NAT@sort@cites[1]{%
  \let\NAT@cite@list\@empty
  \nat@length@list=0%
  \@for\@citeb:=#1\do{%
     \global\advance\nat@length@list\@ne%
     \expandafter\NAT@star@cite\@citeb\@@}%
  \if@filesw
    \expandafter\immediate\expandafter\write\expandafter\@auxout
      \expandafter{\expandafter\string\expandafter\citation\expandafter{\NAT@cite@list}}%
  \fi
  \@ifnum{\NAT@sort>\z@}{%
    \expandafter\NAT@sort@cites@\expandafter{\NAT@cite@list}%
  }{}%
}%

\def\NAT@def@citea@close{%
 \def\@citea{%
     \NAT@@close%
     \ifnum\nat@length@list=2
         \NAT@space and\NAT@space%
      \else
         \NAT@separator\NAT@space%
      \fi
      \advance\nat@length@list-\@ne%
   }%
}

\makeatother

\makeatletter
\def\adl@drawiv#1#2#3{%
        \hskip.5\tabcolsep
        \xleaders#3{#2.5\@tempdimb #1{1}#2.5\@tempdimb}%
                #2\z@ plus1fil minus1fil\relax
        \hskip.5\tabcolsep}
\newcommand{\cdashlinelr}[1]{%
  \noalign{\vskip\aboverulesep
           \global\let\@dashdrawstore\adl@draw
           \global\let\adl@draw\adl@drawiv}
  \cdashline{#1}
  \noalign{\global\let\adl@draw\@dashdrawstore
           \vskip\belowrulesep}}
\makeatother

\newcolumntype{L}[1]{>{\raggedright\let\newline\\arraybackslash\hspace{0pt}}m{#1}}
\newcolumntype{C}[1]{>{\centering\let\newline\\arraybackslash\hspace{0pt}}m{#1}}
\newcolumntype{R}[1]{>{\raggedleft\let\newline\\arraybackslash\hspace{0pt}}m{#1}}

\renewcommand{\thesection}{\Roman{section}}  

\renewcommand{\thesubsection}{\thesection.\Alph{subsection}}  

\renewcommand{\thesubsubsection}{\arabic{subsubsection}.}  

\renewcommand{\theparagraph}{\roman{paragraph}.}

\titleformat{\section}{\normalfont\centering}{\thesection.}{1em}{\MakeUppercase}  

\titleformat{\subsection}{\normalfont\itshape}{\thesubsection}{1em}{}  

\titleformat{\subsubsection}{\normalfont\itshape}{\thesubsubsection}{1em}{}  

\titleformat{\paragraph}[runin]{\normalfont\itshape}{\theparagraph}{1em}{}

\pgfplotsset{compat=1.17}

\begin{document}

\begin{titlepage}

\title{\Large{Skill Substitution, Expectations, and the Business Cycle}\thanks{\protect\doublespacing\footnotesize\noindent I would like to thank David~Autor, Eric~Bettinger, Nick~Bloom, George~Bulman, Aline~Bütikofer, \mbox{Kerwin~Charles}, Rick~Hanushek, Caroline~Hoxby, Steffen~Huck, Jan~Marcus, José~Montalbán~Castilla, Markus~Nagler, Luigi~Pistaferri, Sean~Reardon, and C. Katharina~Spiess as well as seminar and conference participants at DIW Berlin, Stanford University, Federal Ministry of Education and Research, Potsdam University, IAB Nuremberg, ifo Munich, ifo Dresden, Bochum University, as well as EEA 2023, EALE 2023, VfS Annual 2023, VPET-ECON 2024, EWMES 2025, and IAAE 2026 for helpful comments. I further thank Eric~Bettinger and Caroline~Hoxby for their hospitality at Stanford University. Anri Konomi provided excellent research assistance. All errors are mine.}}

\author{Andreas Leibing\thanks{German Centre for Higher Education Research and Science Studies (DZHW), Lange Laube 12, 30159 Hannover, Germany, and IZA@LISER; email: \href{mailto:leibing@dzhw.eu}{leibing@dzhw.eu}}}

\maketitle

\vspace{-2em}
\begin{abstract}
\noindent Using administrative data on more than six million German high school graduates from 1995 to 2018, this paper studies how labor market conditions around high school graduation affect postsecondary skill investments. Exploiting cyclical deviations from secular state-specific trends, I find procyclical overall university enrollment, especially at academic universities. Long-run attainment effects are procyclical for academic universities and countercyclical for applied universities. Using large-scale survey data, I present changes in expected returns to different degrees as a potential mechanism. During downturns, graduates expect lower returns to university degrees, while expected returns to the outside option, apprenticeships, are stable. 
\vspace{-2em} \\

\noindent\textbf{Keywords:} college enrollment, subjective expectations, business cycle, apprenticeship

\noindent\textbf{JEL Codes:} D84, E24, E32, I23, I24, I26 \\


\end{abstract}

\setcounter{page}{0}
\thispagestyle{empty}
\end{titlepage}


\setlength{\parskip}{0.15em}


\newpage
\setcounter{page}{1}

\section{Introduction}
\label{sec:intro}

Skills are a key driver of economic growth \citep[e.g.,][]{hanushek2008growth}, and the type of skills acquired is crucial for individual earnings trajectories. While general skills offer higher long-term returns and greater adaptability, specific skills provide immediate labor market advantages but may become obsolete more quickly \citep[e.g.,][]{hanushek2017general, deming2020stemskills}. Economic shocks can affect human capital investments by shifting expected returns and the value of outside options \citep[e.g.,][]{becker1962investment}. But how do they affect the choice between general and specific skills? While countercyclical college-going is a stylized fact in economics, less is known about how economic conditions shape the skill content of educational choices, particularly the trade-off between general and specific skills.\footnote{I distinguish between general and specific skills (i.e., firm- or occupation-specific) and abstract from the distinction between basic and applied skills \citep[see][for respective overviews]{taber2012specific, woessmann2025skills}.}

Germany's education system provides an ideal case to study how individuals substitute between general and specific skill investments. Academic universities (\textit{Universitäten}) focus on general skills, whereas applied universities (such as \textit{Fachhochschulen}) offer a skill mix and blend general education and vocational training. While varying in skill content, Bachelor's degrees in Germany are formally equivalent across institutions. Apprenticeships serve as the typical outside option and are highly firm-specific in contrast. Germany's centrally funded higher education system and regulated labor market resemble those of many OECD countries. 

This article exploits deviations from secular state-specific trends in the unemployment rate around high school graduation to study how economic conditions affect postsecondary skill investments. Using administrative microdata on German higher education, I construct a commuting-zone (CZ) level panel on the share of more than six million high school graduates who enroll in and graduate from different universities (1995--2018). To distinguish between expected returns and outside options as mechanisms, I support my analysis with administrative data on the apprenticeship system, large-scale survey data on the expected returns to university vs. apprenticeship and study intentions, as well as a decomposition of unemployment shocks.

I find that a 1 percentage point (pp) deviation of the state unemployment rate (UR) from its long-term quadratic trend in the senior high school year reduces overall first-time enrollment by 2.10~pp. Enrollment effects are concentrated at academic universities and result in a long-term reduction in attainment at these institutions (-1.04~pp). In contrast, attainment at applied universities increases significantly by 0.40~pp. Large-scale survey data support an expectations channel, as higher unemployment lowers the subjectively expected returns to higher education and the enrollment intentions measured half a year before graduation. During downturns, especially low-performing young women expect the value of an apprenticeship degree, the main outside option, to increase in both relative and even in absolute terms. A decomposition of the state UR suggests that these effects are mainly driven by changes in aggregate conditions rather than changes in state-specific conditions.

This paper makes three contributions. First, it documents a novel margin of adjustment within the large literature on economic conditions and college-going: Investment in general vs. specific skills.\footnote{Related work has mostly focused on the effects of economic shocks on on-the-job training \citep{fukao1993accumulation, mendez2012cyclicality} and the demand for skills \citep{hershbein2018recessions, blair2020structural}, rather than on investment decisions.} Most related work studies Anglo-Saxon settings \citep[see][for a survey]{graves2022higher} and finds countercyclical enrollment, a pattern documented as far back as the late 1960s \citep{card2001dropout}. My findings contrast with, but do not contradict, this conventional wisdom: for German high school graduates obtaining the highest leaving qualification (\textit{Abitur}), I find procyclical college enrollment.\footnote{However, focusing on the selected group of high school graduates with \textit{Abitur} alone cannot explain procyclical enrollment. For example, \cite{johnson2013graduate} also shows countercyclical enrollment for postgraduate students in the U.S.} In the U.S., cyclical responses are concentrated at community colleges \citep{goodman2025declining, bicakova2025unpacking}, which serve as a ``safe port in a storm'' \citep{betts1995safe}. Yet, Germany has no equivalent two-year college sector, and intensive tracking in secondary school. During downturns, high school graduates shift towards more specific skills: Enrollment declines are concentrated at academic universities, while long-run attainment at applied universities increases.

Second, this paper helps to clarify which level of business cycle fluctuation shapes human capital investment by decomposing market shocks into national and (sub-)state-level shocks. Traditionally, the literature on graduating in a recession has focused on more macroeconomic shocks to study different outcomes.\footnote{For high school graduates, these include college enrollment \citep[e.g.,][]{dellas2003business, dellas2003cyclical, hillman2013community, long2014financial, alessandrini2018port, charles2018housing}, major choice \citep{ersoy2020effects, blom2021investment}, later cognitive skills \citep{hampf2020effects, arellanobover2022skills}, and earnings \citep{bicakova2021cycle, bicakova2023luck}.} Recently, the literature focuses more on local labor markets \citep[e.g.,][]{clark2011recessions, sievertsen2016local, deneault2025teaching}, or sector and occupation-specific shocks \citep{charles2018housing, han2020industry, acton2021community, weinstein2022local, buetikofer2025oil}. I synthesize these strands of the literature by showing that CZ-level unemployment shocks affect outside options, and national shocks affect both outside options and expectations. 

Lastly, this paper contributes to the literature on belief formation and its interplay with the business cycle \citep[e.g.,][]{beaudry2006stock, blanchard2013news, enke2019correlation, nowzohour2020more} and is the first to study its effect on expectations in education \citep[see][for a survey]{giustinelli2023expectations}. I show that, in particular, changes in aggregate labor market conditions reduce the expected value of academic degrees. In contrast, the expected value of an apprenticeship, and thus more specific skills, is stable across the business cycle.\footnote{This belief is at odds with empirical evidence showing that university graduates, especially those with more general skills, are less vulnerable to recessions \citep[see][respectively]{hoynes2012recessions, altonji2016cashier, eggenberger2022specific}.} These responses reflect extrapolation from current conditions rather than forward-looking updating, consistent with experience-based belief formation \citep{malmendier2011nagel, malmendier2016learning, nagel2026experiences} and also matter for eventual choices. I show that these beliefs directly affect enrollment intentions of high school students.

Next, \Cref{sec:institution} discusses the institutional context and \Cref{sec:data} introduces the data. \Cref{sec:setup} outlines the identification strategy. \Cref{sec:admin} presents the main effects on college enrollment and attainment. \Cref{sec:surveydata} presents individual-level evidence from survey data on the expectations channel. \Cref{sec:conclusion} concludes. 

\section{Institutional Background}
\label{sec:institution}

To study substitution patterns in postsecondary investments in general vs. specific skills, this paper abstracts from the exact institutional type and major and distinguishes between three options in students' choice set, which vary in their underlying skill content and length:

\begin{enumerate}
    \item \textbf{Academic university}: All full universities (\textit{Universitäten}) in Germany that are publicly financed. They offer a full range of majors and are known for providing a comprehensive, theoretical education. Degrees at academic universities are typically awarded in full-time programs and emphasize the acquisition of general skills. Internships are allowed but usually not mandatory. Hence, there is greater emphasis on general skills. Before the Bologna Process (see Online Appendix \Cref{sec:bologna} for details), academic universities awarded \textit{Diplom} and \textit{Magister} degrees with a formal duration of four and a half years and a median completion time of six years. Since the reform, the formal duration is three years for a Bachelor's degree and two years for a consecutive Master's degree. Two-thirds of Bachelor's graduates at academic universities enter a Master's program \citep{destatis2022master}. Several classical professions, for example medicine, law, and teaching, require a degree from an academic university.

    \item \textbf{Applied university}: All other institutions, predominantly Universities of Applied Sciences (\textit{Fachhochschulen}), but also e.g., private universities, typically offer a more practice-oriented education. They focus on engineering, business, design, media, and technology majors, and their curriculum often includes internships. Some are privately financed, and many offer bachelor's degrees through dual study programs, in which students spend much of their studies in paid on-the-job training. Depending on the degree type chosen (dual or full-time), the focus on specific skills is much stronger. Formal study duration is the same as in academic universities, but only 29 percent of Bachelor's graduates at applied universities enter a Master's program \citep{destatis2022master}. Typical graduates work, for example, as engineers in industry, business administrators, or social workers.

    \item \textbf{Apprenticeship}: High school graduates who enter the labor market typically start an apprenticeship at a firm. Apprenticeships combine on-the-job training at the firm with theoretical education at a vocational school (\textit{Berufsschule}). These vocational training programs cover various occupations, e.g., traditional crafts, technical fields, and healthcare. By design, skills taught in apprenticeships are highly firm- and occupation-specific. The training content of each occupation is codified in state-approved, nationally standardized apprenticeship plans \citep{langer2023skills}. Depending on the occupation, apprenticeships formally last between two and three and a half years \citep{bibb2022datenreport}. High school graduates with \textit{Abitur} graduate faster, in around two and a half years. Among them, typical training occupations include bank clerk, industrial clerk, or IT specialist.
\end{enumerate}

I focus on high school graduates who obtained the highest leaving qualification (\textit{allgemeine Hochschulreife}, i.e. \textit{Abitur}). It typically takes 12--13 years of schooling, and graduates with \textit{Abitur} represent the largest share (about 75~percent) of high school graduates in the academic track (\textit{Gymnasium}). Their total share has risen over the sample period, from roughly a quarter of a birth cohort in the mid-1990s to about 40 percent by the late 2010s, as participation in the academic track expanded \citep{destatis2020}.\footnote{Further descriptive statistics on high school graduates, enrollment, and unemployment, as well as all relevant educational reforms between 1995 and 2018, are discussed in the Online Appendix \Cref{sec:descriptive}.} Graduates with \textit{Abitur} have the full choice set: They can enroll at any institution, in any degree, and, depending on their GPA, in any major. Direct labor market entry without any of these qualifications is rare. Four and a half years after graduation, only about 1 percent of school leavers with a higher education entrance qualification neither hold nor plan a postsecondary qualification \citep{spangenberg2016studienberechtigte}.

Due to the high degree of tracking in secondary school, enrollment among \textit{Abitur} holders is high. Over 2009--2018, around 76 percent of my sample enrolled in higher education within a year of graduation, compared with a U.S. immediate college enrollment rate of about 68 percent over 2012--2018 \citep{nces2023enrollment}. The selectivity of the academic track is already evident by mid-secondary school. In PISA 2022, 15-year-olds in the academic track (\textit{Gymnasium}) scored 546 points in mathematics \citep{lewalter2023}, well above the German national mean of 475 and the U.S. mean of 465 \citep{oecd2023pisa}.\footnote{PISA assesses 15-year-olds, several years before the \textit{Abitur} at around age 18. The score thus reflects the selectivity of the academic track rather than the skill level of actual graduates.}

\section{Data}
\label{sec:data}

\subsection{Student and Exams Register}

The student register \citep[\textit{Statistik der Studierenden},][]{RDC2019a} covers the universe of students enrolled in higher education in Germany. Each observation represents an individual enrollment spell and contains information on the institution, degree, entry qualification, gender, age, county, and year of high school graduation. Conversely, the exam register \citep[\textit{Statistik der Prüfungen,}][]{RDC2019b} covers the same information for each awarded degree. Due to strict data protection measures, however, they do not contain individual identifiers, and the register data can not be linked internally to understand the exact course of study. I thus link the register data on the number of first-year students and first-time university graduates who graduated from high school in a given county and year to the county-level data on the number of high school graduates from the \textit{Regionaldatenbank} and county-level unemployment rates from the Federal Employment Agency (BA).

To account for the relevance of local labor markets \citep[e.g.,][]{amior2018persistence, hershbein2024local} and commuting across county borders \citep{krebs2023road}, I aggregate all counties to the so-called \textit{Raumordnungsregionen} level. These regions are groups of counties that reflect a larger commuting zone (CZ). Throughout the analysis, I use the current 96 CZ delineations from the Federal Office for Building and Regional Planning (BBR). My final sample spans the years 1995 to 2018 and includes 6,296,662 high school graduates.\footnote{Of all 2,304 CZ $\times$ cohort cells, I have to exclude five cells from the state of Hesse (no data on high school graduates in 2007), two cells for 1996 and 1997 in the CZ of Osnabrück, and four cells from each state Mecklenburg-Western Pomerania and Saxony-Anhalt in 2001 (no high school graduates due to an increase in the mandatory years of schooling), totaling 2,289 cells.}

As main outcome variables, I use the share of high school graduates with \textit{Abitur}, from commuting zone $r$ and graduation cohort $t$, enrolling in different colleges $c$ within one year after high school graduation. Each enrollment share is defined by: 

\vspace{-1em}
\begin{equation}
\label{eq:outcomeext} 
E^{c}_{rt} = \sum_{\tau=t}^{t+1} \sum_{i \in r, t}^{} \text{Enrolling in College}^{\thinspace c,\, y=\tau}_{i} \; / \sum_{i \in r, t}^{} \text{High School Graduates}_{i}, \quad t \leq 2018,
\end{equation}

\noindent where $y$ indicates the year of first-time enrollment. I focus on individuals $i$ in the student register with \textit{Abitur} and use the number of high school graduates from the CZ with \textit{Abitur} as the denominator. Until 2011, male high school graduates had to fulfill one year of civil service. Hence, I consider enrollment within one year of graduation. My final sample thus ends with the 2018 high school cohort, but I still count enrollment in 2019.

To study long-term attainment, I use the same denominator but count all successful first-time college graduations of high school graduates with \textit{Abitur} from a given CZ $r$ and cohort $t$ at any time $y$ after high school graduation. The attainment shares are thus defined by:

\vspace{-1em}
\begin{equation}
\label{eq:outcomegrad}
A^{c}_{rt} = \sum_{\tau=t}^{2019} \sum_{i \in r, t}^{} \text{Graduating from College}^{\thinspace c,\, y=\tau}_{i} \; / \sum_{i \in r, t}^{} \text{High School Graduates}_{i}, \quad t \leq 2016.
\end{equation}

\noindent Because my data cover college enrollment and graduation until 2019, and I want to give all cohorts at least three years of regular study duration to finish their bachelor's degree and want to avoid the COVID-19 pandemic as a confounder, I restrict the attainment sample to high school cohorts 1995--2016, but again count first-time graduations (at any time) until 2019.

\subsection{Panel Study of School Leavers}

To study subjective expectations and enrollment intentions, I rely on the \textit{Panel Study of School Leavers with a Higher Education Entrance Qualification} \citep{daniel2017dzhw} of the German Centre for Higher Education Research and Science Studies (DZHW). The data cover 131,606 high school students scheduled to graduate in 2008, 2012, 2015, and 2018 in a given state. Each cohort is surveyed at least twice: a first wave in December of the senior high school year, half a year before graduation, and a follow-up wave in December of the graduation year. Follow-up waves record students' first realized activity and updated beliefs.

Students are surveyed in the calendar year before (wave 1) and half a year after high school graduation (wave 2) on their beliefs about the returns to (i) a college degree and (ii) an apprenticeship. Students are asked (i): ``In general, how do you value the job perspectives for graduates with a college degree?'' and (ii) ``In general, how do you value the job perspectives for graduates with a vocational degree, without a college degree?'' on a five-point Likert scale, with the option to state ``do not know.''\footnote{In the same battery, students are asked about their own prospects: ``How do you value your personal job perspectives?'', on the same scale.} From these measures, I compute (iii) a proxy for the implicitly expected ``relative return'' of academic vs. vocational education, i.e., the standardized difference between the two. The raw measure ranges from -4 (negative returns) to 4 (very high returns to college), with a mean of 0.52. I consider students who are uncertain about either option to have a weak preference for the other. If a student values the return on a college degree as good (4) and is uncertain about apprenticeships, the difference would be equal to 1.\footnote{Otherwise, students answering ``do not know'' are excluded: the expectations estimates use students answering all three items (N = 104,666), and the weak-preference coding of one-sided ``do not know'' responses enters only the larger sample of the intention estimates in \Cref{sec:expectations}, where it adds 9,496 students.}

The first waves also record enrollment intentions. I code an intention indicator that equals 1 if a student answers ``yes, certainly'' or ``yes, probably'' when asked about enrolling after graduation and names a higher-education institution, and 0 otherwise.\footnote{To match the register data, I count 2,203 students (3 percent of those with enrollment intentions) with \textit{Berufsakademie} as intended institution as not intending to enroll. These institutions are not part of the student register and focus heavily on specific skills.} Because intentions are measured in the first wave, they are available for all four cohorts and are unaffected by panel attrition and delayed enrollment. Among students also observed in the follow-up wave, 63 percent of intenders enroll by December of the graduation year.

The estimation sample on subjective expectations is effectively a repeated cross-section, restricted to 104,666 students with complete information on gender, parental education, grades, both expectation items, and a valid intention response. In the follow-up wave, students report their current activity, which I categorize into higher education, vocational training, employment, civil or military service, and other, to analyze study intentions and choices in a second step and make use of the panel dimension. Until the suspension of conscription in 2011, most male high school graduates completed civil or military service first, so their activity half a year after graduation is uninformative about enrollment. I therefore exclude men of the 2008 cohort wherever realized choices are the outcome.

\section{Identification}
\label{sec:setup}

To estimate the causal effect of labor market conditions on skill investments, I regress enrollment rates $E$ (and analogously attainment shares $A$) of each high school cohort $t$ from commuting zone $r$ at each college type $c$ on the unemployment rate $\text{UR}_{s(r),t-1}$ in the respective state $s(r)$ in the senior year before graduation:

\vspace{-1em}
\begin{equation}
\label{eq:specification}
E_{rt}^c\;=\;
\beta \, \mathrm{UR}_{s(r), \, t-1}
\;+\; \alpha_{s(r)}
\;+\; g_{s(r)}(t)
\;+\; \gamma' Z_{s(r), \, t}
\;+\; \varepsilon_{rt},
\end{equation}

\noindent where $\alpha_{s(r)}$ are state fixed effects and $Z$ contains indicators for state-specific education reforms.\footnote{These are the introduction and abolition of tuition fees across states, as well as inflated high school graduation cohorts (``double-cohorts'') resulting from the reduction of statutory years spent in high school \citep[][]{marcus2019effect}. Cohorts without high school graduates, resulting from increased statutory schooling, are excluded from the sample. These reforms may otherwise affect university enrollment \citep[see, e.g.,][]{bietenbeck2023tuition} and act as supply shocks on the apprenticeship market \citep{muehlemann2024double, dorner2024empty}.} To obtain coefficients equivalent to those from an individual-level regression, I weight each cell by the number of high school graduates. To adjust standard errors for common shocks across cohorts and correlation within states, I allow for two-way clustering by state and cohort \citep{cameron2011robust}. My identification strategy is similar to that of, e.g., \cite{blom2021investment}, which isolates cyclical changes in U.S. college major choice. 

\begin{figure}[!ht]
	\centering
	\begin{minipage}{\textwidth}
 	\caption{Aggregated main variables of interest and quadratic fits}
    \vspace{-1em}
    \label{fig:trendfit}
    \begin{center}
    \includegraphics[width=0.95\textwidth]{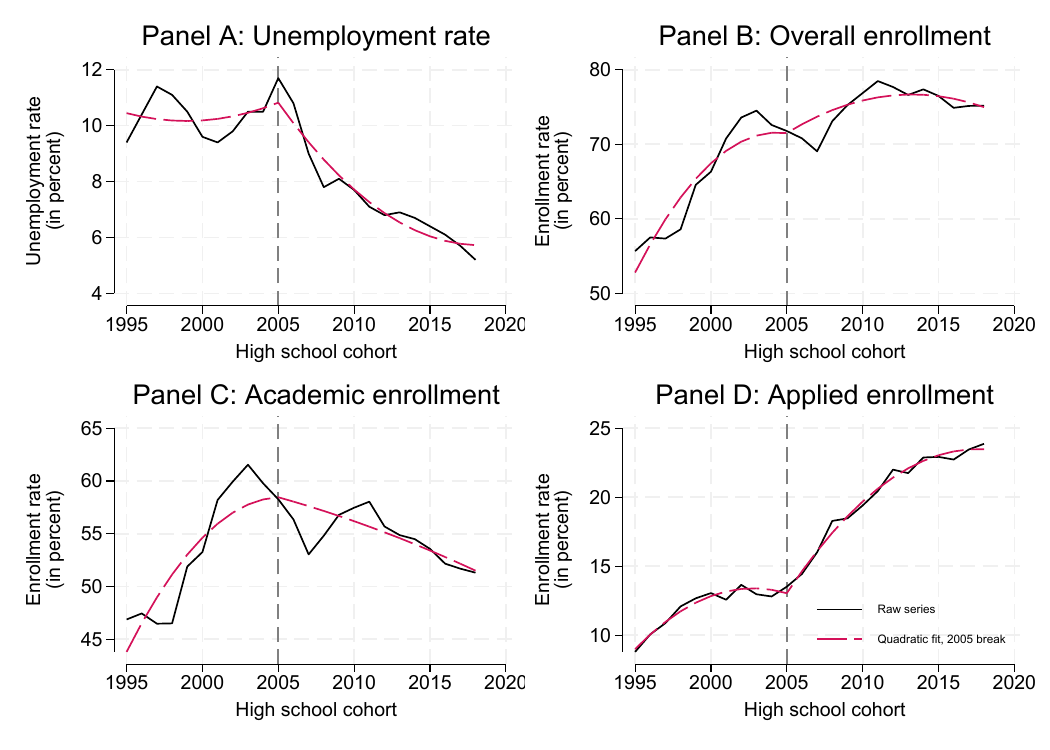} \\
    \end{center}
    \vspace{-1em}
    {\scriptsize \textit{Notes:} Each panel plots the raw series (solid black) and the fitted quadratic trend with a continuous slope break in 2005 (red long dash, with the vertical line marking the break year). Panel A shows the unemployment rate, Panel B the overall share of high school graduates with \textit{Abitur} enrolling at any college, Panel C the share enrolling at academic universities, and Panel D the share enrolling at applied universities. The fitted trend takes the form of \Cref{eq:trend}, shown here at the national level. In the estimating equation, the trend is estimated separately by state. \\ \textit{Source}: Student register, Federal Employment Agency (BA), \textit{Regionaldatenbank}, years 1995--2018.
    \par}
    \end{minipage}

\end{figure}

\begin{figure}[!htbp]
	\centering
	\begin{minipage}{\textwidth}
	\caption{Aggregated business cycle variation in enrollment and unemployment}
    \vspace{-1em}
 	\label{fig:kg2x2_enrol}
	\begin{center}
	\includegraphics[width=0.95\textwidth]{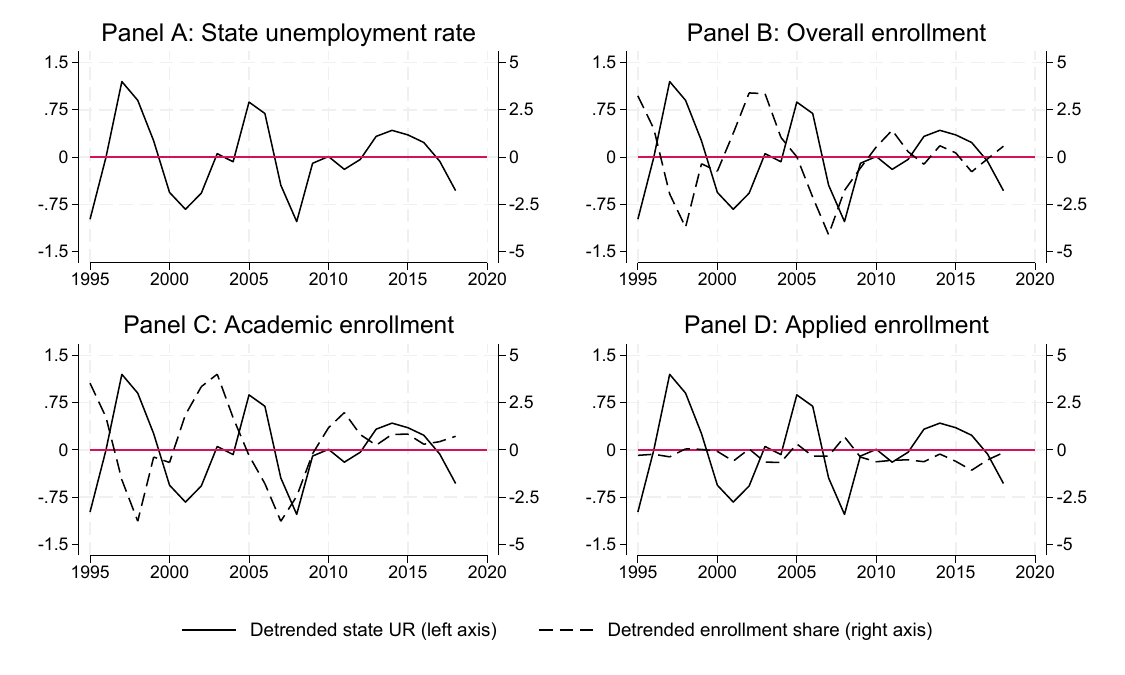} \\
	\end{center}
    \vspace{-1em}
    {\scriptsize \textit{Notes:} Each panel plots, for each high school cohort, residuals after removing state fixed effects and state-specific quadratic trends with a continuous 2005 break (\Cref{eq:trend}). Panel A shows the residual state unemployment rate. Panels B, C, and D overlay the residual state unemployment rate (solid line, left axis) with the residual overall, academic-university, and applied-university enrollment rate (dashed line, right axis). Residuals are aggregated by cohort year and weighted by the number of \textit{Abitur} holders. The red line indicates the zero mean. \\ \textit{Source}: Student register, Federal Employment Agency (BA), Federal Statistical Office (Hochschulstatistik), years 1995--2018.
    \par}
    \end{minipage}
\end{figure}

Against the background of diverging trends across East and West German states in the number of high school graduates, university enrollment, and unemployment rates, as well as structural breaks in these variables (see \Cref{sec:descriptive}), my main specification aims to net out these confounders, isolating cyclical fluctuations in enrollment and unemployment rates, via:

\vspace{-1em}
\begin{equation}
\label{eq:trend}
g_s(t)\;=\;
\theta_{1s}\,\tau_t \;+\;
\theta_{2s}\,\tau_t^2 \;+\;
\theta_{3s}\,{\mathbf{1}\{t\ge 2005\}}\,\tau_t \;+\;
\theta_{4s}\,{\mathbf{1}\{t\ge 2005\}}\,\tau_t^2,
\end{equation}

\noindent which is a function that controls for quadratic time trends within each state $s(r)$, that are allowed to change slope after the year 2005, and $\tau_t = t-2005$ gives the number of years relative to 2005.\footnote{To study the effect of aggregate labor market shocks, the main specification omits year fixed effects. \Cref{sec:decomposition} decomposes the state UR into federal and state-specific variation.} \Cref{fig:trendfit} overlays the fitted trends on the time series of the national unemployment rate and aggregate enrollment rates, both on average and across university types, and \Cref{fig:kg2x2_enrol} plots the resulting residual variation in enrollment and unemployment, i.e., the cyclical variation after detrending, aggregated across states and by cohort.\footnote{Corresponding CZ-level residuals in enrollment and state-level variation in UR, as well as the variation absorbed by alternative specifications, including cohort fixed effects, are shown in Online Appendix \Cref{sec:oa_identifying_variation}.} Overall, the graphs show a procyclical pattern that is particularly pronounced for enrollment at academic universities. Deviations in enrollment from their long-term trend follow fluctuations in unemployment with a delay of about a year, as mirrored in my main specification.

Under the conditional independence assumption, i.e., $E\bigl(\varepsilon_{rt} \mid \text{UR}_{s(r),t-1}, \alpha_{s(r)}, g_{s(r)}(t), Z_{s(r)t}\bigr) = 0$, my estimates reflect the causal effect of a one percentage point deviation from the state-specific UR trend on CZ-level enrollment shares $E_{rt}$ of the following graduation cohort. Explicitly, I assume that, conditional on state fixed effects, state-specific quadratic time trends, and educational reforms, the cyclical variation in the senior-year unemployment rate is exogenous to other changes in the relative returns to certain institutional choices or cohort characteristics. Reverse causality seems implausible, as variation in unemployment likely reflects changes in broad labor demand. Because the preferred specification omits year fixed effects to capture aggregate business-cycle fluctuations, omitted variable bias could arise from unobserved within-year cross-state shocks as well as from common national cohort-year shocks, such as national shifts in educational preferences or unmeasured federal policies, that correlate with detrended unemployment and independently alter enrollment.

A first suspect is technology shocks. For example, high-skill-augmenting technological shocks may simultaneously increase the expected return to general skill investments and affect unemployment. Yet, larger technological shocks often take time to be adopted and cause labor market reallocation \citep[see, e.g.,][]{comin2010exploration}. In turn, adoption speed varies geographically and depends on economic conditions \citep{hershbein2018recessions}. As a result, a substantial share of technology-induced changes in unemployment may be captured by state-specific trends. Moreover, the effects of technology shocks on the labor market are typically ambiguous ex-ante.\footnote{Take artificial intelligence as an example, where the exact impact on labor markets is complex and hard to predict \citep[see, e.g.,][]{frank2019toward, acemoglu2024ai}.} Hence, it is unlikely that technology shocks immediately affect average college-going.\footnote{On an intensive margin, however, specific technology shocks may locally affect field of study choice \citep[see, e.g.,][]{acton2021community, weinstein2022local}.} Overall, while I cannot rule out that technology shocks may affect enrollment choices in specific periods, a systematic bias due to unobserved technology shocks is unlikely.

A second main suspect is firms' expectations. If firms lower their expectations, e.g., due to political uncertainty, they may hire fewer workers and apprentices \citep{bloom2009uncertainty, muehlemann2020expectations}. As a result, the UR would increase, and high school graduates with \textit{Abitur} may fail to find an apprenticeship and instead enroll at university. If this were the case, my finding of procyclical college enrollment would be attenuated and may represent a lower bound. Importantly, however, such expectation-driven changes in firms' hiring and training behavior are arguably part of the fluctuations in labor demand that my analysis intends to capture. 

\section{Evidence from Administrative Records}
\label{sec:admin}

\subsection{Skill Substitution in Enrollment and Attainment}
\label{sec:main}

\Cref{tab:main_results} presents the main results for enrollment and attainment. Each panel reports the effect of the senior-year state UR on one institutional margin. Column (1) uses all high school cohorts (1995--2018) and shows that a 1 pp increase in the state UR in the senior year of high school reduces the overall share of students enrolling in any college by 2.10 pp. This procyclical enrollment is driven entirely by a reduction in enrollment at academic universities (-2.05 pp), and enrollment effects at applied universities are insignificant and close to zero (-0.06), consistent with a substitution towards specific skill investments.\footnote{Symmetric to lower overall university enrollment, higher unemployment rates increase the number of applicants and new contracts on the apprenticeship market. Online Appendix \Cref{sec:outsideopt} studies the effect on the apprenticeship market in more detail.} A 1 SD increase in the net-of-trend senior-year state UR (0.67~pp) thus reduces overall enrollment by 0.32 SD (1.4~pp).

\begin{table}[!ht]
    \begin{center}
	\begin{adjustbox}{max width=0.85\textwidth}
	\begin{threeparttable}
		\caption{Effects of the senior-year state UR on enrollment and attainment}
		\label{tab:main_results}
		\scriptsize{
		        \begin{tabular} {l ccc}
	\toprule \toprule
& \multicolumn{1}{c}{1995--2018} & \multicolumn{2}{c}{1995--2016}  \\
\cmidrule(lr){2-2} \cmidrule(lr){3-4}
& Enrollment & Enrollment & Attainment \\
& (1) & (2) & (3) \\
\midrule
\textit{Panel A: Overall (any college)} & & & \\
State UR (t-1)   &      -2.102***&      -2.143***&      -0.640   \\
                 &     (0.378)   &     (0.363)   &     (0.374)   \\
Outcome mean     &        71.2   &        71.0   &        59.9   \\
\addlinespace
\textit{Panel B: Academic university} & & & \\
State UR (t-1)   &      -2.047***&      -2.107***&      -1.039***\\
                 &     (0.373)   &     (0.344)   &     (0.231)   \\
Outcome mean     &        54.2   &        54.6   &        40.0   \\
\addlinespace
\textit{Panel C: Applied university} & & & \\
State UR (t-1)   &      -0.055   &      -0.037   &       0.399** \\
                 &     (0.063)   &     (0.065)   &     (0.184)   \\
Outcome mean     &        17.0   &        16.4   &        19.9   \\
\midrule
No. CZ-cohort cells       &       2,289   &       2,097   &       2,097   \\
No. high school graduates &   6,296,662   &   5,727,918   &   5,727,918   \\
State FE, trends, policy controls &      yes      &      yes      &      yes      \\
				\bottomrule
		\end{tabular}
		}
		\begin{tablenotes}[flushleft]
				\item \scriptsize{ \textit{Notes:} This table presents estimates from \Cref{eq:specification} for the effect of the state unemployment rate in the senior high school year ($t-1$). The outcome is the share of high school graduates with \textit{Abitur} enrolling at (columns 1--2) or graduating from (column 3) any college (Panel A), academic universities (Panel B), or applied universities (Panel C). Column 1 spans high school cohorts 1995--2018. Columns 2 and 3 span cohorts 1995--2016, with attainment measured by 2019. Policy controls include tuition fee introductions and abolitions, as well as double cohorts. Each cell is weighted by the number of high school graduates with \textit{Abitur}. Standard errors in parentheses allow for two-way clustering at the state and cohort level. * $p$\textless 0.1, ** $p$\textless 0.05, *** $p$\textless 0.01. \\ \textit{Source}: Student register, exam register, Federal Employment Agency, \textit{Regionaldatenbank}.
                }
		\end{tablenotes}
	\end{threeparttable}
	\end{adjustbox}
    \end{center}
\end{table}

Column (2) repeats the enrollment analysis for high school cohorts 1995--2016 for which I can trace back university graduation for at least three years after high school graduation. The results are virtually the same as in the full panel. Based on the restricted sample, column (3) then presents the long-run effects of cyclical UR fluctuations before high school graduation on educational attainment. For overall attainment, the effects are negative but insignificant and much smaller (-0.64 pp) than the enrollment effects. This may be well explained by ``second chance studies'', i.e., students who initially started an apprenticeship directly after graduating from high school, but later enrolled in, e.g., a University of Applied Sciences.\footnote{\cite{dahm2023einfach} show that the share of university graduates with previous vocational education declined from 39 percent in 1997 to 19 percent in 2017, and that applied universities are particularly popular among former apprentices. \cite{biewen2017} document the broader role of such second-chance routes into higher education in Germany.} Consistent with this explanation, there is a positive and significant effect on attainment at applied universities of about 0.4 pp, i.e., an increase of 2 percent. At the same time, the attainment effects on academic universities are still negative and significant (-1.04). 

These results suggest that cyclical fluctuations in labor market conditions at high school graduation not only have short-term effects on investment choices but also affect economies' skill supply in the long run. Online Appendix \Cref{tab:period_split} shows that substitution patterns in enrollment are relatively stable over time, with no significant differences in effect sizes before and after 2008.\footnote{After removing the state-specific trends, the senior-year state UR has a standard deviation of 0.88~pp before 2008 and 0.32~pp after. The aggregate component of the detrended enrollment share falls in step (2.4 vs. 0.8~pp), while its dispersion across commuting zones is acyclical and stable.} For attainment, however, effects significantly change with the completion of the Bologna process. After 2008, they are larger on average (1.47~pp vs. 1.00~pp), and for academic universities (1.67~pp vs. 1.31~pp). This is consistent with the considerably higher completion rates of Bachelor's degrees vs., e.g., \textit{Diploma} degrees \citep[58 percent and 74 percent, respectively; see][]{bietenbeck2023tuition}, with a thus higher transmission of enrollment responses. Online Appendix \Cref{tab:gender} presents results separately for male and female high school graduates, using the student and exam registers. Overall, the effects are somewhat larger for women, but the difference is not statistically significant.\footnote{\Cref{sec:soep_heterogeneity} discusses further heterogeneity by gender and parental education using survey data.}

\subsection{Robustness}
\label{sec:robustness}

\begin{figure}[!h]
	\centering
	\begin{minipage}{\textwidth}
	\caption{Senior-year unemployment effect across trend specifications}
 	\label{fig:spec_curve}
    \begin{subfigure}{\textwidth}
        \centering
        \caption{Panel A: Enrollment}
        \label{fig:spec_curve_a}
        \includegraphics[width=\textwidth]{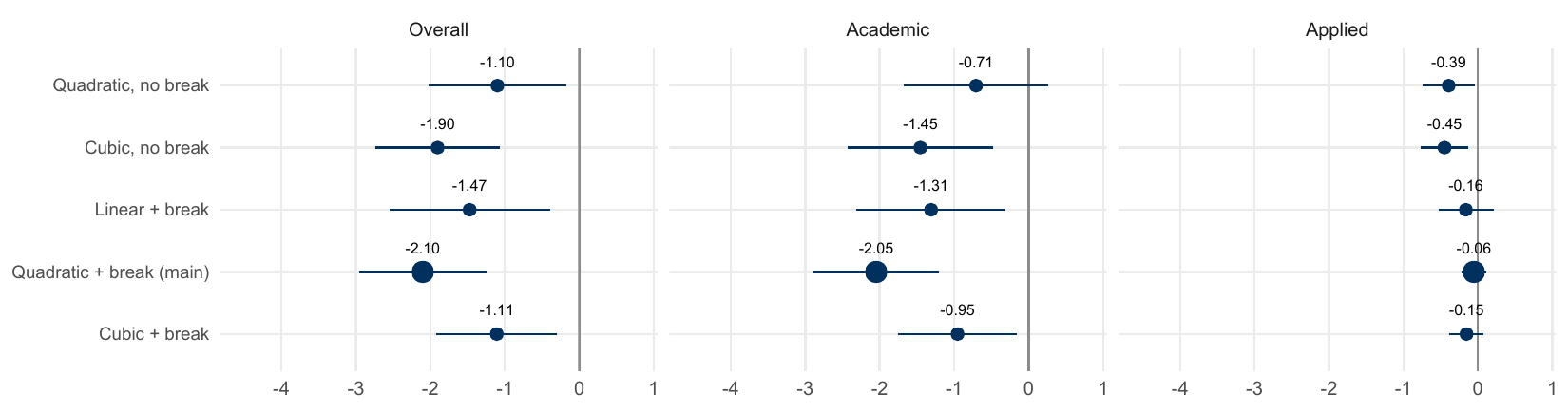}
        \vspace{-0.45em}
    \end{subfigure}

    \begin{subfigure}{\textwidth}
        \centering
        \caption{Panel B: Attainment}
        \label{fig:spec_curve_b}
        \includegraphics[width=\textwidth]{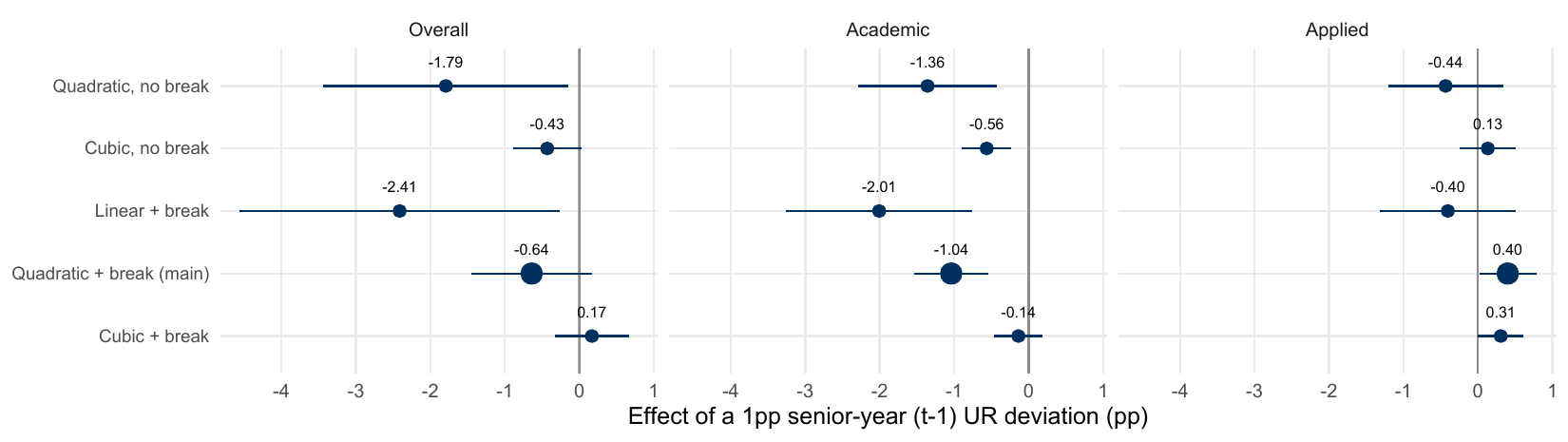}
        \vspace{-0.45em}
    \end{subfigure}
    {\scriptsize \textit{Notes:} This figure plots the estimated effect of a 1 pp deviation of the senior-year ($t-1$) state unemployment rate from its trend on first-time college enrollment (Panel A, high school cohorts 1995--2018) and attainment (Panel B, cohorts 1995--2016, degrees counted through 2019), with 95\% confidence intervals, for enrollment shares overall, at academic universities, and at applied universities. Each row corresponds to a different state-specific detrending choice: a linear trend with a 2005 break, and quadratic and cubic trends, with and without a 2005 break. Main specification indicated by a larger marker. Standard errors allow for two-way clustering at the state and year level. \\ \textit{Source}: Student register, exam register, Federal Employment Agency (BA), \textit{Regionaldatenbank}.
    \par}
    \end{minipage}
\end{figure}

\textit{i. Detrending.} \Cref{fig:spec_curve} reports the senior-year unemployment effect across five trend specifications (i.e., variations of \Cref{eq:trend}) for the three enrollment (panel A) and attainment margins (panel B).\footnote{Online Appendix \Cref{sec:oa_specfigs} shows, for each of these specifications, the fitted trend against the raw series and the residual variation in enrollment and unemployment that remains after detrending.} For overall enrollment, the effect is significantly negative throughout and ranges between roughly -1.1 in the specification with a quadratic trend without break and a cubic trend with a trend break in 2005, and -2.1 pp in my preferred specification, which I interpret as an upper bound in absolute terms. For enrollment at applied universities, the estimates range between -0.4 pp and my estimate of -0.06 pp. Only the two point estimates from specifications that do not account for a trend break are significant.

For attainment in panel B, the overall effects are more sensitive to the exact parametric choice, with in particular the linear trend with a 2005 break and quadratic-no-trend-break specification delivering large confidence intervals. The applied university effect is positive and significant only under the main and the cubic specifications, and insignificant for the other three parametric choices. However, the academic university attainment effect is negative under each specification, but not for cubic trends with a trend break. In sum, only the negative academic-university attainment response is robust to alternative parametric detrending choices.\footnote{The cubic trend including a break in 2005 is an exception to this. However, Online Appendix \Cref{fig:trendfit_grad} shows that the attainment trends are much smoother than enrollment, so a cubic trend with a trend break likely overfits. Similarly, the quadratic trend without trend break, and the linear trend, even with a trend break, are likely to underfit.}

Online Appendix \Cref{tab:rob_trendbreaks} shows that shifting the break year to 2006 or 2004 leaves the enrollment effects similar to the main specification, with overall effects between -1.87 and -2.13~pp, respectively, and small effects at applied universities. Online Appendix \Cref{tab:rob_hp} and \Cref{fig:hp_filter} show that the skill substitution pattern is also robust to replacing the policy-motivated trend specification with the cyclical components from a Hodrick--Prescott (HP) filter.\footnote{The exception is again the applied-university enrollment margin, which turns significantly negative under $\lambda = 100$.} The attainment estimates are also not driven by the unequal completion horizons of early and late cohorts: restricting to cohorts graduating by 2013, which all have at least six years to complete a degree by 2019, yields -0.88~pp at academic and +0.49~pp at applied universities (Online Appendix \Cref{tab:rob_horizon}).

\textit{ii. Treatment Timing.} Online Appendix \Cref{tab:rob_leadlag} assesses the robustness of the main results towards using different leads and lags, contemporaneous values, and moving averages of the state UR.\footnote{$\text{UR}_{s(r),t-1}$ is the annual average unemployment rate of calendar year $t-1$. The senior school year of graduation cohort $t$ runs from the summer of $t-1$ to the early summer of $t$, so calendar year $t-1$ covers its first half, including the autumn and winter in which students form intentions and applications, and in which the panel study interviews take place (December of $t-1$). The graduation-year UR ($t$) instead covers the months around and after graduation, when outside options are realized.} The effects remain negative and significant when using a different lag (t-2) or averages in the unemployment rate during high school as the main regressor. However, when using the lagged unemployment in the year following high school graduation, the effects are positive and largely insignificant. While this could merely reflect serial correlation in unemployment, it may also reflect changes in outside options, assuming that expectations are primarily affected during high school. When using unemployment in the high school graduation year (t) as the regressor, the effects remain negative, but lose significance.

\textit{iii. Compositional Effects.} Lastly, lagged negative effects of labor market conditions on college enrollment may reflect compositional changes of high school graduates. For example, \cite{gaini2013shelter} suggests that during high unemployment, secondary schooling may act ``as a shelter'' \citep[also see][]{petrongolo2002staying}. If, during recessions, low-performing students choose to remain in school and graduate, college enrollment rates in these inflated cohorts may be lower. Using log cohort sizes by gender, Online Appendix \Cref{tab:hsgraduates} shows that adverse labor market conditions two to three years before graduation increase the number of students completing the academic track, consistent with a shelter or continuation effect. Importantly, however, this response occurs before the enrollment decision studied in this paper. Reassuringly, enrollment effects are largest for shocks in t-1. Also, cyclical UR shocks in the senior year do not significantly affect high school cohort sizes. 

Still, as unemployment is serially correlated, the senior-year deviation could still partly reflect the earlier shocks that inflated the cohort.\footnote{Online Appendix \Cref{tab:rob_lagur} adds the state URs two and three years before graduation as controls. Conditional on these lags, the senior-year effect is -2.06~pp overall and -1.91~pp at academic universities, nearly identical to the baseline on the same sample.} Replacing the share denominator with the population aged 18--19 addresses composition directly (Online Appendix \Cref{tab:rob_popdenom}). The share of a birth cohort obtaining the \textit{Abitur} does not respond to the senior-year UR. Per 18--19-year-old, a 1~pp higher senior-year UR reduces first-time enrollment at academic universities by 0.24, about 3 percent of the mean and proportionally the same response as in the share specification, with no effect at applied universities. The enrollment response thus reflects the choices of a given pool of graduates rather than upstream changes in its composition. A related concern is the secular expansion of the \textit{Abitur} itself. With rising participation, the marginal graduate is drawn from further down the achievement distribution, and average cohort quality may erode over time \citep{altonji2012changes}. The state-specific trends absorb this secular selection, so identification relies on cyclical deviations from it.

\subsection{Decomposition}
\label{sec:decomposition}

The literature distinguishes between two main mechanisms of how economic conditions affect enrollment: outside options and expected returns. Under imperfect mobility, more localized economic conditions should determine graduates' outside options. Yet, it remains an open question which type of information students actually use to form their beliefs. To study the distinct effect of national vs. state-specific conditions, I calculate state-specific shocks as $\text{UR}^{\Delta}_{s,t-1} = \text{UR}^\text{state}_{s,t-1} - \text{UR}^\text{national}_{t-1}$. I use the national UR and state-specific residuals as regressors instead of the state UR and estimate:

\vspace{-1em}
\begin{equation}
\label{eq:mechanisms}
E_{rt}^c \;=\;
\beta_1 \, \mathrm{UR}^\text{national}_{t-1}
\;+\; \beta_2 \, \mathrm{UR}^\Delta_{s(r),t-1}
\;+\; \alpha_{s(r)}
\;+\; g_{s(r)}(t)
\;+\; \gamma' Z_{s(r)t}
\;+\; \varepsilon_{rt}.
\end{equation}

\noindent where $\beta_1$ represents the part of the main effect explained by national conditions, and $\beta_2$ represents that part of the effect explained by state-specific residual shocks net of national conditions. 

\Cref{tab:abi_enroll_decomp} column 1 shows that the negative effect of the state UR on enrollment is mostly explained by its national component. A 1~pp deviation of the national UR from its secular trend decreases overall enrollment rates by 2.36~pp. This effect is mostly driven by enrollment at academic universities, while the effect of the national UR for applied universities is zero across samples. The pattern is robust to dropping the high school cohorts of 2017 and 2018 in column 2, for which only graduation below nominal study duration is observable in the exams register.

Column 3 shows that the attainment response follows the same pattern and even turns countercyclical for applied universities. A 1~pp deviation of the national UR reduces the share graduating from academic universities by 1.19~pp while increasing the share graduating from applied universities by 0.47~pp. Unlike for enrollment, state-specific deviations have no significant effect on attainment. For enrollment and attainment at academic universities, as well as for overall attainment, the negative effect of the national UR is significantly larger since 2008 (Online Appendix \Cref{tab:decomp_period}), consistent with the explanation that, following the Bologna Process, applied universities have become a more popular alternative to academic universities (see \Cref{fig:nationalUR_enroll_b}). The positive attainment response at applied universities also weakens after 2008, from 0.35~pp to 0.11~pp, although period-specific coefficients and their change are not statistically significant ($p$ = 0.14). Substitution thus operates at two margins in the scenario of \Cref{fig:statics}: from academic towards applied universities within higher education, and from applied universities towards apprenticeships.

\begin{table}[!h]
    \begin{center}
	\begin{adjustbox}{max width=\textwidth}
	\begin{threeparttable}
		\caption{Decomposed effects on enrollment and attainment}
		\label{tab:abi_enroll_decomp}
		\scriptsize{
		        \begin{tabular} {l ccc}
	\toprule \toprule
& \multicolumn{1}{c}{1995--2018} & \multicolumn{2}{c}{1995--2016}  \\
\cmidrule(lr){2-2} \cmidrule(lr){3-4}
& Enrollment & Enrollment & Attainment \\
& (1) & (2) & (3) \\
\midrule
\textit{Panel A: Overall (any college)} & & & \\
National UR (t-1)          &      -2.364***&      -2.363***&      -0.724*  \\
                           &     (0.433)   &     (0.410)   &     (0.384)   \\
$\text{UR}^\Delta$ (t-1)   &      -0.432   &      -0.672   &      -0.080   \\
                           &     (0.736)   &     (0.665)   &     (0.459)   \\
Outcome mean               &        71.2   &        71.0   &        59.9   \\
\addlinespace
\textit{Panel B: Academic university} & & & \\
National UR (t-1)          &      -2.369***&      -2.397***&      -1.193***\\
                           &     (0.449)   &     (0.419)   &     (0.239)   \\
$\text{UR}^\Delta$ (t-1)   &       0.011   &      -0.164   &      -0.014   \\
                           &     (0.545)   &     (0.482)   &     (0.408)   \\
Outcome mean               &        54.2   &        54.6   &        40.0   \\
\addlinespace
\textit{Panel C: Applied university} & & & \\
National UR (t-1)          &       0.006   &       0.034   &       0.469** \\
                           &     (0.072)   &     (0.076)   &     (0.200)   \\
$\text{UR}^\Delta$ (t-1)   &      -0.444*  &      -0.508** &      -0.066   \\
                           &     (0.235)   &     (0.224)   &     (0.220)   \\
Outcome mean               &        17.0   &        16.4   &        19.9   \\
\midrule
No. CZ-cohort cells       &       2,289   &       2,097   &       2,097   \\
No. high school graduates &   6,296,662   &   5,727,918   &   5,727,918   \\
State FE, trends, policy controls &      yes      &      yes      &      yes      \\
				\bottomrule
		\end{tabular}
		}
		\begin{tablenotes}[flushleft]
				\item \scriptsize{ \textit{Notes:} This table presents estimates from \Cref{eq:mechanisms}, splitting the senior-year ($t-1$) state unemployment rate into the national UR and the state-specific deviation $\text{UR}^\Delta$. The outcome is the share of high school graduates with \textit{Abitur} enrolling at (columns 1--2) or graduating from (column 3) any college (Panel A), academic universities (Panel B), or applied universities (Panel C). Column 1 spans high school cohorts 1995--2018. Columns 2 and 3 span cohorts 1995--2016, with attainment measured by 2019. Policy controls include tuition fee introductions and abolitions, as well as double cohorts. Each cell is weighted by the number of high school graduates with \textit{Abitur}. Standard errors in parentheses allow for two-way clustering at the state and cohort level. * $p$\textless 0.1, ** $p$\textless 0.05, *** $p$\textless 0.01. \\ \textit{Source}: Student register, exam register, Federal Employment Agency, \textit{Regionaldatenbank}.
                }
		\end{tablenotes}
	\end{threeparttable}
	\end{adjustbox}
    \end{center}
\end{table}

Conditional on the national UR, a 1~pp higher state-specific residual $\mathrm{UR}^\Delta_{s(r)t}$ has no significant effect on university enrollment, but significantly decreases enrollment at applied universities by 0.44 to 0.51~pp, depending on the sample and period. A general interpretation is that the marginal academic university student is more mobile and responds to changes in national conditions. Similarly, a marginal high school student indifferent between applied university and an apprenticeship may prefer apprenticeships during times of high unemployment, valuing, e.g., their good school-to-work transition. In contrast, the marginal student at applied universities would be less mobile and respond more to state-specific conditions.

To study effects driven by local, sub-state labor market conditions, Online Appendix \Cref{tab:abi_enroll_CZ} uses the more granular, cyclical variation in the CZ-level UR, with CZ fixed effects and CZ-specific trends in Panels A and B, and with the state fixed effects and trends of the main specification in Panels C and D. Panels B and D add cohort fixed effects. The results are overall smaller and noisier, suggesting that macroeconomic conditions are more salient in belief formation than local conditions. {Panels B and D of \Cref{tab:abi_enroll_CZ} also complete the year-fixed-effects comparison of \Cref{sec:setup}. Once cohort fixed effects absorb the national component of unemployment, the remaining local variation has no significant effect on overall or academic enrollment, and only the small applied margin responds negatively. This mirrors the decomposition result that enrollment response operates through the national component of unemployment, and (sub-)state deviations alone move enrollment little.

\section{Evidence from The Panel Study of School Leavers}
\label{sec:surveydata}

\subsection{Expected Returns}
\label{sec:expectations}

My main finding of procyclical college enrollment, concentrated at academic universities, is consistent with a Roy-type model described in \Cref{sec:model}, in which high school graduates choose between enrollment in academic universities, applied universities, and apprenticeships, based on their relative return. To explicitly measure the expected returns to university degrees vs. apprenticeship degrees as representing general and more specific human capital, respectively, I rely on the \textit{Panel Study of School Leavers with a Higher Education Entrance Qualification} \citep{daniel2017dzhw} of the German Centre for Higher Education Research and Science Studies (DZHW) described above. Unfortunately, the data do not cover beliefs about the returns to academic vs. applied university, so I focus on explaining the extensive margin of any college vs. no college. To mirror my main analysis and to avoid selective attrition\footnote{Attrition between the first two survey waves is severe and ranges from 66.1 percent of first-wave respondents in cohort 2012 to 78.9 percent in cohort 2008. The analysis therefore uses expectations and intentions as measured in the first wave, before graduation.}, I focus on expectations as measured in the calendar year before graduation and estimate the following model on the repeated cross-section of high school students scheduled to graduate in year $t$:

\vspace{-1em}
\begin{equation}
\label{eq:expectations}
Y^{e}_{ist} \;=\;
\beta\,\mathrm{UR}_{s,t-1}
\;+\;
\psi'\,X_{i}
\;+\;
\phi'\,\bigl(\mathrm{UR}_{s,t-1}\cdot X_{i}\bigr)
\;+\;
\alpha_s
\;+\;
h_s(t)
\;+\;
\gamma' Z_{st}
\;+\;
\varepsilon_{ist},
\end{equation}

\noindent where $Y^{e}_{ist}$ is the standardized value of either type of expectation $e$ (college, vocational education, relative return, and personal) for student $i$ in state $s$, $\alpha_s$ are state fixed effects, $\mathrm{UR}_{s,t-1}$ is the state UR, $Z_{st}$ are policy controls and, $X_i$ is a vector of individual-level controls (GPA, GPA squared, gender, and SES) and all their two-way and three-way interactions. A fourth outcome is the standardized value of personally expected job prospects, elicited in the same survey battery. To learn about heterogeneity, I additionally interact the state UR with the full set of interacted controls. In contrast to the main analysis, $h_s(t)=\theta_{1s}\,t+\theta_{2s}\,t^2$ are simple state-specific quadratic trends without a trend break, as the sample starts in 2008. With four cohorts, state fixed effects and state-specific trends leave limited residual variation. The detrended state UR has a standard deviation of 0.14~pp. Standard errors again allow for two-way clustering by state and cohort.

\begin{table}[!h]
    \begin{center}
	\begin{adjustbox}{max width=\textwidth}
	\begin{threeparttable}
		\caption{State UR effects on implicitly expected returns to college and personal job prospects}
		\label{tab:expectations_zreturns}
		\scriptsize{
		        \begin{tabular} {l cc cc cc cc}
	\toprule \toprule
    & \multicolumn{2}{c}{Vocational} & \multicolumn{2}{c}{Academic} & \multicolumn{2}{c}{Relative Returns} & \multicolumn{2}{c}{Personal} \\
  \cmidrule(lr){2-3} \cmidrule(lr){4-5} \cmidrule(lr){6-7} \cmidrule(lr){8-9}
 & (1) & (2) & (3) & (4) & (5) & (6) & (7) & (8)\\
\midrule
\textit{Panel A: State-level UR} & & & & & & & & \\
State UR (t-1)                  &       0.033   &       0.029   &      -0.122** &      -0.120** &      -0.114** &      -0.108**  &      -0.220***&      -0.221*** \\
                                &     (0.021)   &     (0.020)   &     (0.027)   &     (0.026)   &     (0.031)   &     (0.029)    &     (0.014)   &     (0.015)    \\
                                &      [0.230]  &      [0.266]  &      [0.021]  &      [0.020]  &      [0.016]  &      [0.013]   &      [0.002]  &      [0.002]   \\
Female                          &      -0.009   &      -0.039   &      -0.266***&      -0.271***&      -0.184** &      -0.164**  &      -0.292***&      -0.335*** \\
                                &     (0.029)   &     (0.026)   &     (0.034)   &     (0.028)   &     (0.040)   &     (0.035)    &     (0.024)   &     (0.022)    \\
Low SES                         &       0.080***&       0.054** &      -0.054** &      -0.044*  &      -0.100***&      -0.073*** &      -0.052** &      -0.082**  \\
                                &     (0.006)   &     (0.013)   &     (0.012)   &     (0.017)   &     (0.007)   &     (0.011)    &     (0.013)   &     (0.015)    \\
GPA                             &      -0.008   &      -0.022   &       0.095***&       0.103** &       0.074** &       0.091*** &      0.283*** &      0.308***  \\
                                &     (0.019)   &     (0.015)   &     (0.011)   &     (0.021)   &     (0.016)   &     (0.010)    &     (0.020)   &     (0.021)    \\
$\text{GPA}^2$                  &      -0.013** &      -0.017*  &      -0.003   &      -0.008   &       0.008   &       0.008    &      -0.007** &      -0.008    \\
                                &     (0.003)   &     (0.006)   &     (0.005)   &     (0.005)   &     (0.004)   &     (0.005)    &     (0.001)   &     (0.004)    \\
&&&&&&&&\\
\textit{Panel B: Decomposed UR} & & & & & & & & \\
National UR (t-1)               &       0.033   &       0.010   &      -0.190***&      -0.169***&      -0.162** &      -0.129**  &      -0.294***&      -0.296*** \\
                                &     (0.033)   &     (0.032)   &     (0.025)   &     (0.025)   &     (0.030)   &     (0.033)    &     (0.016)   &     (0.022)    \\
$\text{UR}^{\Delta}$ (t-1)      &       0.034   &       0.039   &       0.034   &       0.028   &      -0.002   &      -0.010    &      -0.047   &      -0.051    \\
                                &     (0.067)   &     (0.064)   &     (0.043)   &     (0.043)   &     (0.050)   &     (0.047)    &     (0.031)   &     (0.028)    \\
&&&&&&&&\\
                No. students & \multicolumn{2}{c}{104,666} & \multicolumn{2}{c}{104,666} & \multicolumn{2}{c}{104,666} & \multicolumn{2}{c}{97,316} \\
                State FE, trends, policy controls & yes & yes & yes & yes & yes & yes & yes & yes \\
                Interactions                & no & yes & no & yes & no & yes & no & yes \\
				\bottomrule
		\end{tabular}
		}
		\begin{tablenotes}[flushleft]
				\item \scriptsize{ \textit{Notes:} This table presents estimates from \Cref{eq:expectations} for the effect of the state-level unemployment rate (Panel A) and its components (Panel B) on the standardized value of the expected returns to a vocational degree (apprenticeship), to an academic degree (college), and the implicitly expected relative returns to college, one year before high school graduation. Columns (7) and (8) use the standardized personally expected job prospects as outcome. Item nonresponse on this question reduces its sample to 97,316 students. Controls include gender, SES, and the current high school GPA (standardized), as well as its squared value. Interactions of the UR with the full set of controls are presented in \Cref{fig:margins_returns}. Standard errors in parentheses allow for two-way clustering at the state and cohort level. In columns (2), (4), and (6), the UR is centered at its sample mean, so the coefficients on the controls give their association at the mean UR. Wild cluster bootstrap $p$-values in brackets impose the null and use 9,999 replications with Webb weights, clustering by state. * $p$\textless 0.1, ** $p$\textless 0.05, *** $p$\textless 0.01. \\ \textit{Source}: Federal Employment Agency (BA), and DZHW, years 2008, 2012, 2015, and 2018.
                }
		\end{tablenotes}
	\end{threeparttable}
	\end{adjustbox}
    \end{center}
\end{table}

\Cref{tab:expectations_zreturns} presents the effect of the state UR one year before high school graduation on the expected value of a vocational degree, the expected value of an academic degree, and the implicitly expected relative returns to college. Panel A column 1 shows that a 1~pp increase in the state UR has no significant effect on the expected value of a vocational degree, suggesting that students' beliefs about the value of specific skills are stable over the business cycle. The same state UR fluctuation decreases the expected value of a university degree by 0.12 SD. As a result, the implicitly expected general returns to college are significantly procyclical and thus likely explain the procyclical pattern in college enrollment. Given that students were asked about college in general, the negative effect for academic universities is likely to be even larger, which may explain the smaller enrollment effect for applied universities. 

To examine whether students form their beliefs based on aggregate or more local conditions, Panel B repeats the analysis as in \Cref{sec:decomposition} and includes both the national UR and state-specific deviations. While the effect of state-specific fluctuations is small and insignificant throughout, the effect of the national unemployment rate again follows the same pattern as in Panel A. Conditional on covariates, a 1~pp increase in the national UR decreases the expected value of a college degree by 0.19 SD. It decreases the implicitly expected relative returns by 0.16 SD. This dominant role of aggregate labor market conditions for subjective expectations aligns with the findings for enrollment choices in \Cref{sec:decomposition}.

Personal job prospects respond even more strongly than the beliefs about either educational track. A 1~pp increase in the state UR lowers personally expected job prospects by 0.22 SD (columns 7 and 8), and this response again runs almost entirely through the national component (0.29 SD, Panel B). To the extent that students interpret the survey question about their personal job prospects as an ad-hoc statement about a scenario without further education, smaller coefficients for the general value of degrees may partly capture beliefs about the future state of the economy at (college or apprenticeship) graduation.\footnote{For students observed in both waves, within-person changes in the state UR between the two interviews leave all three expectation measures unchanged. This supports measuring the treatment in the year before graduation, as in \Cref{eq:specification} (Online Appendix \Cref{tab:updating}).} The expectations estimates are unchanged when restricting the sample to students who answer the personal-prospects question (Online Appendix \Cref{tab:expectations_peranswer}).

\begin{figure}[!htbp]
	\centering
	\begin{minipage}{\textwidth} 
	\caption{Marginal state UR effects on implicitly expected returns to college by covariates}
    \vspace{-1em}
 	\label{fig:margins_returns}
	\begin{center}
	\includegraphics[scale=0.7]{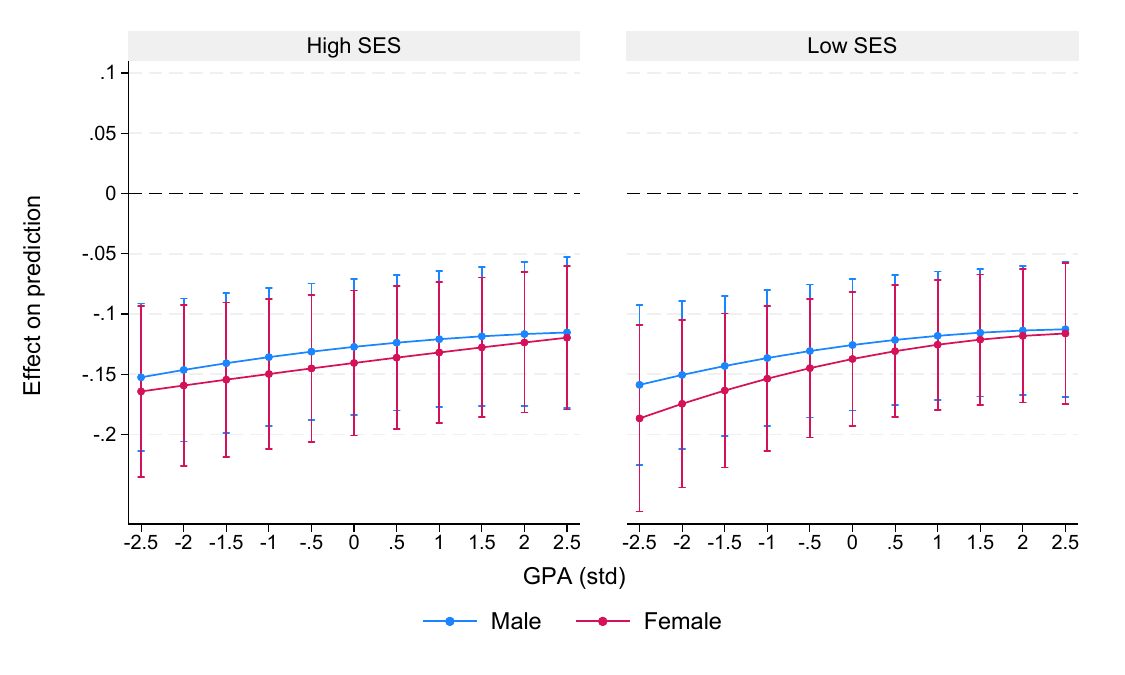} \\
	\end{center}
    \vspace{-1em}
    {\scriptsize \textit{Notes:} This figure plots the marginal effect of a positive 1~pp deviation of the state UR from its trend one year before high school graduation on the implicitly expected relative return to a college degree. Estimates are based on the fully interacted model presented in  \Cref{tab:expectations_zreturns}. \\ \textit{Source}: Federal Employment Agency (BA), and DZHW, years 2007, 2011, 2014, and 2017.
    \par}
    \end{minipage}
\end{figure}

\Cref{fig:margins_returns} plots marginal effects of a state-level labor market shock along the GPA distribution, separately by gender and SES. The effects are larger among female students, who also have lower baseline expected returns, and they decrease in GPA. This suggests that high-performing students are either better informed about the cyclical nature of vocational outside options or more confident about their labor market outcomes, irrespective of the business cycle. The gender pattern is consistent with the differential selection into the \textit{Abitur} discussed in \Cref{sec:robustness}: men who obtain the qualification are more strongly selected on study commitment, so the marginal student is more often female. Online Appendix \Cref{fig:margins_separate} repeats this exercise separately for the expected value of a college degree and of a vocational degree. Low-performing students, and women without college-educated parents in particular, even expect a significantly higher return to a vocational degree in the labor market when unemployment is high. This could imply that they are either less informed about the cyclicality of general vs. specific skills, or are primarily concerned with finding any job after (postsecondary) graduation. If true, they may place greater value on the smooth school-to-work transition typically associated with apprenticeships.

\subsection{Enrollment Intentions}
\label{sec:intentions}

To study changes in expectations as a mechanism for postsecondary skill investments, I further use the Panel Study of School Leavers to (i) study their reduced-form effect on study intentions and (ii) use UR fluctuations as an instrument for expectations with study intentions as second-stage outcome. Panel A of \Cref{tab:intentions} presents estimates of \Cref{eq:expectations} with the intention indicators described in \Cref{sec:data} as outcomes, on the estimation sample of the expectations analysis. A 1~pp increase in the senior-year state UR reduces the probability of intending to enroll by 3.0~pp without individual controls, and by 5.8~pp with individual controls. Both are substantially larger than the enrollment effects estimated using administrative data. A typical cyclical deviation of the state UR, 0.14~pp after removing state-specific trends, moves the intention probability by about 0.8~pp. The response is again concentrated at academic universities, although the institution-type split is less precisely estimated. As for expected returns, the response is larger for students with lower grades, and similar across gender and parental education (Online Appendix \Cref{fig:margins_intentions}). 

Decomposing the state UR into its national and state-specific components attributes the statistically significant part of the intention response to the national component ($-0.060$, wild-cluster-bootstrap $p$ = 0.040), while the state-specific coefficient is of similar size but imprecise ($-0.054$, $p$ = 0.18), in line with \Cref{sec:decomposition}. Online Appendix \Cref{tab:intentions_specs} shows that including year fixed effects and identifying from within-cohort cross-state variation alone yields a smaller but statistically significant response. Moreover, the estimate is robust to linear instead of quadratic state trends. Estimates stay significant when using alternative inference methods that account for the small number of clusters. 

\begin{table}[!h]
    \begin{center}
	\begin{adjustbox}{max width=0.95\textwidth}
	\begin{threeparttable}
		\caption{State UR, expected returns, and enrollment intentions}
		\label{tab:intentions}
		\scriptsize{
		        \begin{tabular} {l cc cc cc}
	\toprule \toprule
    & \multicolumn{6}{c}{Intends to enroll} \\
    \cmidrule(lr){2-7}
    & \multicolumn{2}{c}{Any} & \multicolumn{2}{c}{Academic} & \multicolumn{2}{c}{Applied} \\
    \cmidrule(lr){2-3} \cmidrule(lr){4-5} \cmidrule(lr){6-7}
 & (1) & (2) & (3) & (4) & (5) & (6)\\
\midrule
\textit{Panel A: Reduced form} &&&&&&\\
State UR (t-1)          &      -0.030** &      -0.058***&      -0.013   &      -0.041*  &      -0.010   &      -0.010   \\
                        &     (0.009)   &     (0.008)   &     (0.021)   &     (0.014)   &     (0.020)   &     (0.020)   \\
Wild cluster bootstrap $p$ &    0.018   &       0.010   &       0.624   &       0.116   &       0.728   &       0.719   \\
Outcome mean            &        0.65   &        0.65   &        0.39   &        0.39   &        0.19   &        0.19   \\
No. students            &     104,666   &     104,666   &     101,500   &     101,500   &     101,500   &     101,500   \\
Individual controls     &      No       &      Yes      &      No       &      Yes      &      No       &      Yes      \\
&&&&&&\\
\textit{Panel B: Expected relative returns} &&&&&&\\
OLS                     & \multicolumn{2}{c}{0.099***} & \multicolumn{2}{c}{0.059***} & \multicolumn{2}{c}{0.034***}\\
                        & \multicolumn{2}{c}{(0.011)}  & \multicolumn{2}{c}{(0.007)}  & \multicolumn{2}{c}{(0.005)} \\
2SLS                    & \multicolumn{2}{c}{0.428***} & \multicolumn{2}{c}{0.291}    & \multicolumn{2}{c}{0.090}   \\
                        & \multicolumn{2}{c}{(0.108)}  & \multicolumn{2}{c}{(0.177)}  & \multicolumn{2}{c}{(0.144)} \\
Anderson-Rubin 95\% conf.\ set & \multicolumn{2}{c}{$[0.26, 1.25]$} & \multicolumn{2}{c}{$[-0.07, 3.60]$} & \multicolumn{2}{c}{$[-2.59, 0.43]$} \\
First-stage $F$         & \multicolumn{2}{c}{13.1}     & \multicolumn{2}{c}{12.8}     & \multicolumn{2}{c}{12.8}    \\
No. students            & \multicolumn{2}{c}{114,162}  & \multicolumn{2}{c}{110,778}  & \multicolumn{2}{c}{110,778} \\
				\bottomrule
		\end{tabular}
		}
		\begin{tablenotes}[flushleft]
				\item \scriptsize{ \textit{Notes:} This table presents estimates from \Cref{eq:expectations} with study intention indicators as outcomes. Panel A uses the estimation sample of \Cref{tab:expectations_zreturns}. The intention indicators are 1 if the student intends to enroll certainly or probably (columns 1--2), and among these names an academic (columns 3--4) or applied (columns 5--6) institution. Panel A regresses the indicators on the state UR one year before graduation, without and with individual controls (gender, SES, standardized GPA, and its square). Outcome means refer to the estimation sample. Panel B regresses the indicators on the standardized expected relative return to college, by OLS and by 2SLS with the state UR one year before graduation as the instrument, with individual controls. Panel B additionally keeps students whose relative return is coded from a ``do not know'' response on one expectation item (\Cref{sec:data}). The Anderson-Rubin confidence sets impose the null and use the wild cluster bootstrap by state \citep{davidson2014confidence}. First-stage $F$ statistics are cluster-robust. All specifications include state fixed effects, state-specific quadratic trends, and policy controls. Standard errors in parentheses allow for two-way clustering at the state and cohort level in Panel A and the OLS rows, and for clustering by state in the 2SLS rows. Wild cluster bootstrap $p$-values from a bootstrap by state with 9,999 replications and Webb weights.  * $p$\textless 0.1, ** $p$\textless 0.05, *** $p$\textless 0.01. \\ \textit{Source}: Federal Employment Agency (BA), and DZHW, cohorts 2008, 2012, 2015, and 2018.
                }
		\end{tablenotes}
	\end{threeparttable}
	\end{adjustbox}
    \end{center}
\end{table}

In Panel B, a 1~SD higher expected return is associated with a 9.9~pp higher probability of intending to enroll. Using the senior-year state UR as an instrument for expected returns yields a 2SLS estimate of 0.43, with a first stage of $-0.134$, a cluster-robust $F$ statistic of 13.1, and a weak-instrument-robust Anderson-Rubin confidence set of $[0.26, 1.25]$ \citep{davidson2014confidence, roodman2019fast}. The corresponding estimates for the academic and applied margins carry the expected signs but are not statistically distinguishable from zero. The estimates are similar, and the first stage is stronger among students who also answer the personal-prospects question (Online Appendix \Cref{tab:iv_peranswer}). 

Realized choices point in the same direction as intentions but are estimated with less precision. In the attrition-reduced follow-up sample, and excluding men who are affected by conscription until 2011, a 1~pp higher senior-year state UR reduces the probability of being enrolled half a year after graduation by 2.6~pp. However, the effect is not statistically significant, with similar estimates for women and men (Online Appendix \Cref{tab:sbp_enrolled}). Instrumenting expected returns in this sample yields positive but imprecise estimates (Online Appendix \Cref{tab:iv_expectations}). The intention estimates are therefore the more reliable individual-level evidence.

The exclusion restriction requires that unemployment affects intentions only via expected returns. Even though I cannot explicitly rule out alternative channels, common competing channels do not show the same pattern. Online Appendix \Cref{tab:costinfl} shows that students report a lower influence of study costs on their enrollment decision when unemployment is high, which speaks against a binding financing constraint. This matches the evidence that capacity constraints (\Cref{sec:capacity}) and credit constraints (\Cref{sec:credit}) play a secondary role at best in explaining procyclical enrollment.

The expectations channel states that current unemployment moves beliefs about the future value of a specific degree on the labor market.  However, cyclical fluctuations in labor demand today are not a perfect signal for future labor demand. Even though today's conditions may still be a good proxy for conditions in the near future, after an apprenticeship, or fast Bachelor's studies within three years, they become less reliable after e.g. second-chance studies or a Master's degree (see Online Appendix \Cref{tab:ur_persistence} for details).   Against this background, it is unlikely that high school graduates act as forward-looking ``adolescent econometricians'' \citep{manski1993adolescent} and form rational expectations about the returns to different skills several years after graduation. Instead, my findings are consistent with an extrapolation from current conditions rather than forward-looking updating, in line with experience-based belief formation \citep{malmendier2011nagel, malmendier2016learning, nagel2026experiences}. Given the high degree of uncertainty about the future state of the economy, investing in specific skills and betting on a good school-to-work transition associated with specific skills may seem like a safe choice for students. Yet, over the lifecycle, this belief is at odds with the finding that workers with more general skills and higher levels of education typically suffer less during recessions \citep[see][respectively]{hoynes2012recessions, altonji2016cashier, eggenberger2022specific}.

\section{Conclusion}
\label{sec:conclusion}

This paper examines how labor market conditions at high school graduation affect postsecondary skill investments, distinguishing between general academic education and more specific vocational pathways. Using administrative microdata on the universe of students in German higher education, I find that higher unemployment at graduation reduces enrollment at academic universities and lowers long-run attainment there, while attainment at applied universities rises. Complementary administrative data on the apprenticeship market suggest that apprenticeships absorb part of the foregone college enrollment, although these data cannot be linked to the same graduation cohorts.

My empirical analysis shows that high school graduates' skill investments respond mainly to aggregate fluctuations, consistent with the role of salient macroeconomic conditions in experience-based beliefs updating. Survey data on subjective expectations show that for the average student, downturns reduce the expected value of an academic degree, while low-performing students, and women without college-educated parents in particular, even expect greater returns to a vocational degree in a recession. Capacity and credit constraints only play a negligible role.

My findings contrast with conventional wisdom on countercyclical college-going and demonstrate that downturns can cause a shift to investments in more specific skills. The results highlight that students' adjustment to business cycle conditions depends critically on the available outside options. In systems with established vocational pathways, recessions can shift students toward investments in specific skills, while overall college enrollment declines. Countries that wish to adopt elements of the German apprenticeship system should be aware of how these changes affect the responsiveness of human capital investments to economic shocks. Ultimately, my findings highlight the need to clearly distinguish between the effects of macroeconomic and local conditions on human capital investment.


\FloatBarrier
\newpage

\titlespacing{\section}{0ex}{1.5ex}{1.5ex}
\newpage
\begin{spacing}{2.0}
\bibliography{library}

@article{charles2018housing,
  title={Housing Booms and Busts, Labor Market Opportunities, and College Attendance},
  author={Charles, Kerwin Kofi and Hurst, Erik and Notowidigdo, Matthew J},
  journal={American Economic Review},
  volume={108},
  number={10},
  pages={2947--94},
  year={2018}
}

@article{betts1995safe,
  title={Safe Port in a Storm: The Impact of Labor Market Conditions on Community College Enrollments},
  author={Betts, Julian R and McFarland, Laurel L},
  journal={Journal of Human Resources},
  pages={741--765},
  year={1995},
  publisher={JSTOR}
}

@article{blom2021investment,
  title={Investment Over the Business Cycle: Insights from College Major Choice},
  author={Blom, Erica and Cadena, Brian C and Keys, Benjamin J},
  journal={Journal of Labor Economics},
  volume={39},
  number={4},
  pages={1043--1082},
  year={2021},
  publisher={The University of Chicago Press Chicago, IL}
}

@article{lovenheim2011effect,
  title={The Effect of Liquid Housing Wealth on College Enrollment},
  author={Lovenheim, Michael F},
  journal={Journal of Labor Economics},
  volume={29},
  number={4},
  pages={741--771},
  year={2011},
  publisher={University of Chicago Press Chicago, IL}
}

@article{hampf2020effects,
  title={The Effects of Graduating from High School in a Recession: College Investments, Skill Formation, and Labor-Market Outcomes},
  author={Hampf, Franziska and Piopiunik, Marc and Wiederhold, Simon},
  year={2020},
  publisher={CESifo},
  journal={CESifo Working Paper No. 8252}
}

@article{graves2022higher,
  title={Higher Education Decisions and Macroeconomic Conditions at Age Eighteen},
  author={Graves, Jennifer and Kuehn, Zo{\"e}},
  journal={SERIEs},
  volume={13},
  number={1},
  pages={171--241},
  year={2022},
  publisher={Springer}
}

@article{dellas2003cyclical,
 ISSN = {00307653, 14643812},
 URL = {http://www.jstor.org/stable/3488876},
 author = {Harris Dellas and Plutarchos Sakellaris},
 journal = {Oxford Economic Papers},
 number = {1},
 pages = {148--172},
 publisher = {Oxford University Press},
 title = {On the Cyclicality of Schooling: Theory and Evidence},
 urldate = {2022-11-17},
 volume = {55},
 year = {2003}
}

@article{johnson2013graduate,
title = {The Impact of Business Cycle Fluctuations on Graduate School Enrollment},
journal = {Economics of Education Review},
volume = {34},
pages = {122-134},
year = {2013},
issn = {0272-7757},
doi = {https://doi.org/10.1016/j.econedurev.2013.02.002},
url = {https://www.sciencedirect.com/science/article/pii/S0272775713000290},
author = {Matthew T. Johnson}
}

@article{weinstein2022local,
  title={Local Labor Markets and Human Capital Investments},
  author={Weinstein, Russell},
  journal={Journal of Human Resources},
  volume={57},
  number={5},
  pages={1498--1525},
  year={2022},
  publisher={University of Wisconsin Press}
}

@article{ersoy2020effects,
  title={The Effects of the Great Recession on College Majors},
  author={Ersoy, Fulya Y},
  journal={Economics of Education Review},
  volume={77},
  pages={102018},
  year={2020},
  publisher={Elsevier}
}

@article{acton2021community,
  title={Community College Program Choices in the Wake of Local Job Losses},
  author={Acton, Riley K},
  journal={Journal of Labor Economics},
  volume={39},
  number={4},
  pages={1129--1154},
  year={2021},
  publisher={The University of Chicago Press Chicago, IL}
}

@article{han2020industry,
  title={Industry Fluctuations and College Major Choices: Evidence From an Energy Boom and Bust},
  author={Han, Luyi and Winters, John V},
  journal={Economics of Education Review},
  volume={77},
  pages={101996},
  year={2020},
  publisher={Elsevier}
}

@Misc{middendorff2017wirtschaftliche,
  title={Die wirtschaftliche und soziale Lage der Studierenden in Deutschland 2016. 21},
  author={Middendorff, Elke and Apolinarski, Beate and Becker, Karsten and Bornkessel, Philipp and Brandt, Tasso and Hei{\ss}enberg, Sonja and Poskowsky, Jonas},
  howpublished    = {{Bonn/Berlin: German Federal Ministry of Education and Research}},
  pages={2018},
  year={2017}
}

@article{fidan2022loanaversion,
author = {Muervet Fidan and Christian Manger},
title = {Why do German students reject free money?},
journal = {Education Economics},
volume = {30},
number = {3},
pages = {303-319},
year  = {2022},
publisher = {Routledge},
doi = {10.1080/09645292.2021.1978937},
URL = {https://doi.org/10.1080/09645292.2021.1978937},
eprint = {https://doi.org/10.1080/09645292.2021.1978937}
}

@article{bulman2015returns,
  title={The returns to the federal tax credits for higher education},
  author={Bulman, George and Hoxby, Caroline M},
  journal={Tax Policy and the Economy},
  volume={29},
  number={1},
  pages={13--88},
  year={2015},
  publisher={University of Chicago Press Chicago, IL}
}

@article{deming2020stemskills,
    author = {Deming, David J and Noray, Kadeem},
    title = {Earnings Dynamics, Changing Job Skills, and STEM Careers},
    journal = {The Quarterly Journal of Economics},
    volume = {135},
    number = {4},
    pages = {1965-2005},
    year = {2020},
    month = {06},
    issn = {0033-5533},
    doi = {10.1093/qje/qjaa021},
    url = {https://doi.org/10.1093/qje/qjaa021},
    eprint = {https://academic.oup.com/qje/article-pdf/135/4/1965/33668545/qjaa021.pdf},
}

@article{christian2007liquidity,
 ISSN = {00307653, 14643812},
 URL = {http://www.jstor.org/stable/4500092},
 author = {Michael S. Christian},
 journal = {Oxford Economic Papers},
 number = {1},
 pages = {141--169},
 publisher = {Oxford University Press},
 title = {Liquidity Constraints and the Cyclicality of College Enrollment in the United States},
 urldate = {2022-12-09},
 volume = {59},
 year = {2007}
}

@article{dellas2003business,
  title={Business cycles and schooling},
  author={Dellas, Harris and Koubi, Vally},
  journal={European Journal of Political Economy},
  volume={19},
  number={4},
  pages={843--859},
  year={2003},
  publisher={Elsevier}
}

@article{hazarika2002role,
  title={The role of credit constraints in the cyclicality of college enrolments},
  author={Hazarika, Gautam},
  journal={Education Economics},
  volume={10},
  number={2},
  pages={133--143},
  year={2002},
  publisher={Taylor \& Francis}
}

@article{bietenbeck2023tuition,
title = {Tuition fees and educational attainment},
journal = {European Economic Review},
volume = {154},
pages = {104431},
year = {2023},
issn = {0014-2921},
doi = {https://doi.org/10.1016/j.euroecorev.2023.104431},
url = {https://www.sciencedirect.com/science/article/pii/S0014292123000600},
author = {Jan Bietenbeck and Andreas Leibing and Jan Marcus and Felix Weinhardt}
}

@article{altonji2016cashier,
author = {Altonji, Joseph G. and Kahn, Lisa B. and Speer, Jamin D.},
title = {Cashier or Consultant? Entry Labor Market Conditions, Field of Study, and Career Success},
journal = {Journal of Labor Economics},
volume = {34},
number = {S1},
pages = {S361-S401},
year = {2016},
doi = {10.1086/682938},
URL = {https://doi.org/10.1086/682938},
eprint = {https://doi.org/10.1086/682938}
}

@incollection{card2001dropout,
  title={Dropout and enrollment trends in the postwar period: What went wrong in the 1970s?},
  author={Card, David and Lemieux, Thomas},
  booktitle={Risky behavior among youths: An economic analysis},
  pages={439--482},
  year={2001},
  publisher={University of Chicago Press}
}

@article{gaini2013shelter,
 ISSN = {21154430, 19683863},
 URL = {http://www.jstor.org/stable/23646333},
 author = {Mathilde Gaini and Aude Leduc and Augustin Vicard},
 journal = {Annals of Economics and Statistics},
 volume = {111/112},
 pages = {251--270},
 publisher = {[GENES, ADRES]},
 title = {School as a Shelter? School Leaving-Age and the Business Cycle in France},
 urldate = {2022-12-15},
 year = {2013}
}

@article{hanushek2017general,
  title={General education, vocational education, and labor-market outcomes over the lifecycle},
  author={Hanushek, Eric A and Schwerdt, Guido and Woessmann, Ludger and Zhang, Lei},
  journal={Journal of Human Resources},
  volume={52},
  number={1},
  pages={48--87},
  year={2017},
  publisher={University of Wisconsin Press}
}

@incollection{long2014financial,
  title={The financial crisis and college enrollment: How have students and their families responded?},
  author={Long, Bridget Terry},
  booktitle={How the financial crisis and Great Recession affected higher education},
  pages={209--233},
  year={2014},
  publisher={University of Chicago Press}
}

@article{hillman2013community,
  title={Community colleges and labor market conditions: How does enrollment demand change relative to local unemployment rates?},
  author={Hillman, Nicholas W and Orians, Erica Lee},
  journal={Research in Higher Education},
  volume={54},
  number={7},
  pages={765--780},
  year={2013},
  publisher={Springer}
}

@article{alessandrini2018port,
title = {Is post-secondary education a safe port and for whom? Evidence from Canadian data},
journal = {Economics of Education Review},
volume = {67},
pages = {1-13},
year = {2018},
issn = {0272-7757},
doi = {https://doi.org/10.1016/j.econedurev.2018.09.005},
url = {https://www.sciencedirect.com/science/article/pii/S0272775716301005},
author = {Diana Alessandrini}
}

@Misc{RDC2019a,
  author       = {{RDC}},
  year         = {2019a},
  title        = {{FDZ der Statistischen \"Amter des Bundes und der L\"ander: Statistik der Studierenden, 1995--2019}},
  howpublished = {DOI: 10.21242/21311.1996.12.00.1.1.0--10.21242/21311.2020.12.00.1.1.0 (one DOI per survey year)},
}

@Misc{RDC2019b,
  author       = {{RDC}},
  year         = {2019b},
  title        = {{FDZ der Statistischen \"Amter des Bundes und der L\"ander: Statistik der Pr\"ufungen, 1995--2019}},
  howpublished = {DOI: 10.21242/21321.1996.12.00.1.1.0--10.21242/21321.2020.12.00.1.1.0 (one DOI per survey year)},
}

@article{hodrick1997postwar,
  title={Postwar {U.S.} business cycles: an empirical investigation},
  author={Hodrick, Robert J. and Prescott, Edward C.},
  journal={Journal of Money, Credit and Banking},
  volume={29},
  number={1},
  pages={1--16},
  year={1997}
}

@article{ravn2002adjusting,
  title={On adjusting the {Hodrick--Prescott} filter for the frequency of observations},
  author={Ravn, Morten O. and Uhlig, Harald},
  journal={The Review of Economics and Statistics},
  volume={84},
  number={2},
  pages={371--376},
  year={2002}
}

@article{hamilton2018why,
  title={Why you should never use the {Hodrick--Prescott} filter},
  author={Hamilton, James D.},
  journal={The Review of Economics and Statistics},
  volume={100},
  number={5},
  pages={831--843},
  year={2018}
}

@article{amior2018persistence,
  title={The persistence of local joblessness},
  author={Amior, Michael and Manning, Alan},
  journal={American Economic Review},
  volume={108},
  number={7},
  pages={1942--70},
  year={2018}
}

@article{krebs2023road,
  title={On the road (again): Commuting and local employment elasticities in germany},
  author={Krebs, Oliver and Pfl{\"u}ger, Michael},
  journal={Regional Science and Urban Economics},
  pages={103874},
  year={2023},
  publisher={Elsevier}
}

@article{bicakova2021cycle,
    author = {Bi{\v{c}}áková, Alena and Cortes, Guido Matias and Mazza, Jacopo},
    title = "{Caught in the Cycle: Economic Conditions at Enrolment and Labour Market Outcomes of College Graduates}",
    journal = {The Economic Journal},
    volume = {131},
    number = {638},
    pages = {2383-2412},
    year = {2021},
    month = {01},
    issn = {0013-0133},
    doi = {10.1093/ej/ueab003},
    url = {https://doi.org/10.1093/ej/ueab003},
    eprint = {https://academic.oup.com/ej/article-pdf/131/638/2383/39509660/ueab003.pdf},
}

@article{leibing2023gender,
title = {Gender gaps in early wage expectations},
journal = {Economics of Education Review},
volume = {94},
pages = {102398},
year = {2023},
issn = {0272-7757},
doi = {https://doi.org/10.1016/j.econedurev.2023.102398},
url = {https://www.sciencedirect.com/science/article/pii/S0272775723000456},
author = {Andreas Leibing and Frauke Peter and Sevrin Waights and C. Katharina Spiess}
}

@article{pan2014parents,
title = {The Impact of Parental Layoff on Higher Education Investment},
journal = {Economics of Education Review},
volume = {42},
pages = {53-63},
year = {2014},
issn = {0272-7757},
doi = {https://doi.org/10.1016/j.econedurev.2014.06.006},
url = {https://www.sciencedirect.com/science/article/pii/S0272775714000636},
author = {Weixiang Pan and Ben Ost}
}

@article{becker1962investment,
 ISSN = {00223808, 1537534X},
 URL = {http://www.jstor.org/stable/1829103},
 author = {Gary S. Becker},
 journal = {Journal of Political Economy},
 number = {5},
 pages = {9--49},
 publisher = {University of Chicago Press},
 title = {Investment in Human Capital: A Theoretical Analysis},
 urldate = {2023-02-14},
 volume = {70},
 year = {1962}
}

@article{hoynes2012recessions,
Author = {Hoynes, Hilary and Miller, Douglas L. and Schaller, Jessamyn},
Title = {Who Suffers during Recessions?},
Journal = {Journal of Economic Perspectives},
Volume = {26},
Number = {3},
Year = {2012},
Month = {September},
Pages = {27-48},
DOI = {10.1257/jep.26.3.27},
URL = {https://www.aeaweb.org/articles?id=10.1257/jep.26.3.27}
}

@article{bulman2021lottery,
Author = {Bulman, George and Fairlie, Robert and Goodman, Sarena and Isen, Adam},
Title = {Parental Resources and College Attendance: Evidence from Lottery Wins},
Journal = {American Economic Review},
Volume = {111},
Number = {4},
Year = {2021},
Month = {April},
Pages = {1201-40},
DOI = {10.1257/aer.20171272},
URL = {https://www.aeaweb.org/articles?id=10.1257/aer.20171272}}

@article{manoli2018cash,
Author = {Manoli, Day and Turner, Nicholas},
Title = {Cash-on-Hand and College Enrollment: Evidence from Population Tax Data and the Earned Income Tax Credit},
Journal = {American Economic Journal: Economic Policy},
Volume = {10},
Number = {2},
Year = {2018},
Month = {May},
Pages = {242–71},
DOI = {10.1257/pol.20160298},
URL = {https://www.aeaweb.org/articles?id=10.1257/pol.20160298}}

@article{cameron2004estimation,
  title={Estimation of Educational Borrowing Constraints Using Returns to Schooling},
  author={Cameron, Stephen V and Taber, Christopher},
  journal={Journal of Political Economy},
  volume={112},
  number={1},
  pages={132--182},
  year={2004},
  publisher={The University of Chicago Press}
}

@article{hilger2016parental,
  title={Parental job loss and children's long-term outcomes: Evidence from 7 million fathers' layoffs},
  author={Hilger, Nathaniel G},
  journal={American Economic Journal: Applied Economics},
  volume={8},
  number={3},
  pages={247--283},
  year={2016},
  publisher={American Economic Association 2014 Broadway, Suite 305, Nashville, TN 37203-2425}
}

@article{mendez2012cyclicality,
  title={The cyclicality of skill acquisition: evidence from panel data},
  author={M{\'e}ndez, Fabio and Sep{\'u}lveda, Facundo},
  journal={American Economic Journal: Macroeconomics},
  volume={4},
  number={3},
  pages={128--152},
  year={2012},
  publisher={American Economic Association}
}

@article{malmendier2011nagel,
    author = {Malmendier, Ulrike and Nagel, Stefan},
    title = "{Depression Babies: Do Macroeconomic Experiences Affect Risk Taking?}",
    journal = {The Quarterly Journal of Economics},
    volume = {126},
    number = {1},
    pages = {373-416},
    year = {2011},
    month = {02},
    issn = {0033-5533},
    doi = {10.1093/qje/qjq004},
    url = {https://doi.org/10.1093/qje/qjq004},
    eprint = {https://academic.oup.com/qje/article-pdf/126/1/373/17088890/qjq004.pdf},
}

@incollection{giustinelli2023expectations,
title = {Expectations in Education},
editor = {Rüdiger Bachmann and Giorgio Topa and Wilbert {van der Klaauw}},
booktitle = {Handbook of Economic Expectations},
publisher = {Academic Press},
pages = {193-224},
year = {2023},
chapter = {7},
isbn = {978-0-12-822927-9},
doi = {https://doi.org/10.1016/B978-0-12-822927-9.00014-8},
url = {https://www.sciencedirect.com/science/article/pii/B9780128229279000148},
author = {Pamela Giustinelli},
}

@article{destatis2020,
  Title                    = {{Nichtmonetäre hochschulstatistische Kennzahlen}},
  Author                   = {Destatis},
  Journal                  = {{Fachserie 11 Reihe 4.3.1}},
  Year                     = {2020},
  Note                     = {{Statistisches Bundesamt, Wiesbaden}}
}

@article{bicakova2023luck,
title = {Make your own luck: The wage gains from starting college in a bad economy},
author = {Bi{\v{c}}áková, Alena and Cortes, Guido Matias and Mazza, Jacopo},
journal = {Labour Economics},
volume = {84},
pages = {102411},
year = {2023},
issn = {0927-5371},
doi = {https://doi.org/10.1016/j.labeco.2023.102411},
url = {https://www.sciencedirect.com/science/article/pii/S0927537123000866}
}

@article{sievertsen2016local,
  title={Local unemployment and the timing of post-secondary schooling},
  author={Sievertsen, Hans Henrik},
  journal={Economics of Education Review},
  volume={50},
  pages={17--28},
  year={2016},
  publisher={Elsevier}
}

@article{luthi2020apprenticeships,
  title={Are apprenticeships business cycle proof?},
  author={Lüthi, Samuel and Wolter, Stefan C},
  journal={Swiss Journal of Economics and Statistics},
  volume={156},
  number={1},
  pages={1--11},
  year={2020},
  publisher={SpringerOpen}
}

@article{baldi2014effect,
  title={The effect of the business cycle on apprenticeship training: Evidence from Germany},
  author={Baldi, Guido and Br{\"u}ggemann-Borck, Imke and Schlaak, Thore},
  journal={Journal of Labor Research},
  volume={35},
  pages={412--422},
  year={2014},
  publisher={Springer}
}

@article{goebel2019german,
  title={The German socio-economic panel (SOEP)},
  author={Goebel, Jan and Grabka, Markus M and Liebig, Stefan and Kroh, Martin and Richter, David and Schr{\"o}der, Carsten and Schupp, J{\"u}rgen},
  journal={Jahrb{\"u}cher f{\"u}r National{\"o}konomie und Statistik},
  volume={239},
  number={2},
  pages={345--360},
  year={2019},
  publisher={De Gruyter}
}

@article{brunello2009effect,
  title={The effect of economic downturns on apprenticeships and initial workplace training: a review of the evidence},
  author={Brunello, Giorgio},
  journal={Empirical research in vocational education and training},
  volume={1},
  number={2},
  pages={145--171},
  year={2009},
  publisher={SpringerOpen}
}

@article{buetikofer2025oil,
title = {Natural resources, demand for skills, and schooling choices},
journal = {Journal of Environmental Economics and Management},
volume = {133},
pages = {103212},
year = {2025},
issn = {0095-0696},
doi = {https://doi.org/10.1016/j.jeem.2025.103212},
url = {https://www.sciencedirect.com/science/article/pii/S0095069625000968},
author = {Aline Bütikofer and Antonio Dalla-Zuanna and Kjell G. Salvanes}
}

@article{arellanobover2022skills,
    author = {Arellano-Bover, Jaime},
    title = "{The Effect of Labor Market Conditions at Entry on Workers' Long-Term Skills}",
    journal = {The Review of Economics and Statistics},
    volume = {104},
    number = {5},
    pages = {1028-1045},
    year = {2022},
    month = {09},
    issn = {0034-6535},
    doi = {10.1162/rest_a_01008},
    url = {https://doi.org/10.1162/rest\_a\_01008},
    eprint = {https://direct.mit.edu/rest/article-pdf/104/5/1028/2042685/rest\_a\_01008.pdf},
}

@article{fukao1993accumulation,
  title={Accumulation of human capital and the business cycle},
  author={Fukao, Kyoji and Otaki, Masayuki},
  journal={Journal of Political Economy},
  volume={101},
  number={1},
  pages={73--99},
  year={1993},
  publisher={The University of Chicago Press}
}

@article{daniel2017dzhw,
  title={DZHW Studienberechtigtenpanel 2008},
  author={Daniel, Andreas and Hoffst{\"a}tter, Ute and Hu{\ss}, Bj{\"o}rn and Scheller, Percy},
  journal={Daten-und Methodenbericht zu den Erhebungen des Studienberechtigtenjahrgangs 2008 (1. bis 3. Befragungswelle)},
  year={2017}
}

@Article{breuer2022svr,
  author={Breuer, Sebastian and Elstner, Steffen and Kirsch, Florian and Wieland, Volker},
  title={{Konjunkturzyklen in Deutschland – die Datierung durch den Sachverständigenrat}},
  journal={Perspektiven der Wirtschaftspolitik},
  year={2022},
  volume={23},
  number={3},
  pages={200-240},
  month={September},
  doi={10.1515/pwp-2022-0017},
  url={https://ideas.repec.org/a/bpj/pewipo/v23y2022i3p200-240n5.html}
}

@article{thomsen2021bologna,
  title={Did the ``Bologna Process'' Achieve Its Goals?: 20 Years of Empirical Evidence on Student Enrolment, Study Success and Labour Market Outcomes},
  author={Kroher, Martina and Leuze, Kathrin and Thomsen, Stephan L and Trunzer, Johannes},
  year={2021},
  month={September},
  journal={IZA Discussion Paper No. 14757}
}

@article{goehausen2024housing,
  title={Housing Costs, College Enrollment, and Student Mobility},
  author={Goehausen, Johannes and Thomsen, Stephan L},
  year={2024},
  month={January},
  journal={IZA Discussion Paper No. 16726}
}

@article{hanushek2008growth,
Author = {Hanushek, Eric A. and Woessmann, Ludger},
Title = {The Role of Cognitive Skills in Economic Development},
Journal = {Journal of Economic Literature},
Volume = {46},
Number = {3},
Year = {2008},
Month = {September},
Pages = {607–68},
DOI = {10.1257/jel.46.3.607},
URL = {https://www.aeaweb.org/articles?id=10.1257/jel.46.3.607}}

@article{hershbein2018recessions,
  title={Do recessions accelerate routine-biased technological change? Evidence from vacancy postings},
  author={Hershbein, Brad and Kahn, Lisa B},
  journal={American Economic Review},
  volume={108},
  number={7},
  pages={1737--1772},
  year={2018},
  publisher={American Economic Association 2014 Broadway, Suite 305, Nashville, TN 37203}
}

@inproceedings{blair2020structural,
  title={Structural increases in demand for skill after the great recession},
  author={Blair, Peter Q and Deming, David J},
  booktitle={AEA Papers and Proceedings},
  volume={110},
  pages={362--365},
  year={2020},
  organization={American Economic Association 2014 Broadway, Suite 305, Nashville, TN 37203}
}

@article{taber2012specific,
   author = "Sanders, Carl and Taber, Christopher",
   title = "Life-Cycle Wage Growth and Heterogeneous Human Capital", 
   journal= "Annual Review of Economics",
   year = "2012",
   volume = "4",
   number = "Volume 4, 2012",
   pages = "399-425",
   doi = "https://doi.org/10.1146/annurev-economics-080511-111011",
   url = "https://www.annualreviews.org/content/journals/10.1146/annurev-economics-080511-111011",
   publisher = "Annual Reviews",
   issn = "1941-1391",
   type = "Journal Article"
  }

@article{woessmann2025skills,
   author = "Woessmann, Ludger",
   title = "Skills and Earnings: A Multidimensional Perspective on Human Capital",
   journal = "Annual Review of Economics",
   issn = "1941-1383",
   year = "2025",
   publisher = "Annual Reviews",
   url = "https://www.annualreviews.org/content/journals/10.1146/annurev-economics-081324-081733",
   doi = "https://doi.org/10.1146/annurev-economics-081324-081733"
}

@article{hershbein2024local,
Author = {Hershbein, Brad and Stuart, Bryan A.},
Title = {The Evolution of Local Labor Markets after Recessions},
Journal = {American Economic Journal: Applied Economics},
Volume = {16},
Number = {3},
Year = {2024},
Month = {July},
Pages = {399–435},
DOI = {10.1257/app.20220132},
URL = {https://www.aeaweb.org/articles?id=10.1257/app.20220132}
}

@article{cameron2011robust,
  title={Robust inference with multiway clustering},
  author={Cameron, A Colin and Gelbach, Jonah B and Miller, Douglas L},
  journal={Journal of Business \& Economic Statistics},
  volume={29},
  number={2},
  pages={238--249},
  year={2011},
  publisher={Taylor \& Francis}
}

@article{marcus2019effect,
  title={The effect of increasing education efficiency on university enrollment: Evidence from administrative data and an unusual schooling reform in Germany},
  author={Marcus, Jan and Zambre, Vaishali},
  journal={Journal of Human Resources},
  volume={54},
  number={2},
  pages={468--502},
  year={2019},
  publisher={University of Wisconsin Press}
}

@article{ditzen2025multiple,
  title={Multiple structural breaks in interactive effects panel data models},
  author={Ditzen, Jan and Karavias, Yiannis and Westerlund, Joakim},
  journal={Journal of Applied Econometrics},
  volume={40},
  number={1},
  pages={74--88},
  year={2025},
  publisher={Wiley Online Library}
}

@article{ditzen2025testing,
  title={Testing and estimating structural breaks in time series and panel data in Stata},
  author={Ditzen, Jan and Karavias, Yiannis and Westerlund, Joakim},
  journal={The Stata Journal},
  volume={25},
  number={3},
  pages={526--560},
  year={2025},
  publisher={SAGE Publications Sage CA: Los Angeles, CA}
}

@article{dustmann2014sick,
Author = {Dustmann, Christian and Fitzenberger, Bernd and Schönberg, Uta and Spitz-Oener, Alexandra},
Title = {From Sick Man of Europe to Economic Superstar: Germany's Resurgent Economy},
Journal = {Journal of Economic Perspectives},
Volume = {28},
Number = {1},
Year = {2014},
Month = {February},
Pages = {167–88},
DOI = {10.1257/jep.28.1.167},
URL = {https://www.aeaweb.org/articles?id=10.1257/jep.28.1.167}}

@article{muehlemann2024double,
title = {Supply Shocks in the Market for Apprenticeship Training},
journal = {Economics of Education Review},
volume = {86},
pages = {102197},
year = {2022},
issn = {0272-7757},
doi = {https://doi.org/10.1016/j.econedurev.2021.102197},
url = {https://www.sciencedirect.com/science/article/pii/S0272775721001126},
author = {Samuel Mühlemann and Hans Dietrich and Gerard Pfann and Harald Pfeifer}
}

@article{dorner2024empty,
title = {The impact of a missing school graduation cohort on the training market},
journal = {Economics of Education Review},
volume = {103},
pages = {102580},
year = {2024},
issn = {0272-7757},
doi = {https://doi.org/10.1016/j.econedurev.2024.102580},
url = {https://www.sciencedirect.com/science/article/pii/S0272775724000748},
author = {Matthias Dorner and Katja Görlitz and Elke J. Jahn}
}

@article {deneault2025teaching,
	author = {Deneault, Christa},
	title = {Local Labor Markets and Selection into the Teaching Profession},
	elocation-id = {0424-13535R2},
	year = {2025},
	doi = {10.3368/jhr.0424-13535R2},
	publisher = {University of Wisconsin Press},
	issn = {0022-166X},
	URL = {https://jhr.uwpress.org/content/early/2025/10/02/jhr.0424-13535R2},
	eprint = {https://jhr.uwpress.org/content/early/2025/10/02/jhr.0424-13535R2.full.pdf},
	journal = {Journal of Human Resources}
}

@article{clark2011recessions,
  title={Do recessions keep students in school? The impact of youth unemployment on enrolment in post-compulsory education in England},
  author={Clark, Damon},
  journal={Economica},
  volume={78},
  number={311},
  pages={523--545},
  year={2011},
  publisher={Wiley Online Library}
}

@article{enke2019correlation,
  title={Correlation Neglect in Belief Formation},
  author={Enke, Benjamin and Zimmermann, Florian},
  journal={The Review of Economic Studies},
  volume={86},
  number={1},
  pages={313--332},
  year={2019},
  publisher={Oxford University Press}
}

@article{blanchard2013news,
Author = {Blanchard, Olivier J. and L'Huillier, Jean-Paul and Lorenzoni, Guido},
Title = {News, Noise, and Fluctuations: An Empirical Exploration},
Journal = {American Economic Review},
Volume = {103},
Number = {7},
Year = {2013},
Month = {December},
Pages = {3045–70},
DOI = {10.1257/aer.103.7.3045},
URL = {https://www.aeaweb.org/articles?id=10.1257/aer.103.7.3045}}

@article{beaudry2006stock,
  title={Stock Prices, News, and Economic Fluctuations},
  author={Beaudry, Paul and Portier, Franck},
  journal={American Economic Review},
  volume={96},
  number={4},
  pages={1293--1307},
  year={2006},
  publisher={American Economic Association}
}

@incollection{manski1993adolescent,
  title={Adolescent econometricians: How do youth infer the returns to schooling?},
  author={Manski, Charles F},
  booktitle={Studies of supply and demand in higher education},
  pages={43--60},
  year={1993},
  publisher={University of Chicago Press}
}

@article{nowzohour2020more,
  title={More than a feeling: Confidence, uncertainty, and macroeconomic fluctuations},
  author={Nowzohour, Laura and Stracca, Livio},
  journal={Journal of Economic Surveys},
  volume={34},
  number={4},
  pages={691--726},
  year={2020},
  publisher={Wiley Online Library}
}

@techreport{deming2017budget,
 title = "The Impact of Price Caps and Spending Cuts on U.S. Postsecondary Attainment",
 author = "Deming, David J and Walters, Christopher R",
 institution = "National Bureau of Economic Research",
 type = "Working Paper",
 series = "Working Paper Series",
 number = "23736",
 year = "2017",
 month = "August",
 doi = {10.3386/w23736},
 URL = "http://www.nber.org/papers/w23736"
}

@misc{eurydice2025funding,
  author       = {Eurydice},
  title        = {Higher Education Funding -- Germany},
  year         = {2025},
  institution  = {European Education and Culture Executive Agency (EACEA)},
  howpublished = {\url{https://eurydice.eacea.ec.europa.eu/eurypedia/germany/higher-education-funding}},
  note         = {Accessed November 27, 2025},
  language     = {English}
}

@misc{gwk2022zukunftsvertrag,
  author       = {GWK},
  title        = {Verwaltungsvereinbarung zwischen Bund und L{\"a}ndern gem{\"a}{\ss} Artikel 91b Absatz 1 des Grundgesetzes {\"u}ber den Zukunftsvertrag Studium und Lehre st{\"a}rken},
  year         = {2022},
  institution  = {Gemeinsame Wissenschaftskonferenz},
  howpublished  = {\url{https://www.gwk-bonn.de/fileadmin/Redaktion/Dokumente/Papers/Verwaltungsvereinbarung_Zukunftsvertrag_2022.pdf}},
  note      = {Accessed November 27, 2025},
  language     = {German}
}

@misc{hrk2025funding,
  author       = {HRK},
  title        = {Hochschulfinanzierung},
  year         = {2025},
  institution  = {Hochschulrektorenkonferenz},
  howpublished = {\url{https://www.hrk.de/themen/hochschulsystem/hochschulfinanzierung/}},
  urldate      = {2025-11-27},
  note         = {Accessed November 27, 2025},
  language     = {German}
}

@book{crosier2013bologna,
  title={The Bologna process: Its impact in Europe and beyond},
  author={Crosier, David and Parveva, Teodora},
  year={2013},
  publisher={{UNESCO}}
}

@incollection{dahm2023einfach,
  title={Einfach anders oder vielf{\"a}ltig verschieden? Ein differenzierter Blick auf Hochschulabsolvent*innen mit beruflicher Vorqualifikation},
  author={Dahm, Gunther and Peter, Frauke},
  booktitle={Vielfalt von hochschulischen Bildungsverl{\"a}ufen: Wege in das, durch das und nach dem Studium},
  pages={223--262},
  year={2023},
  publisher={Springer}
}

@article{comin2010exploration,
  title={An Exploration of Technology Diffusion},
  author={Comin, Diego and Hobijn, Bart},
  journal={American Economic Review},
  volume={100},
  number={5},
  pages={2031--2059},
  year={2010},
  publisher={American Economic Association}
}

@article{frank2019toward,
  title={Toward understanding the impact of artificial intelligence on labor},
  author={Frank, Morgan R and Autor, David and Bessen, James E and Brynjolfsson, Erik and Cebrian, Manuel and Deming, David J and Feldman, Maryann and Groh, Matthew and Lobo, Jos{\'e} and Moro, Esteban and others},
  journal={Proceedings of the National Academy of Sciences},
  volume={116},
  number={14},
  pages={6531--6539},
  year={2019},
  publisher={National Academy of Sciences}
}

@article{acemoglu2024ai,
    author = {Acemoglu, Daron},
    title = {The simple macroeconomics of AI},
    journal = {Economic Policy},
    volume = {40},
    number = {121},
    pages = {13-58},
    year = {2024},
    month = {08},
    issn = {0266-4658},
    doi = {10.1093/epolic/eiae042},
    url = {https://doi.org/10.1093/epolic/eiae042},
    eprint = {https://academic.oup.com/economicpolicy/article-pdf/40/121/13/59091970/eiae042.pdf},
}

@article{bloom2009uncertainty,
  author  = {Bloom, Nicholas},
  title   = {The Impact of Uncertainty Shocks},
  journal = {Econometrica},
  year    = {2009},
  volume  = {77},
  number  = {3},
  pages   = {623--685},
  doi     = {10.3982/ECTA6248}
}

@article{muehlemann2020expectations,
  author  = {Mühlemann, Samuel and Pfeifer, Harald and Wittek, Bernhard H.},
  title   = {The effect of business cycle expectations on the {G}erman apprenticeship market: estimating the impact of {COVID}-19},
  journal = {Empirical Research in Vocational Education and Training},
  year    = {2020},
  volume  = {12},
  number  = {8},
  doi     = {10.1186/s40461-020-00094-9}
}

@article{petrongolo2002staying,
  title={Staying-on at school at 16: the impact of labor market conditions in Spain},
  author={Petrongolo, Barbara and San Segundo, María J},
  journal={Economics of Education Review},
  volume={21},
  number={4},
  pages={353--365},
  year={2002},
  publisher={Elsevier}
}

@article{eggenberger2022specific,
title = {The value of specific skills under shock: High risks and high returns},
journal = {Labour Economics},
volume = {78},
pages = {102187},
year = {2022},
issn = {0927-5371},
doi = {https://doi.org/10.1016/j.labeco.2022.102187},
url = {https://www.sciencedirect.com/science/article/pii/S0927537122000781},
author = {Christian Eggenberger and Simon Janssen and Uschi Backes-Gellner}
}

@article{biewen2017,
  author  = {Biewen, Martin and Tapalaga, Madalina},
  title   = {Life-Cycle Educational Choices in a System with Early Tracking and Second-Chance Options},
  journal = {Economics of Education Review},
  volume  = {56},
  pages   = {80--94},
  year    = {2017}
}

@techreport{goodman2025declining,
  author      = {Goodman, Joshua and Winkelmann, Joseph},
  title       = {Labor Market Strength and Declining Community College Enrollment},
  institution = {National Bureau of Economic Research},
  type        = {NBER Working Paper},
  number      = {34498},
  year        = {2025}
}

@techreport{bicakova2025unpacking,
  author      = {Bi{\v c}{\'a}kov{\'a}, Alena and Cortes, Guido Matias and Foley, Kelly and Mazza, Jacopo and McHenry, Peter},
  title       = {Unpacking the Countercyclicality of Post-Secondary Enrollment in the United States},
  institution = {Canadian Labour Economics Forum (CLEF), University of Waterloo},
  type        = {CLEF Working Paper},
  number      = {092-2025},
  year        = {2025}
}

@misc{nces2023enrollment,
  author       = {{National Center for Education Statistics}},
  title        = {Immediate College Enrollment Rate},
  year         = {2024},
  howpublished = {Condition of Education, U.S. Department of Education, Institute of Education Sciences},
  note         = {Digest of Education Statistics}
}

@techreport{oecd2023pisa,
  author      = {{OECD}},
  title       = {{PISA 2022 Results (Volume I): The State of Learning and Equity in Education}},
  institution = {OECD Publishing, Paris},
  year        = {2023}
}

@book{lewalter2023,
  editor    = {Lewalter, Doris and Diedrich, Jennifer and Goldhammer, Frank and K{\"o}ller, Olaf and Rei{\ss}, Kristina},
  title     = {{PISA 2022: Analyse der Bildungsergebnisse in Deutschland}},
  year      = {2023},
  publisher = {Waxmann},
  address   = {M{\"u}nster}
}

@article{cameron2008bootstrap,
  author  = {Cameron, A. Colin and Gelbach, Jonah B. and Miller, Douglas L.},
  title   = {Bootstrap-Based Improvements for Inference with Clustered Errors},
  journal = {The Review of Economics and Statistics},
  year    = {2008},
  volume  = {90},
  number  = {3},
  pages   = {414--427}
}

@article{roodman2019fast,
  author  = {Roodman, David and Nielsen, Morten {\O}rregaard and MacKinnon, James G. and Webb, Matthew D.},
  title   = {Fast and Wild: Bootstrap Inference in {Stata} Using boottest},
  journal = {The Stata Journal},
  year    = {2019},
  volume  = {19},
  number  = {1},
  pages   = {4--60}
}

@article{mackinnon2023cluster,
  author  = {MacKinnon, James G. and Nielsen, Morten {\O}rregaard and Webb, Matthew D.},
  title   = {Cluster-Robust Inference: A Guide to Empirical Practice},
  journal = {Journal of Econometrics},
  year    = {2023},
  volume  = {232},
  number  = {2},
  pages   = {272--299}
}

@article{goldin2006homecoming,
  title={The Homecoming of American College Women: The Reversal of the College Gender Gap},
  author={Goldin, Claudia and Katz, Lawrence F. and Kuziemko, Ilyana},
  journal={Journal of Economic Perspectives},
  volume={20},
  number={4},
  pages={133--156},
  year={2006}
}

@article{riphahn2015reversal,
  title={What Drives the Reversal of the Gender Education Gap? Evidence from Germany},
  author={Riphahn, Regina T. and Schwientek, Caroline},
  journal={Applied Economics},
  volume={47},
  number={53},
  pages={5748--5775},
  year={2015}
}

@article{malmendier2016learning,
    author = {Malmendier, Ulrike and Nagel, Stefan},
    title = "{Learning from Inflation Experiences}",
    journal = {The Quarterly Journal of Economics},
    volume = {131},
    number = {1},
    pages = {53-87},
    year = {2016}
}

@article{nagel2026experiences,
  title={Experiences, expectations, and asset prices},
  author={Nagel, Stefan},
  journal={Journal of Business Economics},
  pages={1--24},
  year={2026},
  publisher={Springer}
}

@article{davidson2014confidence,
  author  = {Davidson, Russell and MacKinnon, James G.},
  title   = {Confidence Sets Based on Inverting {Anderson--Rubin} Tests},
  journal = {The Econometrics Journal},
  year    = {2014},
  volume  = {17},
  number  = {2},
  pages   = {S39--S58}
}

@article{altonji2012changes,
  author  = {Altonji, Joseph G. and Bharadwaj, Prashant and Lange, Fabian},
  title   = {Changes in the Characteristics of American Youth: Implications for Adult Outcomes},
  journal = {Journal of Labor Economics},
  year    = {2012},
  volume  = {30},
  number  = {4},
  pages   = {783--828}
}

@misc{destatis2022master,
  Title  = {{Prüfungsjahr 2019: 45\,\% der Bachelorabsolventinnen und -absolventen begannen ein Masterstudium}},
  Author = {Destatis},
  Year   = {2022},
  Note   = {{Press release No. 201, 12 May 2022. Statistisches Bundesamt, Wiesbaden}}
}

@misc{spangenberg2016studienberechtigte,
  Title  = {{Bildungsentscheidungen und Umorientierungen im nachschulischen Verlauf: Dritte Befragung der Studienberechtigten 2010 viereinhalb Jahre nach Schulabschluss}},
  Author = {Spangenberg, Heike and Quast, Heiko},
  Year   = {2016},
  Note   = {{Forum Hochschule No. 5/2016. DZHW, Hannover}}
}

@misc{bibb2022datenreport,
  Title  = {{Datenreport zum Berufsbildungsbericht 2022}},
  Author = {{Bundesinstitut für Berufsbildung}},
  Year   = {2022},
  Note   = {{BIBB, Bonn}}
}

@techreport{langer2023skills,
  author      = {Langer, Christina and Wiederhold, Simon},
  title       = {The Value of Early-Career Skills},
  institution = {CESifo},
  type        = {CESifo Working Paper},
  number      = {10288},
  year        = {2023}
}
\end{spacing}
\newpage

\appendix


\clearpage
\pagenumbering{arabic}\setcounter{page}{1}

\setcounter{section}{0}
\renewcommand{\thesection}{OA.\Roman{section}}
\renewcommand{\theHsection}{OA.\Roman{section}}  
\setcounter{figure}{0}
\renewcommand{\thefigure}{OA.\arabic{figure}}
\renewcommand{\theHfigure}{OA.\arabic{figure}}
\setcounter{table}{0}
\renewcommand{\thetable}{OA.\arabic{table}}
\renewcommand{\theHtable}{OA.\arabic{table}}
\setcounter{footnote}{0}

\pdfbookmark[1]{Online Appendix}{oa-bookmark-root}
\makeatletter
\def\toclevel@section{2}
\def\toclevel@subsection{3}
\def\toclevel@subsubsection{4}
\makeatother

\begin{center}
{\Large\textbf{Online Appendix}}\\[12pt]
{\Large{Skill Substitution, Expectations, and the Business Cycle}}\\[20pt]
{\large Andreas Leibing\footnote{German Centre for Higher Education Research and Science Studies (DZHW), Lange Laube 12, 30159 Hannover, Germany, and IZA@LISER; email: \href{mailto:leibing@dzhw.eu}{leibing@dzhw.eu}.}}\\[5pt]
\textit{German Centre for Higher Education Research and Science Studies (DZHW) and IZA@LISER}\\[5pt]
\end{center}
\label{online-appendix}
\vspace{1.5cm}


\startcontents[oa]
\makeatletter
\let\OA@addcontentsline\addcontentsline
\def\OA@loftag{lof}\def\OA@lottag{lot}
\renewcommand{\addcontentsline}[3]{%
  \OA@addcontentsline{#1}{#2}{#3}%
  \edef\OA@tmp{#1}
  \begingroup
    \def\Hy@bookmarkstype{}%
    \ifx\OA@tmp\OA@loftag\OA@addcontentsline{toc}{oafig}{Figure~#3}\fi
    \ifx\OA@tmp\OA@lottag\OA@addcontentsline{toc}{oatab}{Table~#3}\fi
  \endgroup}
\makeatother
\begin{singlespace}
{\makeatletter
\renewcommand*\l@section{\@dottedtocline{1}{0em}{4.2em}}
\renewcommand*\l@subsection{\@dottedtocline{2}{4.2em}{5.4em}}
\providecommand*\l@oafig{}\renewcommand*\l@oafig{\@dottedtocline{3}{6em}{3.9em}}
\providecommand*\l@oatab{}\renewcommand*\l@oatab{\@dottedtocline{3}{6em}{3.9em}}
\makeatother
\hypersetup{linkcolor=black}
\printcontents[oa]{}{1}{\subsection*{Contents}\setcounter{tocdepth}{3}}
}
\end{singlespace}
\clearpage

\FloatBarrier
\clearpage
\section{Institutional Context and Descriptives}
\label{sec:descriptive}

\subsection{Descriptive Statistics}

\begin{table}[H]
    \begin{center}
	\begin{adjustbox}{max width=\textwidth}
        \begin{threeparttable}
		   \caption{Summary statistics}
		   \label{tab:descriptives}
		    \scriptsize{
		        \begin{tabular} {l c c c c c c} 
                    \toprule \toprule
                   &\multicolumn{1}{c}{Mean}&\multicolumn{1}{c}{SD}&\multicolumn{1}{c}{min}&\multicolumn{1}{c}{max}&\multicolumn{1}{c}{N} \\
		    	\midrule
                      \textbf{Students (1995--2018, regions)}  &      &         &     &      &     \\       
High School Graduates (\textit{Abitur})&        4,942&        3,662&         357&       19,182&        2,289 \\
Enrollment Share (Any College)   &       71.36&       10.38&       26.27&       97.82&        2,289 \\
Enrollment Share (academic university)   &       54.25&        9.15&       16.68&       78.78&        2,289 \\
Enrollment Share (Applied university)    &       17.11&        6.07&        3.51&       41.30&        2,289 \\
Detrended Enrollment Rate (Any College)&        0.02&        4.37&      -21.47&       17.59&        2,289\\
Detrended Enrollment Rate (academic university)&        0.36&        4.74&      -19.69&       19.54&        2,289\\
Detrended Enrollment Rate (Applied university)&       -0.34&        2.91&      -12.14&       11.75&        2,289\\
&&&&& \\
\textbf{Unemployment (1995--2018, regions)}  &      &         &     &      &     \\
Detrended State UR (t-1)      &        0.00&        0.67&       -1.74&        3.16&        2,289\\
State UR (t-1)                &        9.28&        4.19&        3.20&       20.50&        2,289\\
National UR (t-1)             &        8.73&        1.84&        5.70&       11.70&        2,289\\
CZ-level UR (t-1)            	&        8.94&        4.38&        2.10&       23.98&        1,907\\
&&&&& \\
\textbf{SOEP v39 (1995--2018, individuals)} &     &       &    &     &     \\
Female             &        0.56&        0.50&        0.00&        1.00&        1,851\\
Low SES            &        0.69&        0.46&        0.00&        1.00&        1,851\\
Any college          &        0.61&        0.49&        0.00&        1.00&        1,851\\
Vocational training &        0.28&        0.45&        0.00&        1.00&        1,851\\
University          &        0.57&        0.50&        0.00&        1.00&        1,567\\
Applied university       &        0.10&        0.30&        0.00&        1.00&        1,567\\
Household Income (EUR)   &     3,678&     1,804&      614&    10,000&        1,447\\
&&&&& \\
\textbf{DZHW (2008--2018, individuals)} &     &       &    &     &     \\
Female      &        0.54&        0.50&        0.00&        1.00&      104,666\\
Low SES     &        0.48&        0.50&        0.00&        1.00&      104,666\\
GPA         &        2.41&        0.59&        0.70&        6.00&      104,666\\
Expected Returns     &        0.52 &        1.04 &       -4.00&        4.00&      104,666 \\
Academic Expectations     &        3.91&        0.75&        1.00&        5.00&      104,666\\
Vocational Expectations     &        3.39&        0.81&        1.00&        5.00&      104,666\\
Personal Expectations     &        3.82&        0.77&        1.00&        5.00&      97,316\\
Enrolled (Dec of graduation year)      &        0.52&        0.50&        0.00&        1.00&      33,735\\
Vocational training (Dec of graduation year)      &        0.14&        0.34&        0.00&        1.00&      33,735\\
			    	\bottomrule
	        	\end{tabular}}
	    	\begin{tablenotes}[flushleft] \scriptsize{
			\item { \textit{Notes:} This table shows summary statistics for the main data sources used in the analysis. Statistics of the two region panels (students and unemployment) are weighted by the number of high school graduates with \textit{Abitur} per cell. Unemployment rates refer to the senior year ($t-1$), the treatment timing of the analysis, and detrended values are residuals of the main specification's state-specific quadratic trends with a 2005 break. GPA: In the German grading system, the best grade is effectively a 0.7, and the worst is a 6.0. In the empirical analysis, I multiply the current GPA by -1 to ease interpretation (higher = better). In this table, ``above median GPA'' also means better GPA. \\ \textit{Source:} Student register, Federal Employment Agency, SOEP v39, DZHW, \textit{Regionaldatenbank}.
   }}
		\end{tablenotes}
	    \end{threeparttable}
	    \end{adjustbox}
    \end{center}
\end{table}

\FloatBarrier
\clearpage
\subsection{Trends and Breaks in Enrollment and Unemployment}

Panel A of \Cref{fig:nationalUR_enroll} plots the national unemployment rate and the aggregate share of high school graduates enrolling at any college. After the German reunification in 1990, the unemployment rate rose continuously until the late 1990s. A short recovery period was interrupted by the dot-com crash in 2000, after which unemployment peaked at around 12\% in 2005, making Germany ``the sick man of Europe'' \citepsupp{dustmann2014sick}. 

\begin{figure}[H]
  \centering
  \begin{minipage}{\textwidth}

    \caption{Time Series of Enrollment and Unemployment}
    \label{fig:nationalUR_enroll}


    \begin{subfigure}{\textwidth}
        \centering
        \caption{Panel A: Overall College Enrollment and National UR}
        \label{fig:nationalUR_enroll_a}
        \includegraphics[scale=0.55]{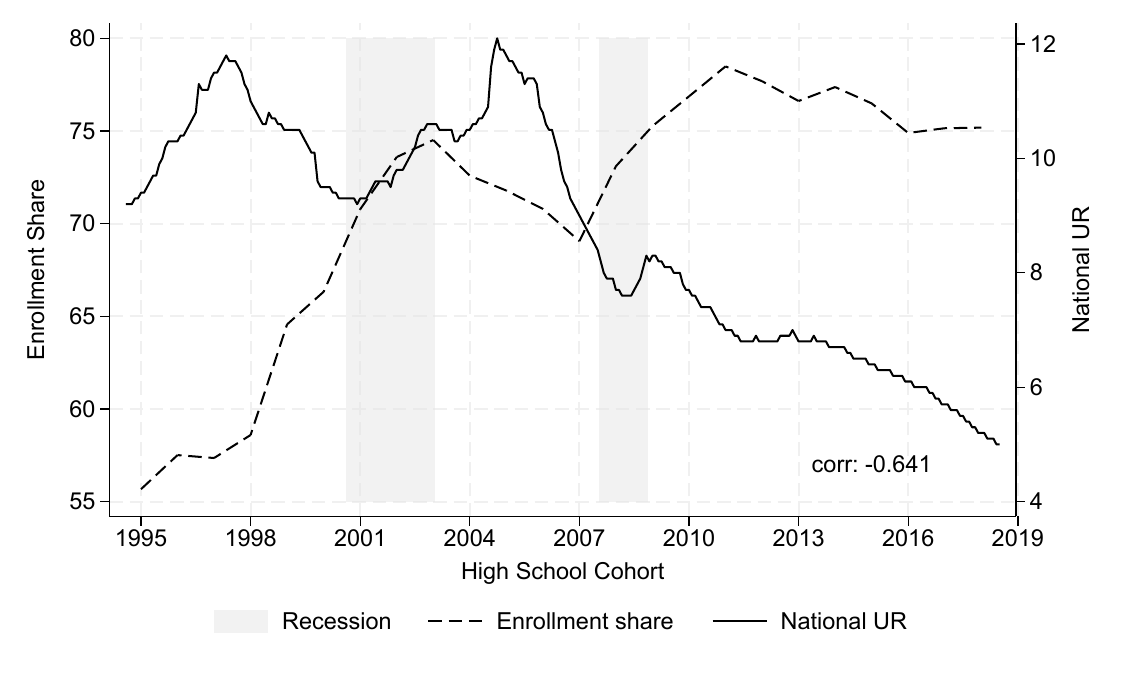}
        \vspace{-0.45em}
    \end{subfigure}


    \begin{subfigure}{\textwidth}
        \centering
        \caption{Panel B: Enrollment by College Type}
        \label{fig:nationalUR_enroll_b}
        \includegraphics[scale=0.55]{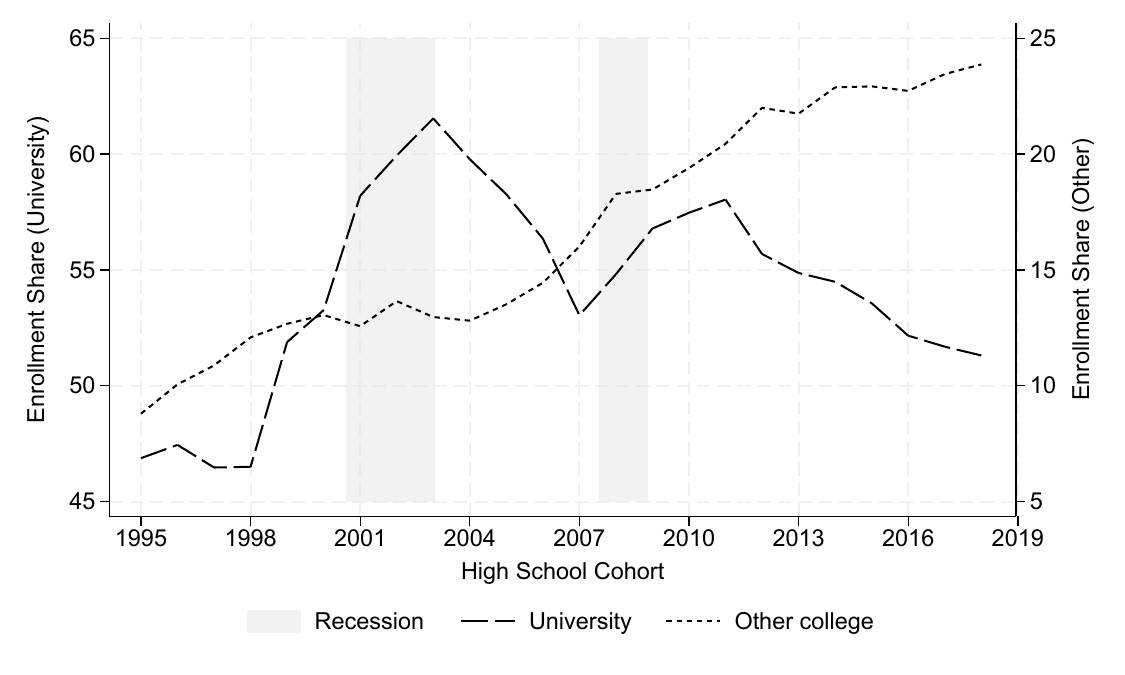}
        \vspace{-0.45em}
    \end{subfigure}


    {\scriptsize \textit{Notes:} Share of high school graduates with \textit{Abitur} enrolling at different college types (yearly), national unemployment rate (monthly), and recession periods (grey, following \citesupp{breuer2022svr}). \\ \textit{Source}: Student register, Federal Employment Agency, \textit{Regionaldatenbank}, German Council of Economic Experts, years 1995--2018.\par}
  \end{minipage}
\end{figure}


The 2005 peak coincides with the final and most far-reaching wave of these reforms, \textit{Hartz~IV}, which took effect in January 2005. It merged unemployment and social assistance into a single, lower, means-tested benefit, tightened job-search and availability requirements, and expanded activation measures, raising both the incentive and the pressure to take up work. After the implementation of the last wave of the Hartz reforms in that year, the unemployment rate fell rapidly until the COVID-19 pandemic, which is excluded from the sample, with only a minor setback during the financial crisis. How much of the subsequent decline in unemployment the reforms themselves account for, relative to the earlier decentralization of wage bargaining and to favorable external demand, remains debated \citepsupp{dustmann2014sick}. What matters for my identification strategy, however, is that the reforms mark a structural break in the level and dynamics of unemployment around 2005, which I accommodate with a state-specific trend break in that year (\Cref{sec:setup}).

Due to the high degree of employment protection (see \Cref{fig:cross-country-oecd}), unemployment and official recessions are therefore largely decoupled in Germany over this period, at least in comparison to the US. The unemployment peaks fall outside the shaded GDP recessions, and the sharp 2008/09 contraction left unemployment almost unchanged. The unemployment rate, rather than GDP, is thus the relevant measure of the labor market conditions students observe.

\begin{figure}[!h]
	\centering
	\begin{minipage}{\textwidth} 
    \caption{Labor market and higher education institutions in the OECD}
    \label{fig:cross-country-oecd}
	\begin{center}
	\includegraphics[scale=0.7]{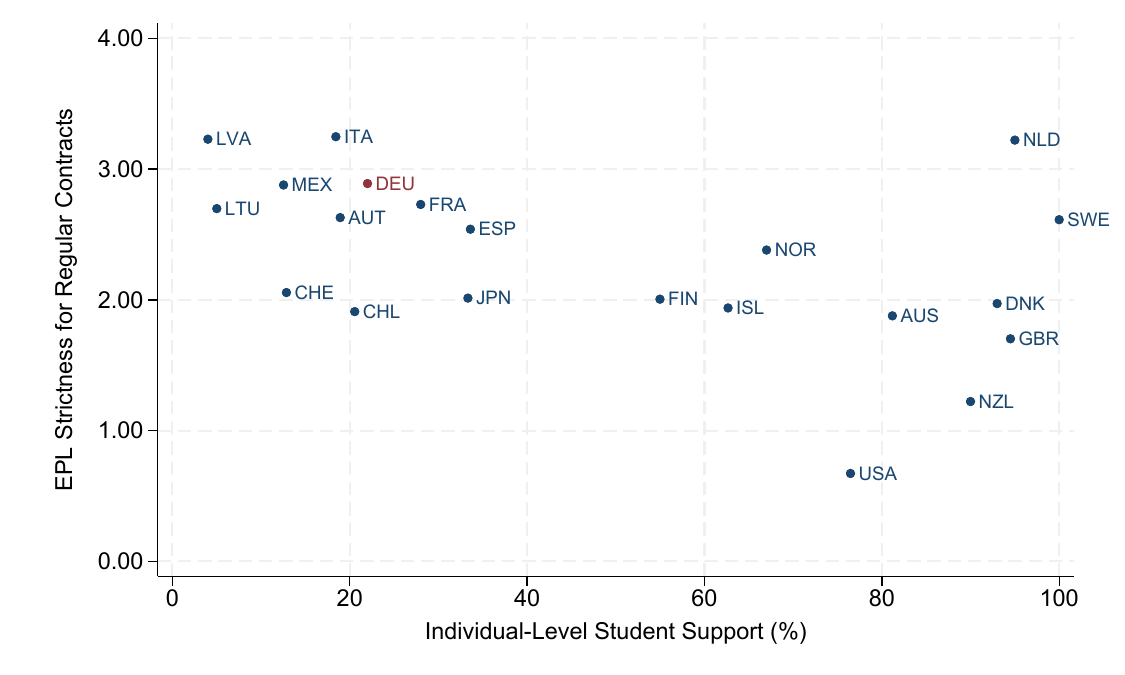} \\
	\end{center}
    \vspace{-1em}
    \vspace{0.3em}
    {\scriptsize \textit{Notes:} Share of college students benefiting from student loans or grants and the summary indicator for individual and collective dismissal of regular workers (1998-2018 average) as a proxy for labor market cyclicality. \\ \textit{Source}: OECD Employment Protection Legislation Database (2020), OECD Education at a Glance (2011, 2022)
    \par}
    \end{minipage}
\end{figure}
During the same 24-year period, enrollment rates rose from around 55 percent in 1995 to almost 80 percent in 2011, after which they slowly decreased. The previous rise in enrollment rates was interrupted during the mid-2000s, coinciding with the Bologna process and the introduction of general tuition fees in 2007 in seven German states. Overall, there is a strong procyclical relationship ($\rho = -0.641$) between unemployment and college enrollment. This negative correlation is driven by the period before 2010. Since 2011, unemployment and college enrollment have fallen simultaneously.

Splitting enrollment rates by college type in Panel B suggests that these educational reforms during the mid-2000s also triggered distinct trends at academic universities and all applied universities. While enrollment at academic universities has been decreasing on average since 2003, enrollment rates at applied universities have surged. In two separate growth periods, from 1995 to 2003, and 2004 to 2018, enrollment at colleges other than academic universities rose from below 10 to around 24 percent. Structural break tests in \Cref{tab:xtbreak} confirm that unemployment and enrollment rates experience structural breaks during the mid-2000s. 

\begin{table}[H]
    \begin{center}
	\begin{adjustbox}{max width=\textwidth}
	\begin{threeparttable}
		\caption{Structural break estimation and tests}
		\label{tab:xtbreak}
		\scriptsize{
		        \begin{tabular} {l cc cc}
	\toprule \toprule
& \multicolumn{2}{c}{Estimated break} &  \multicolumn{2}{c}{\begin{tabular}{@{}c@{}}$H_0$: No Break \\ $H_1$: Break in 2005\end{tabular}}  \\
\cmidrule(lr){2-3} \cmidrule(lr){4-5} 
& Year & 95\% CI & F-statistic & p-value  \\ 
& (1) & (2) & (3) & (4)  \\ 
\midrule
\textit{Panel A: Quadratic trend (levels)} & &    &   &   \\
State UR 			        & 2006 & [2005, 2007] & 152.57 & 0.00 \\
Any college 		        & 2003 & [2002, 2004] & 28.53  & 0.00 \\
University                  & 2003 & [2002, 2004] & 62.06  & 0.00 \\
Applied university               & 2006 & [2005, 2007] & 140.35 & 0.00 \\
& &   &  &     \\
\textit{Panel B: State UR coefficients } & &   &   &   \\
Any college                 & 2006 & [2005, 2007] & 102.96 & 0.00 \\
University                  & 2006 & [2005, 2007] & 158.08 & 0.00 \\
Applied university               & 2007 & [2002, 2012] & 10.73  & 0.00 \\
\bottomrule
		\end{tabular}
		}
		\begin{tablenotes}[flushleft] 
				\item \scriptsize{\textit{Notes:} Panel A reports estimated break dates and tests for a structural break at 2005. Panel B uses detrended series (state FE and state-specific quadratic trends with a trend break at 2005) and estimates a break in the UR-enrollment relationship. All estimates and tests based on N=2,289 cells and implemented via \texttt{xtbreak} for paneldata, based on \citesupp{ditzen2025multiple,ditzen2025testing}. * $p$\textless 0.1, ** $p$\textless 0.05, *** $p$\textless 0.01. \\ \textit{Source}: Student register, Federal Employment Agency, \textit{Regionaldatenbank}.}
		\end{tablenotes}
	\end{threeparttable}
	\end{adjustbox}
    \end{center}
\end{table}

Time series of unemployment rates across states (\Cref{fig:eastwest_UR}) and enrollment rates across commuting zones (\Cref{fig:eastwest_enroll}) show that the structural breaks are particularly pronounced in East Germany. Except for 2005, in West Germany, state-level unemployment fell more or less constantly between 1995 and 2018, with overall levels below 10 percent. In East Germany, however, they peaked in 2005 before falling sharply afterwards, converging to West German levels. Conversely, enrollment rates in West Germany increased in the 1990s and have fluctuated around 75 percent since 2001. Except for the mid-2000s, enrollment rates in East Germany have been increasing from below 50 percent and have converged to West German levels.

\begin{figure}[H]
	\centering
	\begin{minipage}{\textwidth} 
	\caption{Unemployment rates over time}
    \vspace{-1em}
    \label{fig:eastwest_UR}
    
    \begin{center}

    \begin{subfigure}[b]{0.75\textwidth} 
        \centering
        \small{Panel A: By state}
        \includegraphics[width=\textwidth]{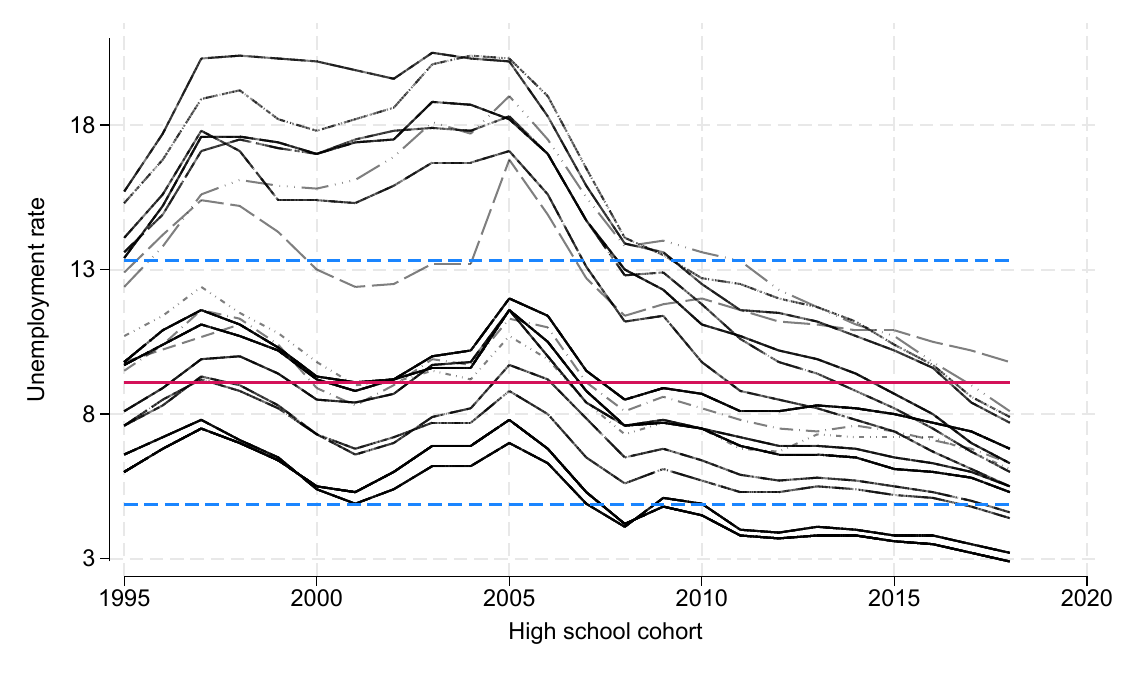}
    \end{subfigure}
    \vspace{-1.5em}    
    \vspace{1em} 
    \begin{subfigure}[b]{0.75\textwidth}
        \centering
        \small{Panel B: Separately for West and East Germany}
        \includegraphics[width=\textwidth]{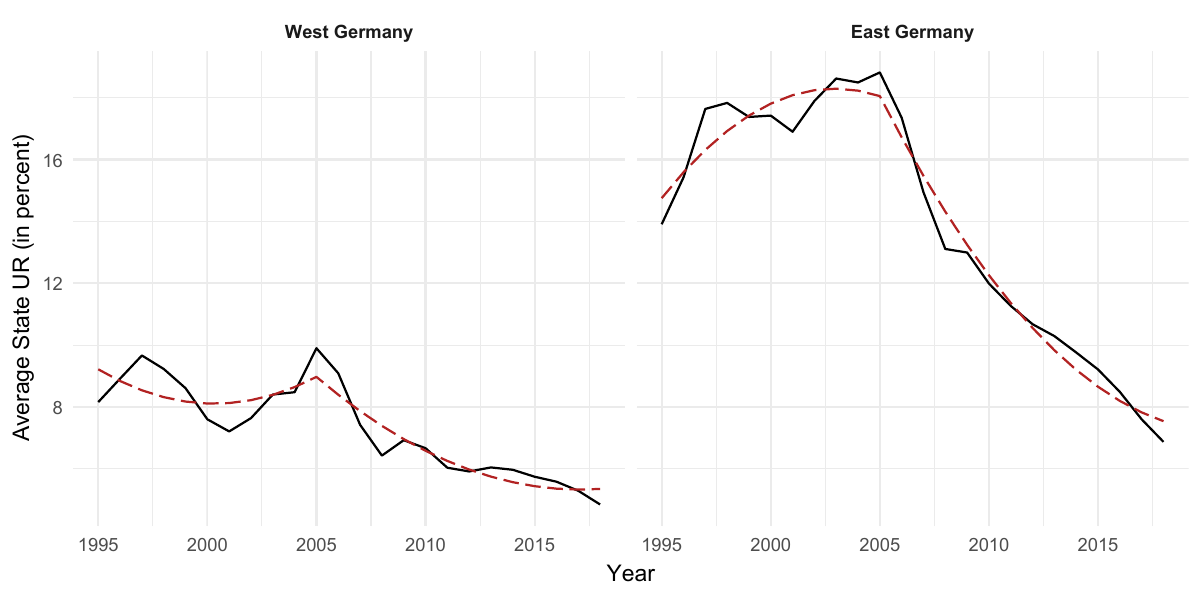}
    \end{subfigure}
    \end{center}
    
    \vspace{-1em}
    {\scriptsize \textit{Notes:} Panel A plots the unemployment rate for each state over time, as well as the average state UR (in red) and the standard deviation (in blue). Panel B plots the average state UR of West German states and East German states (including Berlin) for each year, as well as the respective quadratic trend (in red), allowing for a trend break in 2005. \\ \textit{Source}: Federal Employment Agency, years 1995--2018.
    \par}
    \end{minipage}
\end{figure}

\begin{figure}[H]
	\centering
	\begin{minipage}{\textwidth} 
	\caption{Overall college enrollment over time}
    \vspace{-1em}
    \label{fig:eastwest_enroll}
    
    \begin{center}

    \begin{subfigure}[b]{0.75\textwidth} 
        \centering
        \small{Panel A: By commuting zone}
        \includegraphics[width=\textwidth]{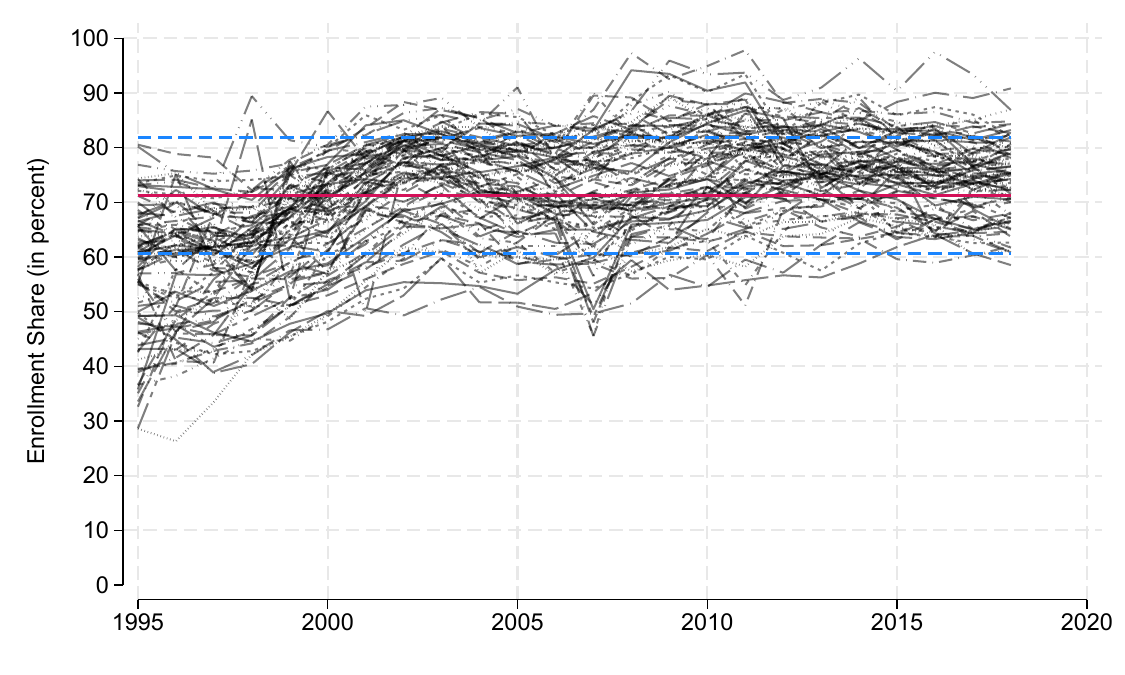}
    \end{subfigure}
    \vspace{-1.5em}    
    \vspace{1em} 
    \begin{subfigure}[b]{0.75\textwidth}
        \centering
        \small{Panel B: Separately for West and East Germany}
        \includegraphics[width=\textwidth]{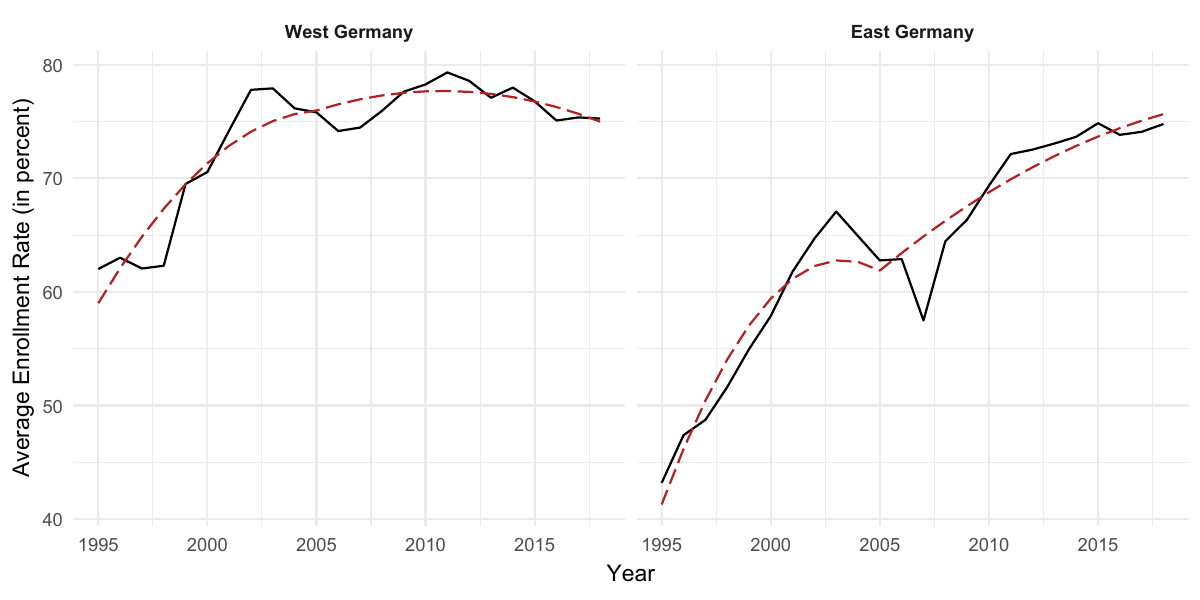}

    \end{subfigure}

    \end{center}
    
    \vspace{-1em}
    {\scriptsize \textit{Notes:} Panel A plots the overall enrollment rate of high school graduates with \textit{Abitur} for each commuting zone over time, as well as the average enrollment rate (in red) and the standard deviation (in blue). Panel B plots the average enrollment rate for West German states and East German states (including Berlin) for each year, as well as the respective quadratic trend (in red), allowing for a trend break in 2005. \\ \textit{Source}: Student register, \textit{Regionaldatenbank}, years 1995--2018.
    \par}
    \end{minipage}
\end{figure}

\FloatBarrier
\clearpage
\subsection{Bologna Reform and International Comparison}
\label{sec:bologna}

The largest reform in the European Higher Education sector was arguably the Bologna Process, which began in 1999 \citep{crosier2013bologna}. In Germany, the main change, alongside the introduction of the European Credit Transfer and Accumulation System (ECTS), was the transition from a single-tier to a two-tier degree system. \Cref{fig:bolognareform} shows that adoption accelerated in the mid-2000s: by 2008, bachelor’s programs had become the dominant entry route, with the sharpest compositional shift around 2005. Relative to the \textit{Diplom} and \textit{Magister} degrees, the median time-to-degree for the first cycle was reduced from 6 to 4 years \citep{bietenbeck2023tuition}, while apprenticeships typically take around 2.5 years. This shift plausibly altered substitution patterns across educational options and the horizon over which students map current macroeconomic conditions to expected returns. 

\begin{figure}[H]
	\centering
	\begin{minipage}{\textwidth} 
    \caption{Bologna Reform and degree choice of first-year students}
 	\label{fig:bolognareform}
	\begin{center}
    \includegraphics[scale=0.5]{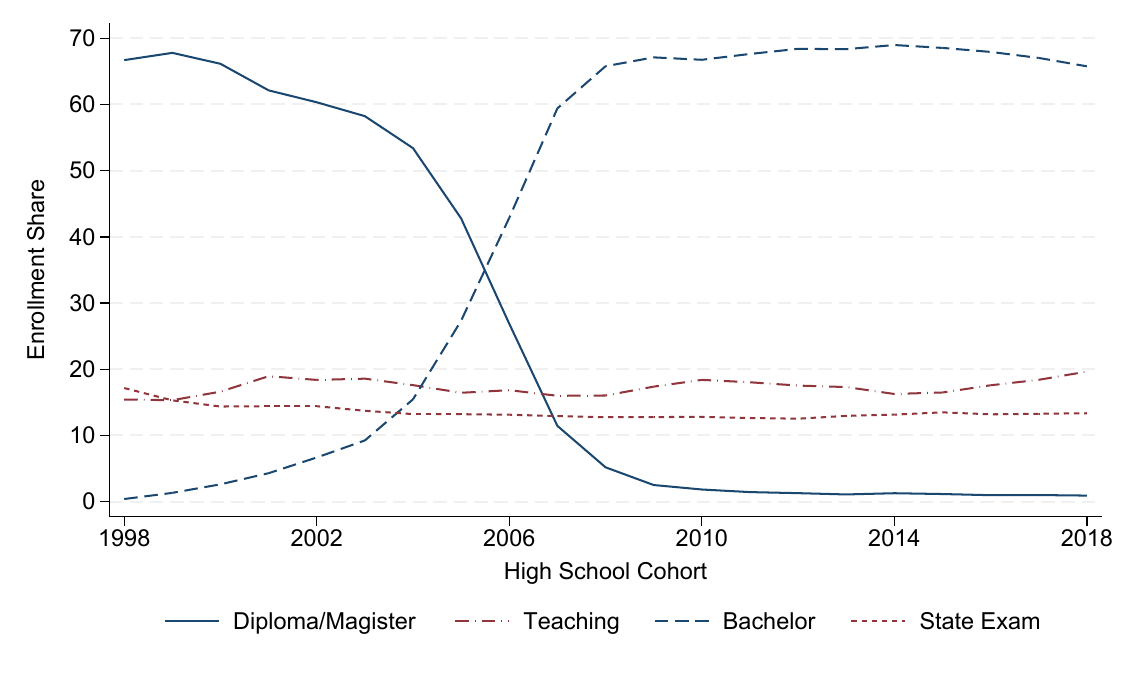} \\
	\end{center}
    \vspace{-1em}
    {\scriptsize \textit{Notes:} Share of first-year university students enrolling across degree types. \\ \textit{Source}: Federal Statistical Office (Student Register).
    \par}
    \end{minipage}
\end{figure}

A further institutional change associated with the Bologna Reforms was the disappearance of the ``Diplom (FH)'' suffix for degrees from Universities of Applied Sciences (\textit{Fachhochschulen}) with the two-tier harmonization, which reduced visible credential segmentation between institution types and facilitated second-chance studies (e.g., combining apprenticeship and BA) within a similar overall time window. Existing evidence on enrollment effects of the Bologna Process is, however, inconclusive \citepsupp[see][for a literature review]{thomsen2021bologna}. 

\Cref{fig:cross-country-oecd} illustrates that Germany's student financial aid coverage and regulated labor market resemble those of many OECD countries. These institutional features may affect not only the cyclicality of outside options but also how economic shocks influence the perceived risks and returns of different skill investments.


\FloatBarrier
\clearpage
\subsection{High School Graduates}

\Cref{fig:abitur_east_west} traces the expansion of the \textit{Abitur} separately for West and East Germany. The share of school leavers obtaining the \textit{Abitur} (dashed) rises in both regions, while the absolute number (solid) grows in the West but falls in the East. The decline in the East reflects the post-reunification fertility collapse, the so-called \textit{Wendeknick}. East German births roughly halved between 1990 and 1994, so the cohorts reaching school-leaving age around 2008--2013 were far smaller, and the falling number of graduates there is a demographic echo rather than a decline in attainment, as the rising share makes clear. The pronounced spikes, most visible in the West, are the double graduating cohorts created when states shortened the academic track from nine to eight years (G8), which I control for throughout (\Cref{sec:setup}). As a larger share of each birth cohort obtains the \textit{Abitur}, and the marginal graduate is drawn from further down the achievement distribution, in line with US evidence on changing cohort characteristics \citepsupp{altonji2012changes}. The state-specific trends absorb this secular selection, and \Cref{tab:hsgraduates} examines its cyclical component.

\begin{figure}[H]
	\centering
	\begin{minipage}{\textwidth}
	\caption{Number of \textit{Abitur} holders and the \textit{Abitur} share over time, West vs.\ East Germany}
	\label{fig:abitur_east_west}
	\begin{center}
	\includegraphics[width=0.8\textwidth]{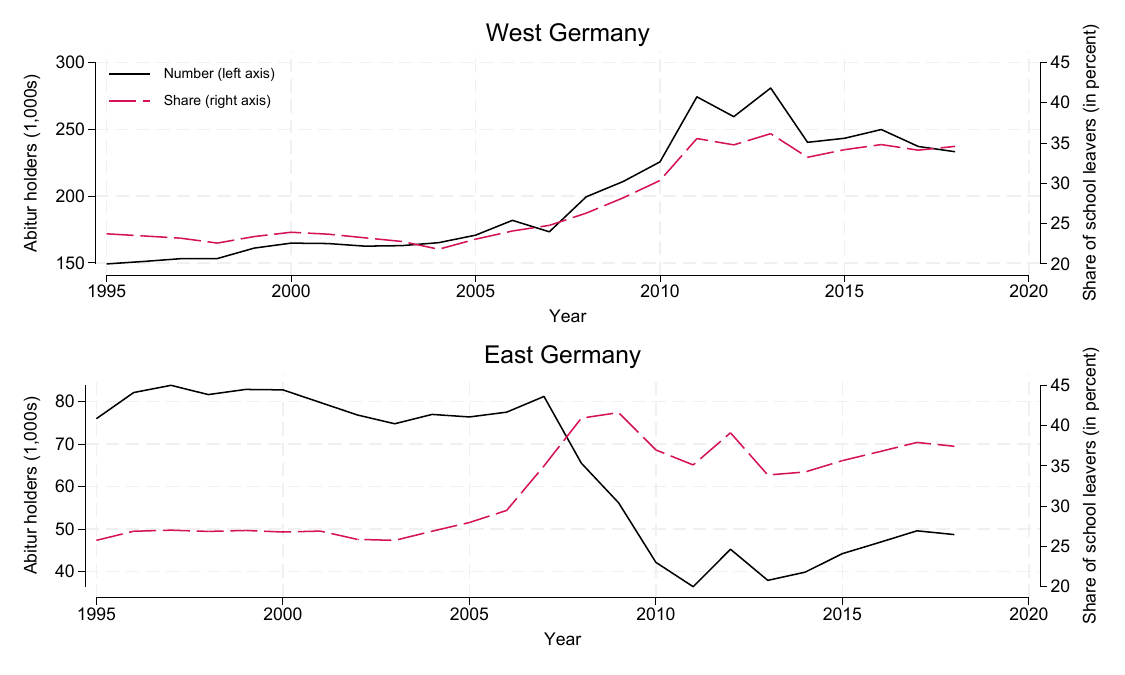}
	\end{center}
	\vspace{-1em}
	{\scriptsize \textit{Notes:} Each panel plots the number of school leavers obtaining the general \textit{Abitur} (solid, left axis, in 1{,}000s) and their share of all school leavers (dashed, right axis), aggregated over the West German and East German states (the latter including Berlin), 1995--2018. The spikes are double graduating cohorts from the staggered shortening of the academic track (G8). The general-school \textit{Fachhochschulreife} is excluded throughout. The 2001 value for East Germany omits Mecklenburg-West Pomerania and Saxony-Anhalt, where a school-reform \textit{Nullkohorte} produced no \textit{Abitur} cohort that year. \\ \textit{Source}: Statistics of general schools, \textit{Regionaldatenbank}.
	\par}
	\end{minipage}
\end{figure}

\FloatBarrier
\clearpage
\subsection{University Funding}

In Germany, academic universities are financed through state budgets and multi-year framework agreements that use upper-bound projections of student numbers \citepsupp{eurydice2025funding}. Around 90 percent of university funding comes from public sources, with roughly three-quarters provided by the states and the remainder by the federal government through joint programs such as the Higher Education Pact (\textit{Hochschulpakt}) and its successor, the Contract for the Future of Higher Education and Teaching (\textit{Zukunftsvertrag Studium und Lehre stärken}) \citepsupp{gwk2022zukunftsvertrag, hrk2025funding}. It is explicitly aimed at maintaining need-based study capacity, and provides dynamically increasing federal co-financing, largely independent of short-run state-level unemployment shocks.

\FloatBarrier
\clearpage
\section{Conceptual Framework}
\label{sec:model}

I extend a Roy-type model from \citesupp{charles2018housing} to incorporate the effect of UR shocks at different levels of aggregation, as well as parental income, and derive testable comparative statics at two different enrollment margins. Consider high school students $i$ with academic ability $\theta_i \in [0,1]$ that maximize lifetime payoff by choosing between three postsecondary options: enrolling at an academic university (college type $c=A$, general skills), enrolling at an applied university (college type $c=B$, skill mix), or starting an apprenticeship ($c=0$, specific skills), before working until retirement $L$.

Students finance their studies via parental income $\omega_i$.\footnote{Parental transfers constitute 66 percent of the average monthly income of a first-year student in Germany, and loan-aversion is high \citepsupp{middendorff2017wirtschaftliche}.} As college type B allows one to allocate more time to work besides studying (e.g., via dual studies), its average monetary net costs are lower ($F_A > F_B$). At college, potential students also face psychic costs of learning $\kappa_c$. The curriculum at college type A focuses on general skills and is thus more challenging ($\kappa_A > \kappa_B$), depending on students' academic ability. The expected college premium at the current period $t$ for type-c college graduates equals $\pi_{t}^{c} = Y_{t}^{c} - Y_{t}^{0} \geq 0$, where $Y_{t}^{c}$ and $Y_{t}^{0}$ denote earnings of college and non-college educated workers, respectively. The lifetime payoff for student $i$ at period $t$ from attending a type-c college thus equals:

\begin{equation}
R_i^c(\theta_i,\omega_i) \;=\; \sum^{L-t}_{k=1} E_{t}[\pi^{c}_{t+k}]- Y^{0}_{t} - (F_c - \omega_i) - \kappa_c (1-\theta_{i}).
\end{equation}

\noindent Here, the term $-Y^{0}_{t}$ is the immediate opportunity cost of forgone earnings while at college, whereas the premia $\pi^{c}_{t+k}$ capture the earnings differential after graduation. When the payoff of the outside option, apprenticeships, $R_{i}^0$, is normalized to zero and under the assumption that changes in unemployment rates do not affect $F_c$ and $\kappa_c$, changes in the state $s$ unemployment rate $d \text{UR}_{st}$ around graduation affect the expected returns to college via: 

\begin{equation}
    \frac{d R_{i}^{c}(\theta_{i}, \omega_{i})}{d \text{UR}_{st}} =
    \underbrace{\frac{d \sum^{L-t}_{k=1} E_{t}[\pi^{c}_{t+k}]}{d  \text{UR}_{st}}}_{\text{(i)}} -  
    \underbrace{\frac{d Y_{t}^{0}}{d \text{UR}_{st}}}_{\text{(ii)}} + 
    \underbrace{\frac{d \omega_{i}}{d \text{UR}_{st}}}_{\text{(iii)}}.
\end{equation}

\noindent  \Cref{fig:statics} Panel (a) shows the equilibrium sorting where students with ability up to threshold $\theta^B$ choose an apprenticeship, students with ability between $\theta^B$ and $\theta^{AB}$ enroll at college type B, and students, with $\theta_i > \theta^{AB}$ enroll at college type A. Panel~(b) shows a scenario where a state-level unemployment shock reduces overall college enrollment ($\theta^{B'} > \theta^B$) while keeping constant the share of high school graduates enrolling at applied universities ($\theta^{AB'} - \theta^{B'} = \theta^{AB} - \theta^{B}$ and $\Bar{R} = \Bar{R}'$). A necessary condition for this is that the negative effect on expected returns to college type A exceeds the negative effect on expected returns to college type B.

\begin{figure}[!h]
    \centering
    \caption{A Roy model of skill investment}
    \begin{minipage}[t]{0.45\linewidth}
        \centering
        \begin{subfigure}[t]{\linewidth}
        \centering
\begin{tikzpicture}
\begin{axis}[
ylabel={$R_{i}^c(\theta_i, \Bar{\omega}_i)$},
ylabel style={xshift=18mm},
axis lines=left,
xmin=0,xmax=1,
ymin=0,ymax=1,
xtick={0,1},
ytick=\empty,
legend style={at={(0.6,0.6)},anchor=north west},
clip=false,
enlargelimits={upper=0.1}
]
\addplot[Blue,domain=0:1] {0.4 + 0.3*(\x)};
\addplot[BrickRed,domain=0:1] {0.1 + 1.0*(\x)};
\draw[dashed] (axis cs: 0.2,0) -- (axis cs: 0.2,0.46);             
\draw[dashed] (axis cs: 0,0.46) -- (axis cs: 1,0.46);              
\draw[dashed] (axis cs: 30/70,0) -- (axis cs: 30/70,0.52857);      
\draw[dashed] (axis cs: 0,0.52857) -- (axis cs: 30/70,0.52857);    
\node[below] at (axis cs: 30/70,0) {$\theta^{AB}$};
\node[below] at (axis cs: 0.2,0) {$\theta^{B}$};
\node[left] at (axis cs: 0,0.46) {0};
\node[left] at (axis cs: 0,0.54) {$\Bar{R}$};
\node[right] at (axis cs: 1,1.1) {$R_{i}^{A}(\theta_{i})$};
\node[right] at (axis cs: 1,0.7) {$R_{i}^{B}(\theta_{i})$};
\draw[decorate,decoration={brace,amplitude=5pt,mirror,raise=2mm}]
(axis cs: 0.0,-0.07) -- (axis cs: 0.2,-0.07);
\node[below=5mm] at (axis cs: 0.1,-0.07) {$c = 0$};
\draw[decorate,decoration={brace,amplitude=5pt,mirror,raise=2mm}]
(axis cs: 0.2,-0.07) -- (axis cs: 30/70,-0.07);
\node[below=5mm] at (axis cs: 0.32,-0.07) {$c = B$};
\draw[decorate,decoration={brace,amplitude=5pt,mirror,raise=2mm}]
(axis cs: 30/70,-0.07) -- (axis cs: 1,-0.07);
\node[below=5mm] at (axis cs: 50/70,-0.07) {$c = A$};
\end{axis}
\end{tikzpicture}
    \caption{Postsecondary enrollment by ability}
    \end{subfigure}
    \end{minipage}\hfill
    \begin{minipage}[t]{0.45\linewidth}
        \centering
        \begin{subfigure}[t]{\linewidth}
        \centering
\begin{tikzpicture}
\begin{axis}[
xlabel={$\theta_i$},
xlabel style={yshift=6mm, xshift=35mm},
axis lines=left,
xmin=0,xmax=1,
ymin=0,ymax=1,
xtick={0,1},
ytick=\empty,
legend style={at={(0.6,0.6)},anchor=north west},
clip=false,
enlargelimits={upper=0.1}
]

\addplot[dashed,Blue,domain=0:1] {0.4 + 0.3*(\x)};
\addplot[solid,Blue,domain=0:1] {0.36 + 0.3*(\x)};
\addplot[dashed,BrickRed,domain=0:1] {0.1 + 1.0*(\x)};
\addplot[solid,BrickRed,domain=0:1,restrict y to domain=0:1,] {-0.037 + 1.0*(\x)};

\draw[dashed] (axis cs: 0.336,0) -- (axis cs: 0.336,0.46);  
\draw[dashed] (axis cs: 0,0.46) -- (axis cs: 1,0.46);       
\draw[dashed] (axis cs: 0.564571429,0) -- (axis cs: 0.564571429, 0.53);  
\draw[dashed] (axis cs: 0,0.53) -- (axis cs: 0.564571429,0.53);  
\node[below] at (axis cs: 0.564571429,0) {$\theta^{AB'}$};
\node[below] at (axis cs: 0.336,0) {$\theta^{B'}$};
\node[left] at (axis cs: 0,0.46) {0};
\node[left] at (axis cs: 0,0.54) {$\Bar{R}'$};
\node[right] at (axis cs: 1,0.96) {$R_{i}^{A}(\theta_{i})'$};
\node[right] at (axis cs: 1,0.66) {$R_{i}^{B}(\theta_{i})'$};
\draw[decorate,decoration={brace,amplitude=5pt,mirror,raise=2mm}]
(axis cs: 0.0,-0.07) -- (axis cs: 0.336,-0.07);
\node[below=5mm] at (axis cs: 0.17,-0.07) {$c = 0$};
\draw[decorate,decoration={brace,amplitude=5pt,mirror,raise=2mm}]
(axis cs: 0.336,-0.07) -- (axis cs: 0.564571429,-0.07);
\node[below=5mm] at (axis cs: 0.46,-0.07) {$c = B$};
\draw[decorate,decoration={brace,amplitude=5pt,mirror,raise=2mm}]
(axis cs: 0.564571429,-0.07) -- (axis cs: 1,-0.07);
\node[below=5mm] at (axis cs: 55/70,-0.07) {$c = A$};
\end{axis}
\end{tikzpicture}
    \caption{Post-shock scenario ($d \text{UR}$ > 0)}
\end{subfigure}
    \end{minipage}

    \label{fig:statics}
    \begin{minipage}{\linewidth}
    \scriptsize \textit{Notes:} Panel (a) shows the lifetime payoff of academic universities (college type $c=A$) and applied universities (college type $c=B$) by ability, normalizing the lifetime payoff of apprenticeships ($c=0$) to zero. Segments on the x-axis indicate enrollment shares. Panel (b) shows a new equilibrium in a scenario where the negative effect of an economic shock at high school graduation on expected returns is stronger for college type A and outweighs the effect on outside options for college type A and B. \\ \textit{Source}: Own representation based on \citesupp{charles2018housing}.
    \end{minipage}
\end{figure}

The effect of changes in labor demand depends on how changes in the state UR affect (i) the expected average returns to college type $c$, (ii) outside options, and (iii) parental income. Conceptually, each change in the unemployment rate $\text{UR}_{s(r)t}$, in state $s(r)$ of commuting zone $r$ at time $t$ can be into a national and a state-specific component:
\begin{equation}
\vspace{-1em}
\label{eq:decomposition}
d\text{UR}_{s(r)t} = d\text{UR}_t^{\text{nat}} + d\text{UR}^{\Delta}_{s(r)t},
\end{equation}
where $\text{UR}_t^{\text{nat}}$ is the national unemployment rate (common across states) and $\text{UR}^{\Delta}_{s(r)t}$ is the deviation of the state unemployment rate from the national rate.\footnote{Note that the state-specific deviation could also be represented as a weighted average across commuting zones $r$ in state $s$. \Cref{sec:decomposition} decomposes the state UR into a national component and the state-specific deviation, and uses more granular variation on the CZ-level.} Empirically, it remains an open question which level of variation dominates for each separate channel.


\FloatBarrier
\clearpage
\section{Identifying Variation}
\label{sec:oa_identifying_variation}

\Cref{fig:trendfit_grad} overlays the fitted trends on the attainment series, the analogue of \Cref{fig:trendfit}, and \Cref{fig:kg2x2_grad} plots the residual variation underlying these estimates, the attainment analogue of \Cref{fig:kg2x2_enrol}. The last observed cohorts show declining raw attainment and large negative attainment residuals, reflecting that they are observed for fewer years by 2019. Lastly, \Cref{sec:oa_specfigs} collects, for each trend specification, the fitted trend against the raw national series and the resulting residual variation.

\begin{figure}[!h]
	\centering
	\begin{minipage}{\textwidth}
	\caption{Aggregated attainment variables and quadratic fits}
    \vspace{-1em}
    \label{fig:trendfit_grad}
    \begin{center}
    \includegraphics[width=0.95\textwidth]{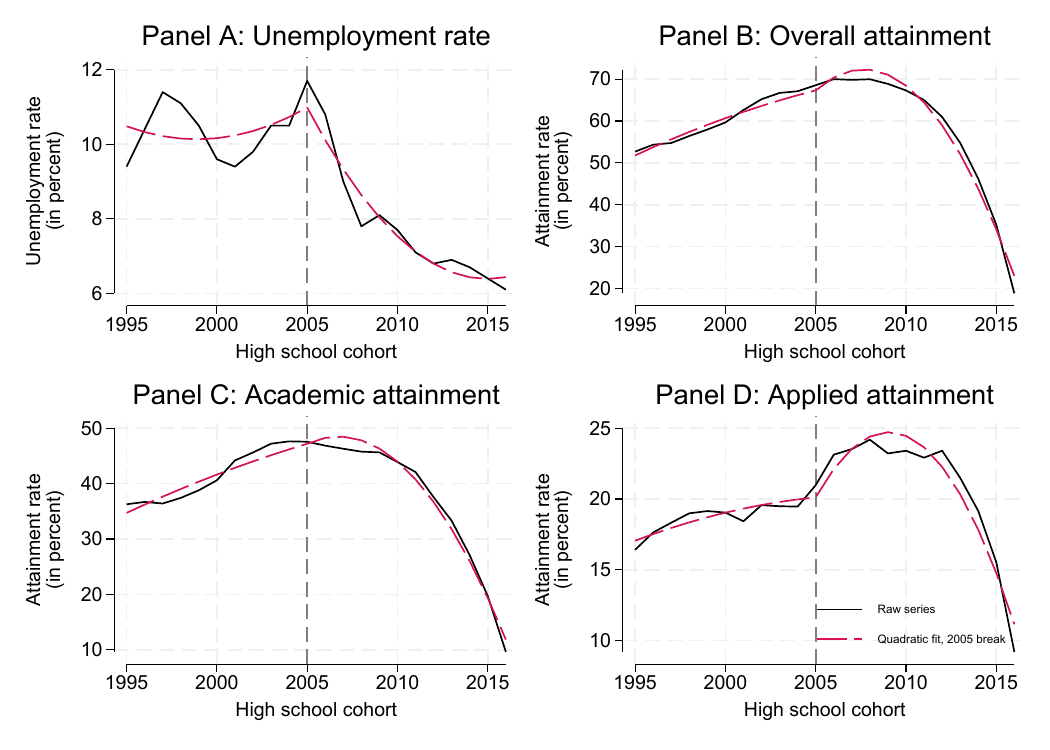} \\
    \end{center}
    \vspace{-1em}
    {\scriptsize \textit{Notes:} Each panel plots the raw series (solid black) and the fitted quadratic trend with a continuous slope break in 2005 (red long dash, with the vertical line marking the break year). Panel A shows the unemployment rate, Panel B the overall share of high school graduates with \textit{Abitur} obtaining a degree by 2019, Panel C the share obtaining an academic-university degree, and Panel D the share obtaining an applied-university degree. The fitted trend takes the form of \Cref{eq:trend}, shown here at the national level. In the estimating equation the trend is estimated separately by state. \\ \textit{Source}: Exam register, Federal Employment Agency (BA), \textit{Regionaldatenbank}, high school cohorts 1995--2016.
    \par}
    \end{minipage}
\end{figure}

\begin{figure}[!h]
	\centering
	\begin{minipage}{\textwidth}
	\caption{Aggregated business cycle variation in attainment and unemployment}
    \vspace{-1em}
 	\label{fig:kg2x2_grad}
	\begin{center}
	\includegraphics[width=0.95\textwidth]{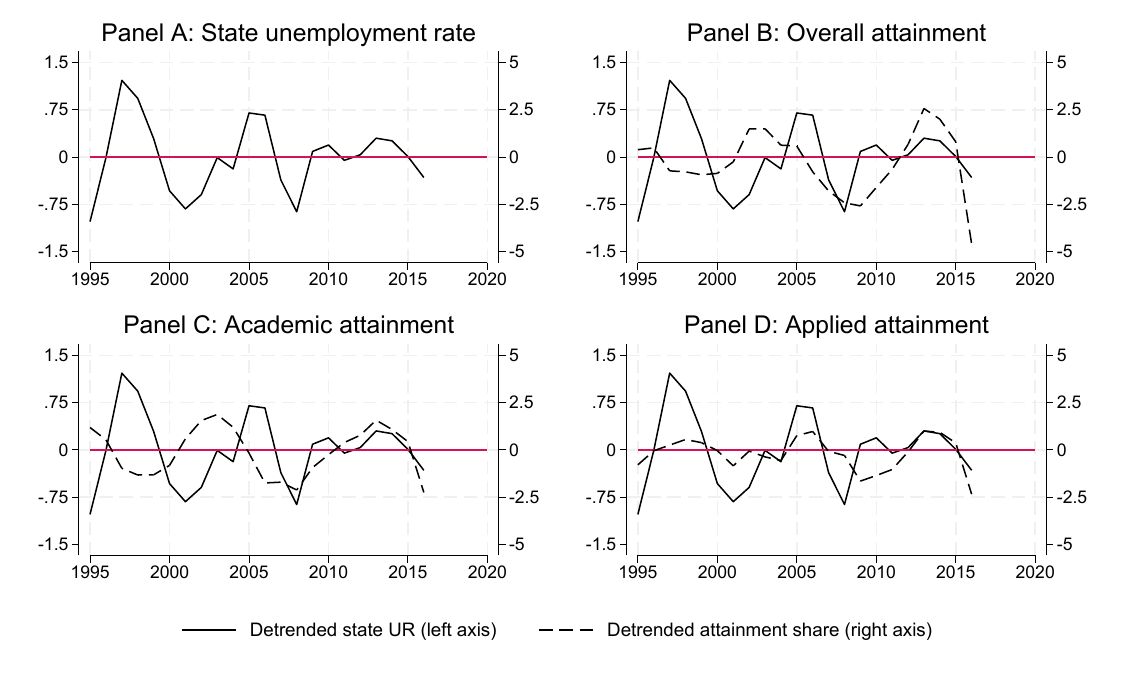} \\
	\end{center}
    \vspace{-1em}
    {\scriptsize \textit{Notes:} Each panel plots, for each high school cohort, residuals after removing state fixed effects and state-specific quadratic trends with a continuous 2005 break (\Cref{eq:trend}), estimated on the 1995--2016 attainment sample. Panel A shows the residual state unemployment rate. Panels B, C, and D overlay the residual state unemployment rate (solid line, left axis) with the residual share of high school graduates with \textit{Abitur} obtaining any, an academic-university, and an applied-university degree by 2019 (dashed line, right axis). Residuals are aggregated by cohort year and weighted by the number of \textit{Abitur} holders. The red line marks the zero mean. \\ \textit{Source}: Exam register, Federal Employment Agency (BA), \textit{Regionaldatenbank}, high school cohorts 1995--2016.
    \par}
    \end{minipage}
\end{figure}

\Cref{tab:ur_persistence} regresses the detrended state UR $k$ years ahead on the current deviation. A 1~pp deviation predicts a 0.47~pp deviation in the following year, when the first postsecondary investment decision is made. Over the horizon at which a university degree would be completed, the deviation fades and reverses, and reaches -0.56~pp after three years. Part of this reversal is mechanical, as residuals from a fitted trend must average to zero within each state. 

Current deviations remain statistically significant predictors at all horizons, but the sign changes after two years, i.e., one year after graduation in my main analysis. The reversal is most pronounced at $k=3$ and $k=4$, where the coefficients exceed the $k=1$ coefficient in absolute value, and fades toward the end of the window, where the point estimates at $k=5$ and $k=6$ are less than half as large as at $k=1$, with 40--50 percent larger standard errors. A student extrapolating from current conditions thus receives a strong signal about entry conditions in the immediate future. The signal about conditions at the horizon of university graduation, however, is of opposite sign over most of that horizon and weaker toward its end. The horizon of the sign reversal, negative from $k=2$ through $k=5$ and positive again at $k=6$, is consistent with the dated chronology of German business cycles, in which the complete cycles of the 2000s lasted six to seven years \citepsupp{breuer2022svr}.

\begin{table}[H]
    \begin{center}
	\begin{adjustbox}{max width=0.9\textwidth}
	\begin{threeparttable}
		\caption{Persistence of cyclical unemployment deviations}
		\label{tab:ur_persistence}
		\scriptsize{
		        \begin{tabular} {l cccccc}
	\toprule \toprule
    & \multicolumn{6}{c}{Detrended state UR in $t+k$} \\
    \cmidrule(lr){2-7}
    & $k=1$ & $k=2$ & $k=3$ & $k=4$ & $k=5$ & $k=6$ \\
 & (1) & (2) & (3) & (4) & (5) & (6)\\
\midrule
Detrended state UR ($t$)    &       0.467***&      -0.264***&      -0.555***&      -0.513***&      -0.213***&       0.188***\\
                            &     (0.018)   &     (0.018)   &     (0.022)   &     (0.026)   &     (0.025)   &     (0.027)   \\
&&&&&&\\
No. state-year cells        &         365   &         349   &         333   &         317   &         301   &         285   \\
				\bottomrule
		\end{tabular}
		}
		\begin{tablenotes}[flushleft]
				\item \scriptsize{ \textit{Notes:} This table presents estimates from regressions of the detrended state unemployment rate $k$ years ahead on the current detrended state unemployment rate. Deviations are residuals from state fixed effects and state-specific quadratic trends with a 2005 break, estimated on the state-year panel 1995--2018. Standard errors in parentheses allow for clustering at the state level. All coefficients have wild-cluster-bootstrap $p$-values below 0.001 (Rademacher weights, 9,999 replications). * $p$\textless 0.1, ** $p$\textless 0.05, *** $p$\textless 0.01. \\ \textit{Source}: Federal Employment Agency (BA), years 1995--2018.
                }
		\end{tablenotes}
	\end{threeparttable}
	\end{adjustbox}
    \end{center}
\end{table}

\Cref{fig:detrended_ur_global} plots the residual state unemployment rate and college enrollment rate over high school cohorts, after removing state fixed effects and the state-specific quadratic trends with a 2005 break of the main specification. These residuals are the cyclical deviations effectively used in \Cref{eq:specification}.

\begin{figure}[!h]
	\centering
	\begin{minipage}{\textwidth}
 	\caption{Detrended state unemployment and college enrollment over time}
    \vspace{-0.5em}
    \label{fig:detrended_ur_global}
    \begin{center}
    \begin{subfigure}[b]{\textwidth}
        \centering
        \small{Panel A: Detrended state UR}
        \includegraphics[width=0.8\textwidth]{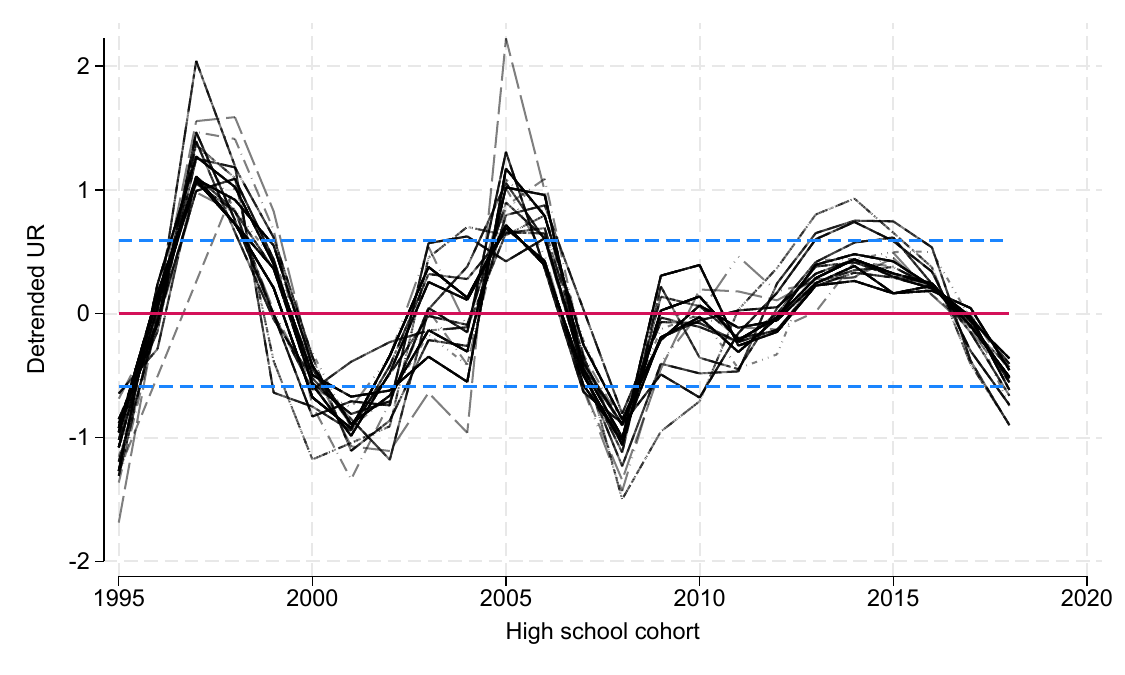}
    \end{subfigure}
    \vspace{0.5em}
    \begin{subfigure}[b]{\textwidth}
        \centering
        \small{Panel B: Detrended CZ-level enrollment (any college)}
        \includegraphics[width=0.8\textwidth]{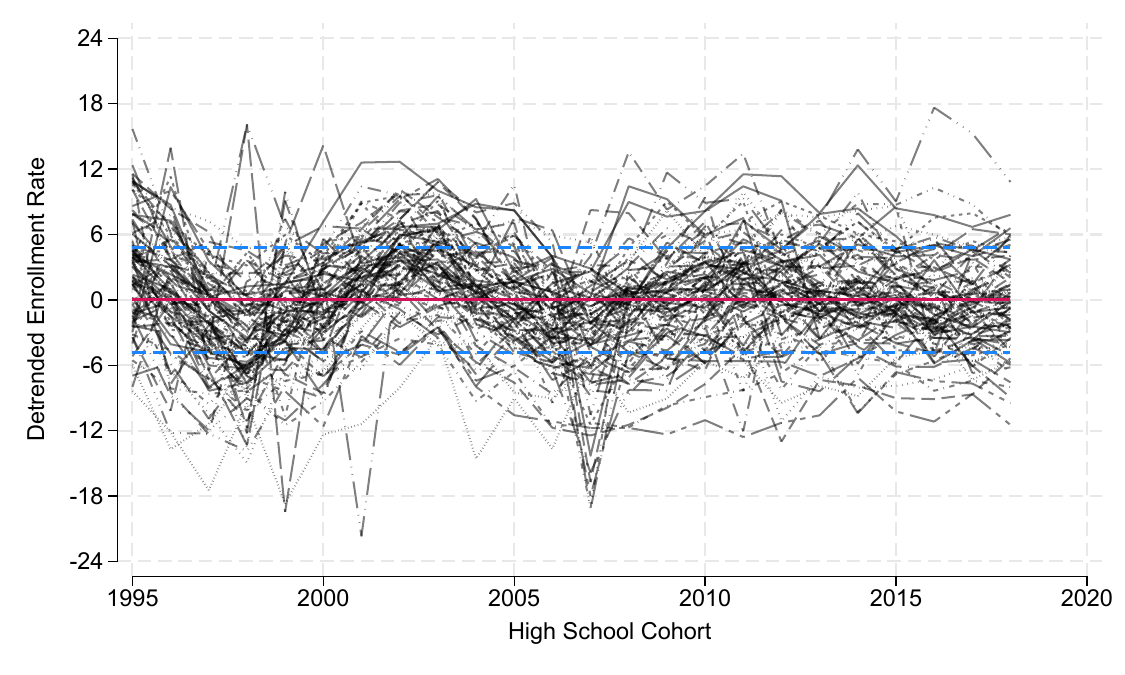}
    \end{subfigure}
    \end{center}
    \vspace{-1em}
    {\scriptsize \textit{Notes:} This figure plots, by high school cohort, the residual state UR (Panel A) and CZ-level college enrollment rate (Panel B) after controlling for state fixed effects and state-specific quadratic time trends with a 2005 break. Each thin line is a region. The zero mean is in red and the $\pm 1$ standard deviation bands in blue.\\ \textit{Source}: Student register, Federal Employment Agency (BA), years 1995--2018.
    \par}
    \end{minipage}
\end{figure}

\Cref{fig:oa_distributions_UR} and \Cref{fig:oa_distributions_enroll} plot kernel densities of the residualized state-level unemployment rate and college enrollment share after progressively saturating the regression specification. Starting from the residual that absorbs only state fixed effects (solid), each subsequent series adds state-specific quadratic trends with a 2005 break, year fixed effects, or both. State-specific trends absorb substantially more variation in the state UR than year fixed effects do on top of the state-FE baseline. Combining the two leaves very little variation in the treatment.

\begin{figure}[!h]
    \centering
    \begin{minipage}{\textwidth}
    \caption{Distributions of state-level UR}
    \vspace{-1em}
    \label{fig:oa_distributions_UR}
    \begin{center}
    \includegraphics[scale=0.73]{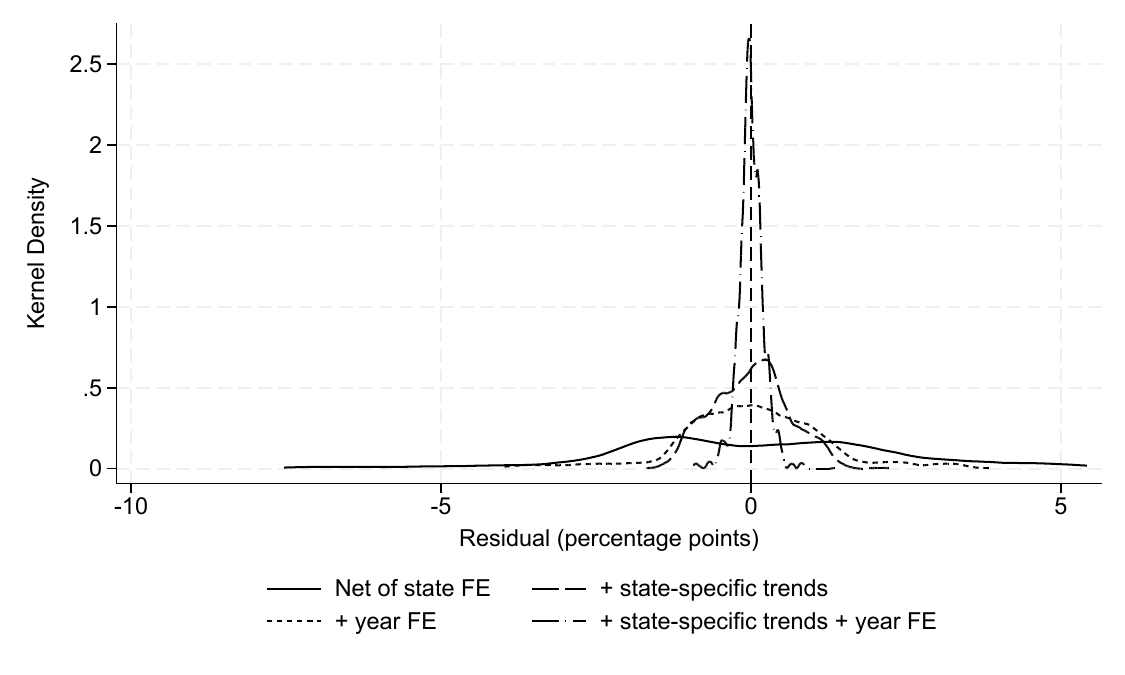} \\
    \end{center}
    \vspace{-1em}
    {\scriptsize \textit{Notes:} Kernel densities of state-level unemployment rate residuals. The baseline (solid line) absorbs state fixed effects. The remaining three series additionally absorb state-specific quadratic trends with a 2005 break (long dashes), year fixed effects (short dashes), or both (long dashes with dots). \\ \textit{Source}: Federal Employment Agency, years 1995--2018.
    \par}
    \end{minipage}
\end{figure}


\begin{figure}[!h]
    \centering
    \begin{minipage}{\textwidth}
    \caption{Distributions of college enrollment share}
    \vspace{-1em}
    \label{fig:oa_distributions_enroll}
    \begin{center}
    \includegraphics[scale=0.73]{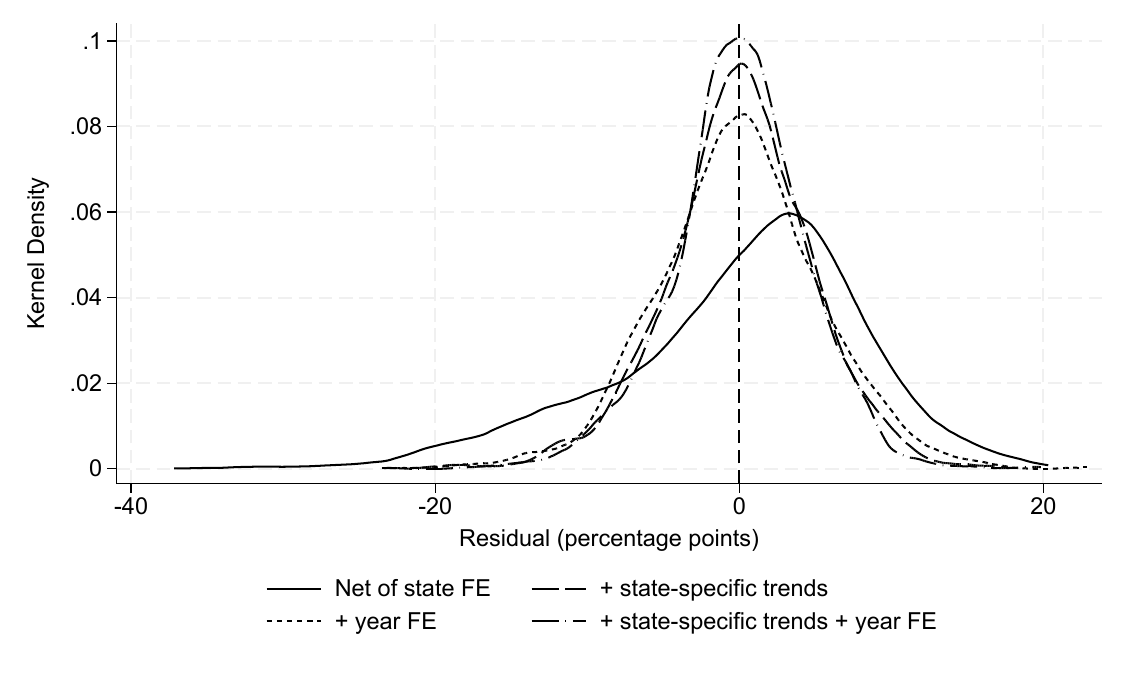} \\
    \end{center}
    \vspace{-1em}
    {\scriptsize \textit{Notes:} Kernel densities of college enrollment share residuals across regions and years. The baseline (solid line) absorbs state fixed effects. The remaining three series additionally absorb state-specific quadratic trends with a 2005 break (long dashes), year fixed effects (short dashes), or both (long dashes with dots). \\ \textit{Source}: Federal Statistical Office (Hochschulstatistik), years 1995--2018.
    \par}
    \end{minipage}
\end{figure}

\FloatBarrier
\clearpage
\subsection{Fitted Trends and Identifying Variation by Specification}
\label{sec:oa_specfigs}

\begin{figure}[H]
	\centering
	\begin{minipage}{\textwidth}
	\caption{Fitted trend and identifying variation: quadratic trend without a break}
 	\label{fig:fit_quad_nb}
	\begin{center}
	\includegraphics[width=0.7\textwidth]{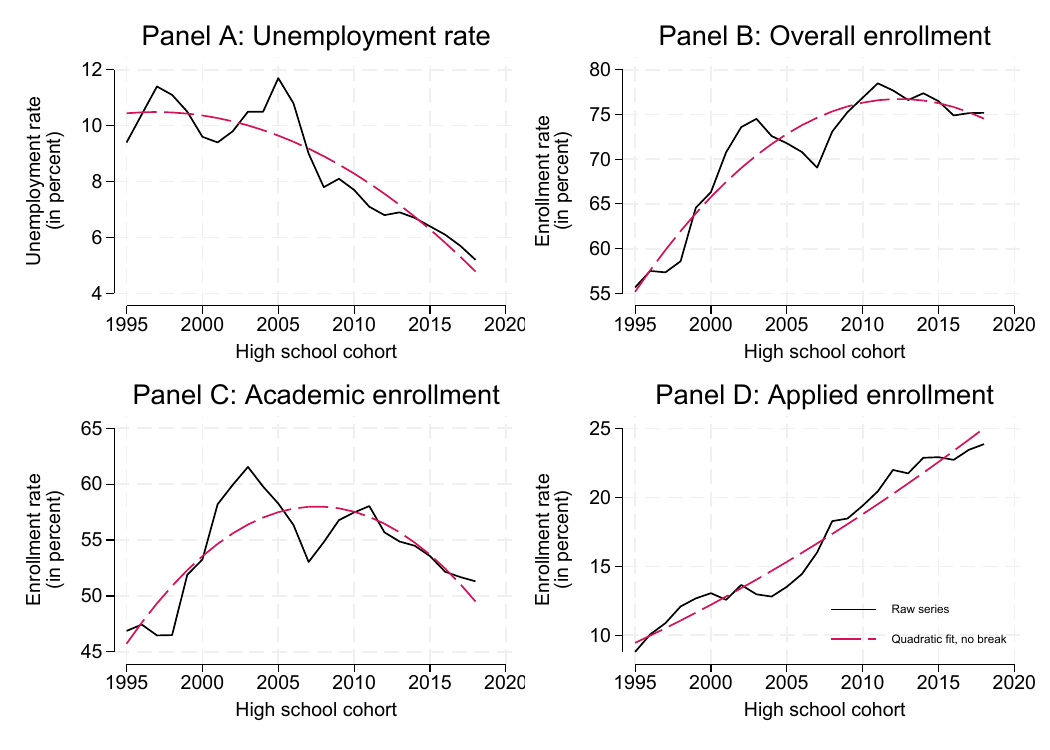} \\
	\vspace{0.3em}
	\includegraphics[width=0.46\textwidth]{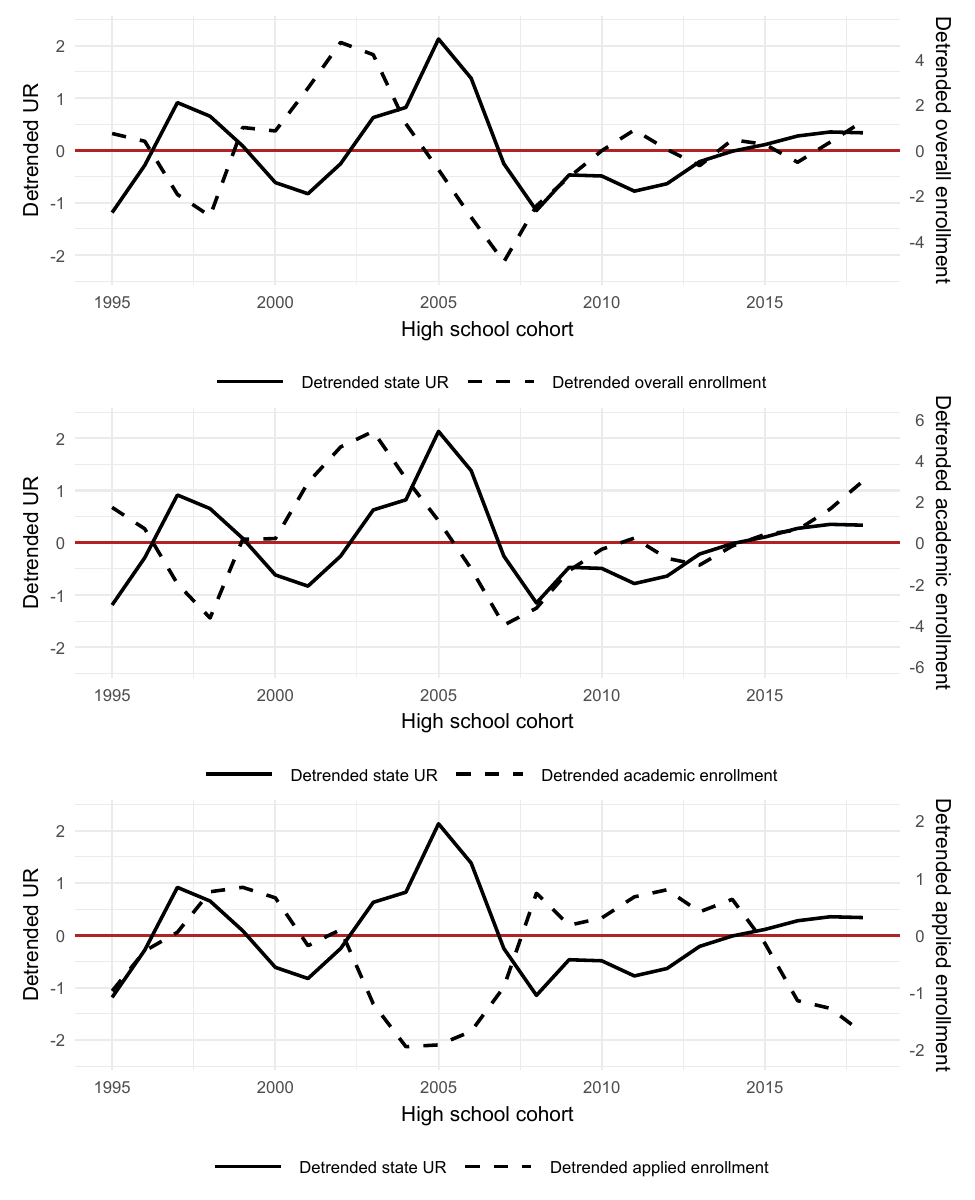}
	\end{center}
    {\scriptsize \textit{Notes:} The top panel shows the national series (solid line) and the fitted quadratic trend without a break (long dashes) by high school cohort, weighted by the number of \textit{Abitur} holders, for the unemployment rate and the overall, academic-university, and applied-university enrollment rates. The bottom panel shows, for each high school cohort, the residual state unemployment rate (solid line) and the residual enrollment rates (dashed line) after removing state fixed effects and the same trend. Residuals are aggregated by cohort year and weighted by the number of \textit{Abitur} holders. The red line marks the zero mean. \\ \textit{Source}: Student register, Federal Employment Agency (BA), Federal Statistical Office (Hochschulstatistik).
    \par}
    \end{minipage}
\end{figure}

\begin{figure}[H]
	\centering
	\begin{minipage}{\textwidth}
	\caption{Fitted trend and identifying variation: cubic trend without a break}
 	\label{fig:fit_cub_nb}
	\begin{center}
	\includegraphics[width=0.7\textwidth]{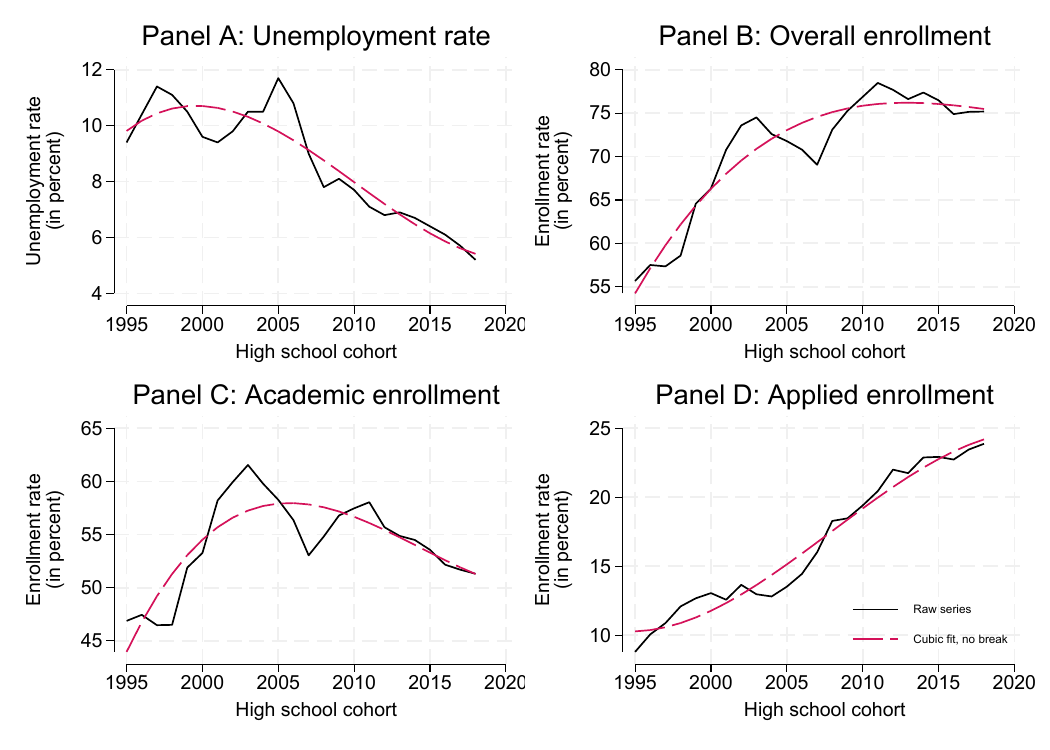} \\
	\vspace{0.3em}
	\includegraphics[width=0.46\textwidth]{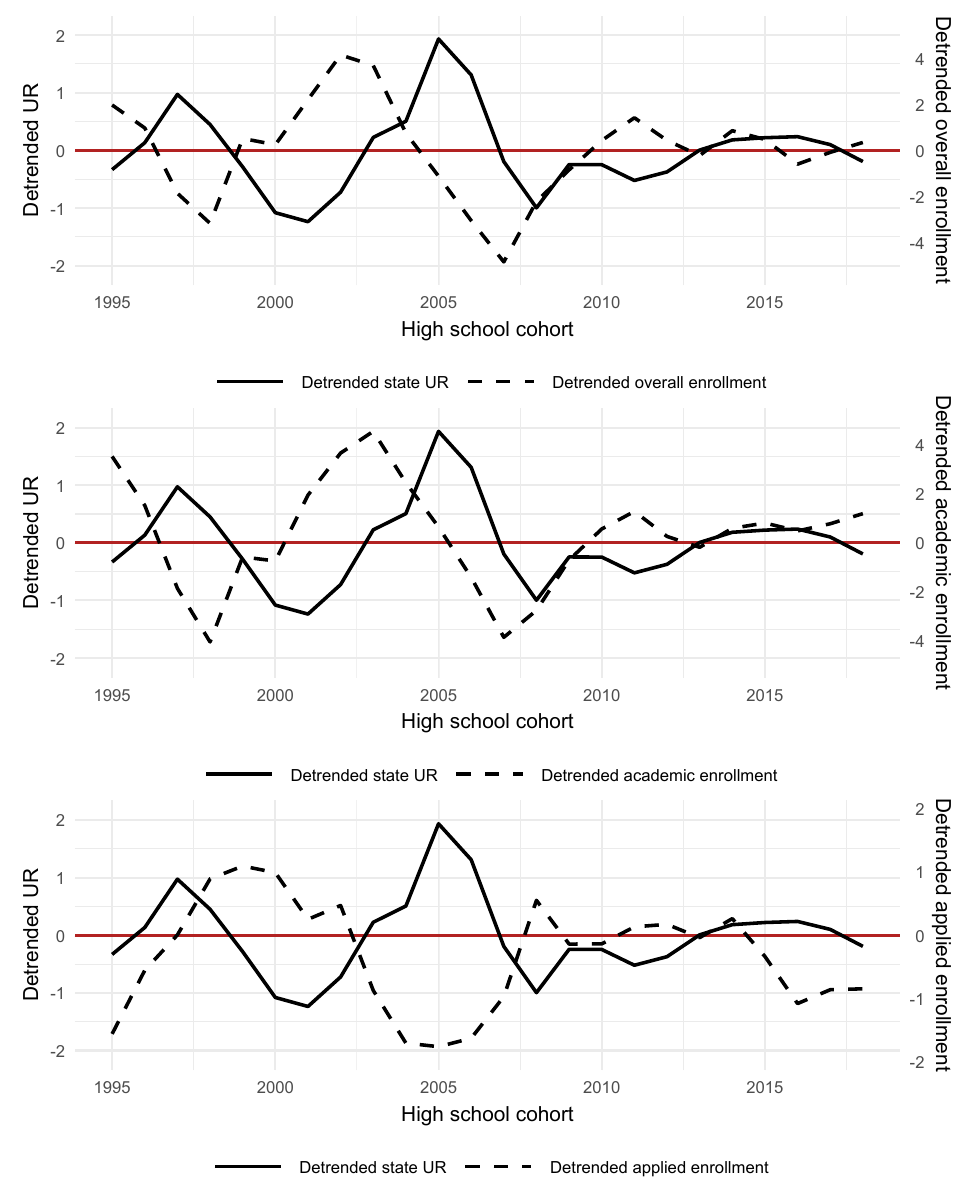}
	\end{center}
    {\scriptsize \textit{Notes:} The top panel shows the national series (solid line) and the fitted cubic trend without a break (long dashes) by high school cohort, weighted by the number of \textit{Abitur} holders, for the unemployment rate and the overall, academic-university, and applied-university enrollment rates. The bottom panel shows, for each high school cohort, the residual state unemployment rate (solid line) and the residual enrollment rates (dashed line) after removing state fixed effects and the same trend. Residuals are aggregated by cohort year and weighted by the number of \textit{Abitur} holders. The red line marks the zero mean. \\ \textit{Source}: Student register, Federal Employment Agency (BA), Federal Statistical Office (Hochschulstatistik).
    \par}
    \end{minipage}
\end{figure}

\begin{figure}[H]
	\centering
	\begin{minipage}{\textwidth}
	\caption{Fitted trend and identifying variation: linear trend with a 2005 break}
 	\label{fig:fit_lin_br}
	\begin{center}
	\includegraphics[width=0.7\textwidth]{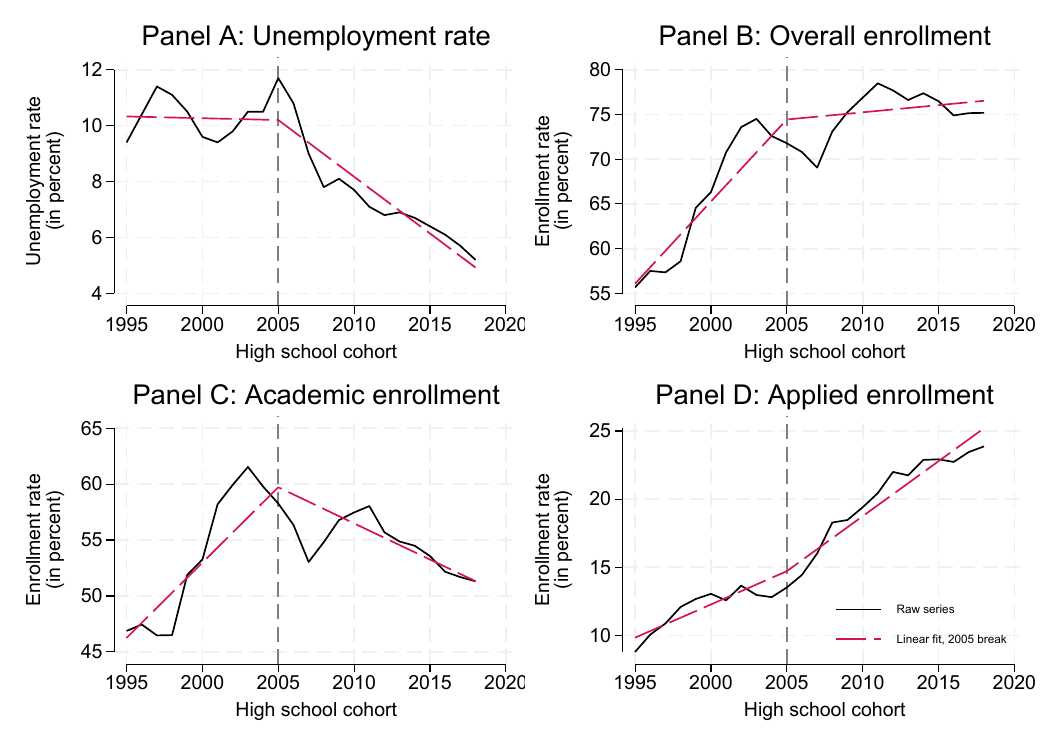} \\
	\vspace{0.3em}
	\includegraphics[width=0.46\textwidth]{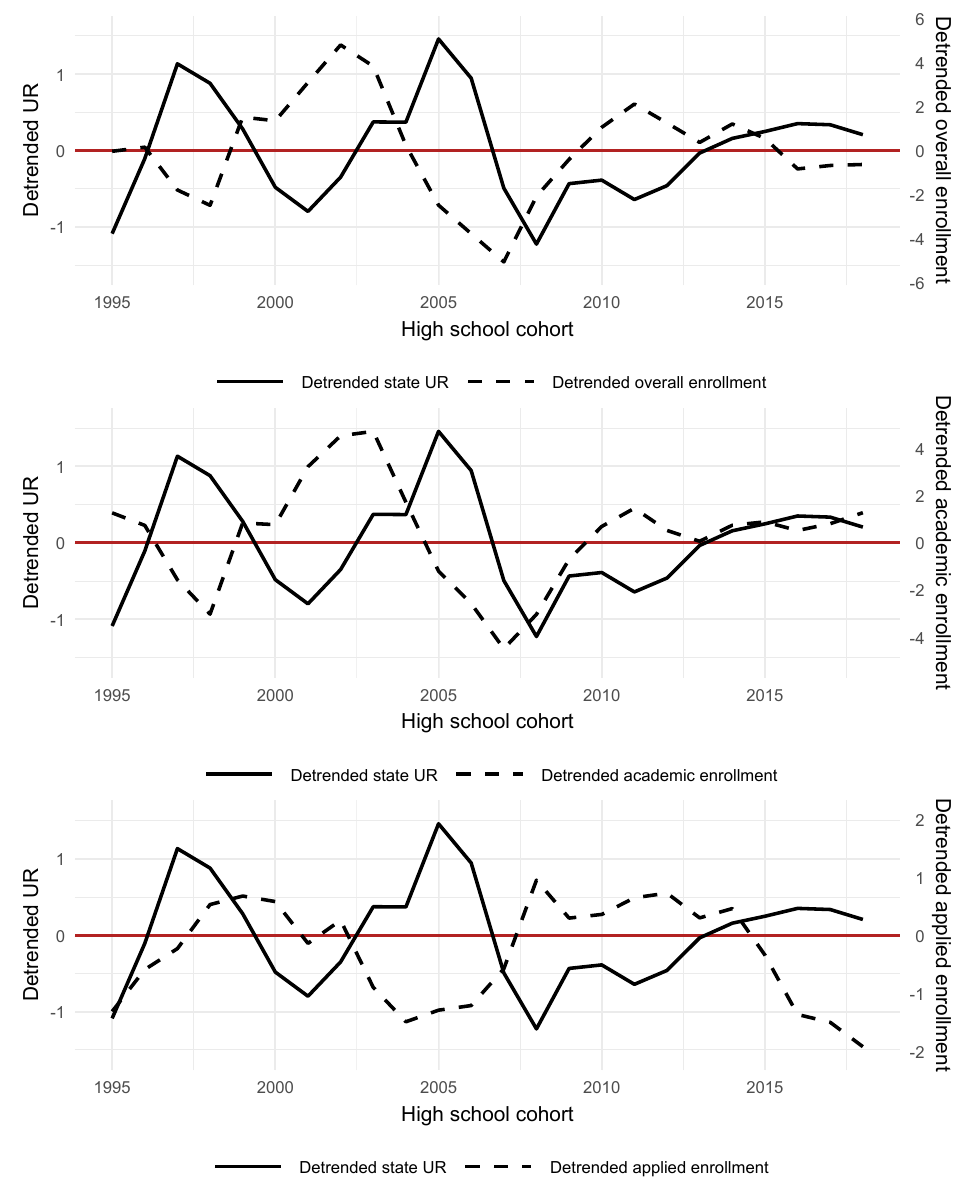}
	\end{center}
    {\scriptsize \textit{Notes:} The top panel shows the national series (solid line) and the fitted linear trend with a 2005 break (long dashes) by high school cohort, weighted by the number of \textit{Abitur} holders, for the unemployment rate and the overall, academic-university, and applied-university enrollment rates. The bottom panel shows, for each high school cohort, the residual state unemployment rate (solid line) and the residual enrollment rates (dashed line) after removing state fixed effects and the same trend. Residuals are aggregated by cohort year and weighted by the number of \textit{Abitur} holders. The red line marks the zero mean. \\ \textit{Source}: Student register, Federal Employment Agency (BA), Federal Statistical Office (Hochschulstatistik).
    \par}
    \end{minipage}
\end{figure}

\begin{figure}[H]
	\centering
	\begin{minipage}{\textwidth}
	\caption{Fitted trend and identifying variation: cubic trend with a 2005 break}
 	\label{fig:fit_cub_br}
	\begin{center}
	\includegraphics[width=0.7\textwidth]{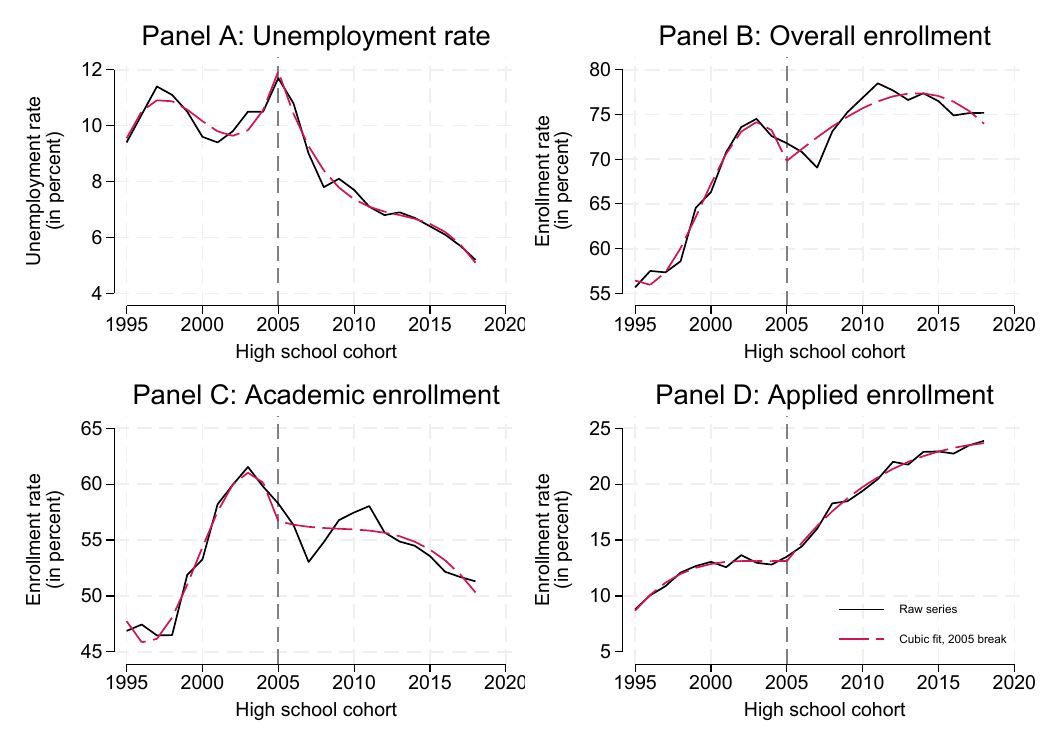} \\
	\vspace{0.3em}
	\includegraphics[width=0.46\textwidth]{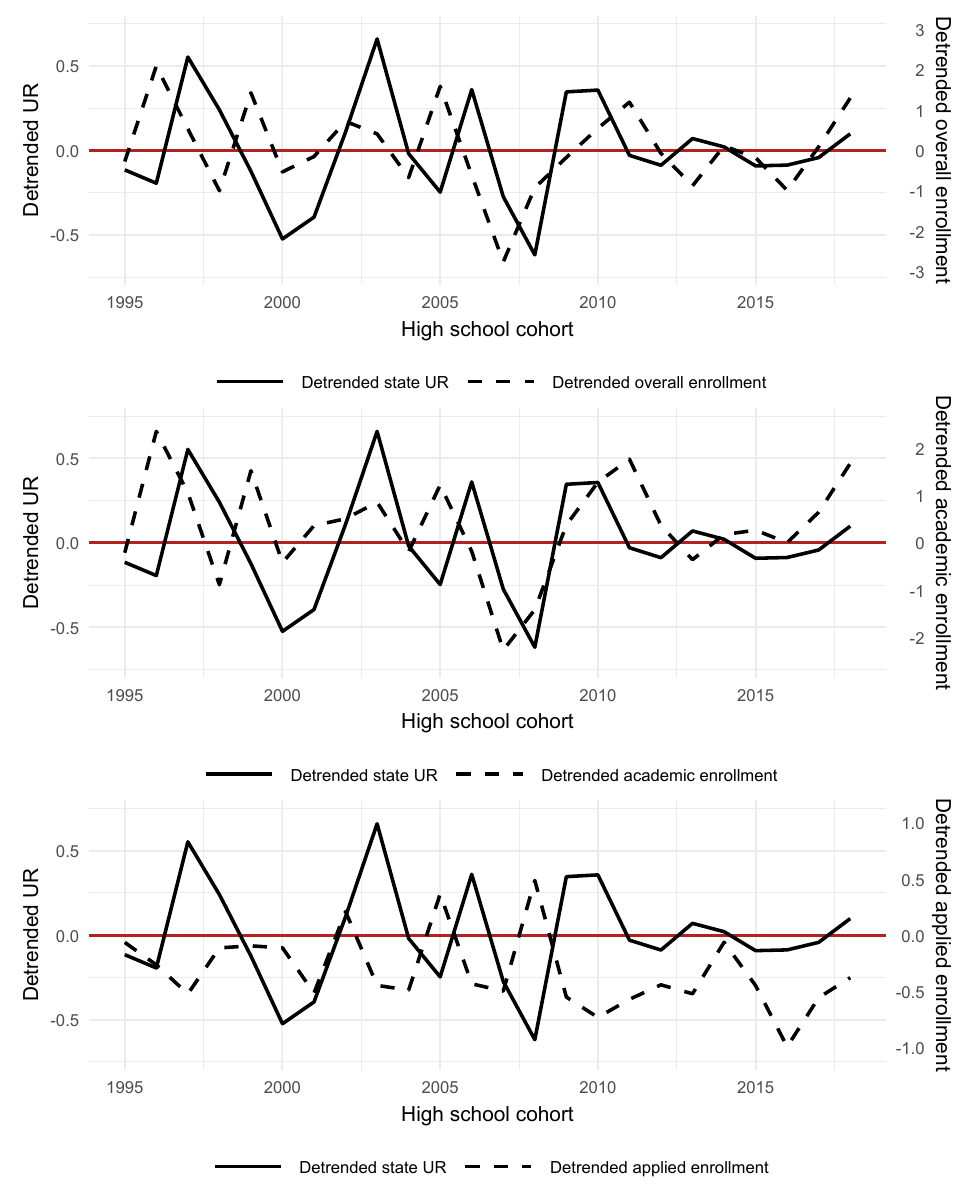}
	\end{center}
    {\scriptsize \textit{Notes:} The top panel shows the national series (solid line) and the fitted cubic trend with a 2005 break (long dashes) by high school cohort, weighted by the number of \textit{Abitur} holders, for the unemployment rate and the overall, academic-university, and applied-university enrollment rates. The bottom panel shows, for each high school cohort, the residual state unemployment rate (solid line) and the residual enrollment rates (dashed line) after removing state fixed effects and the same trend. Residuals are aggregated by cohort year and weighted by the number of \textit{Abitur} holders. The red line marks the zero mean. \\ \textit{Source}: Student register, Federal Employment Agency (BA), Federal Statistical Office (Hochschulstatistik).
    \par}
    \end{minipage}
\end{figure}
\FloatBarrier
\clearpage
\section{Heterogeneity}
\label{sec:heterogeneity}

This section reports the main enrollment and attainment effects by gender in the administrative data, and the heterogeneity of the intention response in the panel study. Heterogeneity of the expectation measures by gender, SES, and GPA is presented in \Cref{sec:expectations}, and by gender and SES in the SOEP in \Cref{sec:soep_heterogeneity}.

\Cref{tab:period_split} reports the senior-year enrollment and attainment effects separately for the periods before and after the completion of the Bologna Process in 2008, \Cref{tab:decomp_period} splits the decomposed national and state-specific components of \Cref{tab:abi_enroll_decomp} by the same periods, and \Cref{tab:gradyear_ur} reports the effects of the unemployment rate in the graduation year.

\begin{table}[H]
    \begin{center}
	\begin{adjustbox}{max width=0.85\textwidth}
	\begin{threeparttable}
		\caption{Enrollment and attainment effects before and after 2008}
		\label{tab:period_split}
		\scriptsize{
		        \begin{tabular} {l ccc ccc}
	\toprule \toprule
& \multicolumn{3}{c}{Enrollment} & \multicolumn{3}{c}{Attainment}  \\
\cmidrule(lr){2-4} \cmidrule(lr){5-7}
& pre 2008 & post 2008 & p-value ($\beta_{\Delta}$) & pre 2008 & post 2008 & p-value ($\beta_{\Delta}$) \\
& (1) & (2) & (3) & (4) & (5) & (6) \\
\midrule
\textit{Panel A: Overall (any college)} & & & & & & \\
State UR (t-1)   &      -2.243***&      -2.416***&       [0.466] &      -0.999** &      -1.473** &       [0.051] \\
                 &     (0.416)   &     (0.556)   &               &     (0.413)   &     (0.526)   &               \\
\addlinespace
\textit{Panel B: Academic university} & & & & & & \\
State UR (t-1)   &      -2.249***&      -2.498***&       [0.294] &      -1.309***&      -1.667***&       [0.019] \\
                 &     (0.363)   &     (0.475)   &               &     (0.251)   &     (0.328)   &               \\
\addlinespace
\textit{Panel C: Applied university} & & & & & & \\
State UR (t-1)   &       0.052   &       0.142   &       [0.195] &       0.332*  &       0.210   &       [0.294] \\
                 &     (0.095)   &     (0.159)   &               &     (0.179)   &     (0.245)   &               \\
\midrule
No. CZ-cohort cells       &       2,289   &       2,289   &  &       2,097   &       2,097   &  \\
No. high school graduates &   6,296,662   &   6,296,662   &  &   5,727,918   &   5,727,918   &  \\
State FE, trends, policy controls &      yes      &      yes      &  &      yes      &      yes      &  \\
				\bottomrule
		\end{tabular}
		}
		\begin{tablenotes}[flushleft]
				\item \scriptsize{ \textit{Notes:} This table presents estimates from \Cref{eq:specification} with the senior-year state unemployment rate interacted with dummies for the period before and after 2008. The outcome is the share of high school graduates with \textit{Abitur} enrolling at (columns 1--3) or graduating from (columns 4--6) any college (Panel A), academic universities (Panel B), or applied universities (Panel C). Enrollment spans high school cohorts 1995--2018. Attainment spans cohorts 1995--2016 and is measured by 2019. P-values in brackets refer to tests of equal coefficients across periods. Policy controls include tuition fee introductions and abolitions, as well as double cohorts. Each cell is weighted by the number of high school graduates with \textit{Abitur}. Standard errors in parentheses allow for two-way clustering at the state and cohort level. * $p$\textless 0.1, ** $p$\textless 0.05, *** $p$\textless 0.01. \\ \textit{Source}: Student register, exam register, Federal Employment Agency, \textit{Regionaldatenbank}.
                }
		\end{tablenotes}
	\end{threeparttable}
	\end{adjustbox}
    \end{center}
\end{table}

\Cref{tab:gender} estimates the main enrollment and attainment effects separately for men and women in the administrative data. Point estimates are slightly larger for women for enrollment and academic-university attainment, and the increase in applied-university attainment is statistically significant among women only. None of the gender differences is statistically significant (columns 3 and 6).

\begin{table}[!h]
    \begin{center}
	\begin{adjustbox}{max width=0.85\textwidth}
	\begin{threeparttable}
		\caption{Enrollment and attainment effects by gender}
		\label{tab:gender}
		\scriptsize{
		        \begin{tabular} {l ccc ccc}
	\toprule \toprule
& \multicolumn{3}{c}{Enrollment} & \multicolumn{3}{c}{Attainment}  \\
\cmidrule(lr){2-4} \cmidrule(lr){5-7}
& Men & Women & p-value ($\beta_{\Delta}$) & Men & Women & p-value ($\beta_{\Delta}$) \\
& (1) & (2) & (3) & (4) & (5) & (6) \\
\midrule
\textit{Panel A: Overall (any college)} & & & & & & \\
State UR (t-1)   &      -1.943***&      -2.134***&       [0.746] &      -0.570   &      -0.714   &       [0.495] \\
                 &     (0.609)   &     (0.381)   &               &     (0.356)   &     (0.464)   &               \\
Outcome mean     &        74.8   &        68.9   &               &        59.1   &        57.7   &               \\
\addlinespace
\textit{Panel B: Academic university} & & & & & & \\
State UR (t-1)   &      -1.912***&      -2.075***&       [0.768] &      -0.925***&      -1.157***&       [0.257] \\
                 &     (0.600)   &     (0.359)   &               &     (0.218)   &     (0.310)   &               \\
Outcome mean     &        55.3   &        53.7   &               &        37.0   &        39.7   &               \\
\addlinespace
\textit{Panel C: Applied university} & & & & & & \\
State UR (t-1)   &      -0.032   &      -0.059   &       [0.812] &       0.355   &       0.443** &       [0.272] \\
                 &     (0.122)   &     (0.083)   &               &     (0.205)   &     (0.201)   &               \\
Outcome mean     &        19.5   &        15.3   &               &        22.2   &        18.1   &               \\
\midrule
No. CZ-cohort cells       &       2,289   &       2,289   &  &       2,097   &       2,097   &  \\
No. high school graduates &   2,711,722   &   3,497,691   &  &   2,453,019   &   3,187,650   &  \\
State FE, trends, policy controls &      yes      &      yes      &  &      yes      &      yes      &  \\
				\bottomrule
		\end{tabular}
		}
		\begin{tablenotes}[flushleft]
				\item \scriptsize{ \textit{Notes:} This table presents estimates from \Cref{eq:specification} for the share of male and female high school graduates with \textit{Abitur} enrolling at (columns 1--3) or graduating from (columns 4--6) each college type. Enrollment spans high school cohorts 1995--2018. Attainment spans cohorts 1995--2016 and is measured by 2019. P-values in brackets refer to tests of equal coefficients across genders, based on a fully interacted regression on the stacked gender-specific cells. Policy controls include tuition fee introductions and abolitions, as well as double cohorts. Each cell is weighted by the number of male or female high school graduates with \textit{Abitur}. The sex-specific graduate counts do not add up to the pooled totals of \Cref{tab:main_results}. The sex-specific \textit{Regionaldatenbank} series miss 87,249 graduates (1.4 percent), almost entirely in Saxony and Saxony-Anhalt before 2008, where some graduates appear in neither sex-specific series. Standard errors in parentheses allow for two-way clustering at the state and cohort level. * $p$\textless 0.1, ** $p$\textless 0.05, *** $p$\textless 0.01. \\ \textit{Source}: Student register, exam register, Federal Employment Agency, \textit{Regionaldatenbank}.
                }
		\end{tablenotes}
	\end{threeparttable}
	\end{adjustbox}
    \end{center}
\end{table}


\begin{table}[H]
    \begin{center}
	\begin{adjustbox}{max width=0.85\textwidth}
	\begin{threeparttable}
		\caption{Decomposed effects before and after 2008}
		\label{tab:decomp_period}
		\scriptsize{
		        \begin{tabular} {l ccc ccc}
	\toprule \toprule
& \multicolumn{3}{c}{Enrollment} & \multicolumn{3}{c}{Attainment}  \\
\cmidrule(lr){2-4} \cmidrule(lr){5-7}
& pre 2008 & post 2008 & p-value ($\beta_{\Delta}$) & pre 2008 & post 2008 & p-value ($\beta_{\Delta}$) \\
& (1) & (2) & (3) & (4) & (5) & (6) \\
\midrule
\textit{Panel A: Overall (any college)} & & & & & & \\
National UR (t-1)        &      -2.606***&      -2.937***&       [0.104] &      -1.220***&      -2.002***&       [0.011] \\
                         &     (0.436)   &     (0.539)   &               &     (0.395)   &     (0.479)   &               \\
$\text{UR}^\Delta$ (t-1) &      -0.396   &      -0.265   &       [0.653] &      -0.039   &       0.397   &       [0.029] \\
                         &     (0.736)   &     (0.794)   &               &     (0.424)   &     (0.514)   &               \\
\addlinespace
\textit{Panel B: Academic university} & & & & & & \\
National UR (t-1)        &      -2.711***&      -3.187***&       [0.020] &      -1.574***&      -2.112***&       [0.002] \\
                         &     (0.398)   &     (0.449)   &               &     (0.244)   &     (0.279)   &               \\
$\text{UR}^\Delta$ (t-1) &       0.067   &       0.310   &       [0.225] &       0.026   &       0.080   &       [0.476] \\
                         &     (0.531)   &     (0.575)   &               &     (0.405)   &     (0.438)   &               \\
\addlinespace
\textit{Panel C: Applied university} & & & & & & \\
National UR (t-1)        &       0.104   &       0.250   &       [0.256] &       0.354   &       0.111   &       [0.138] \\
                         &     (0.131)   &     (0.251)   &               &     (0.204)   &     (0.283)   &               \\
$\text{UR}^\Delta$ (t-1) &      -0.463** &      -0.576** &       [0.489] &      -0.064   &       0.317   &       [0.005] \\
                         &     (0.208)   &     (0.234)   &               &     (0.205)   &     (0.248)   &               \\
\midrule
No. CZ-cohort cells       &       2,289   &       2,289   &  &       2,097   &       2,097   &  \\
No. high school graduates &   6,296,662   &   6,296,662   &  &   5,727,918   &   5,727,918   &  \\
State FE, trends, policy controls &      yes      &      yes      &  &      yes      &      yes      &  \\
				\bottomrule
		\end{tabular}
		}
		\begin{tablenotes}[flushleft]
				\item \scriptsize{ \textit{Notes:} This table presents estimates from \Cref{eq:mechanisms}, splitting the senior-year ($t-1$) state unemployment rate into the national UR and the state-specific deviation $\text{UR}^\Delta$, each interacted with dummies for the period before and after 2008. The outcome is the share of high school graduates with \textit{Abitur} enrolling at (columns 1--3) or graduating from (columns 4--6) any college (Panel A), academic universities (Panel B), or applied universities (Panel C). Enrollment spans high school cohorts 1995--2018. Attainment spans cohorts 1995--2016 and is measured by 2019. P-values in brackets refer to tests of equal coefficients across periods. Policy controls include tuition fee introductions and abolitions, as well as double cohorts. Each cell is weighted by the number of high school graduates with \textit{Abitur}. Standard errors in parentheses allow for two-way clustering at the state and cohort level. * $p$\textless 0.1, ** $p$\textless 0.05, *** $p$\textless 0.01. \\ \textit{Source}: Student register, exam register, Federal Employment Agency, \textit{Regionaldatenbank}.
                }
		\end{tablenotes}
	\end{threeparttable}
	\end{adjustbox}
    \end{center}
\end{table}

\FloatBarrier
\clearpage
\section{Apprenticeships}
\label{sec:outsideopt}

For German high school graduates, the outside option to college is apprenticeships. I use administrative state-level data from the \textit{Berufsbildungsstatistik} (1995--2018) on the universe of new apprenticeship contracts by school degree provided by the Federal Institute for Vocational Education and Training (BIBB), and complement this with CZ-level data from the \textit{Ausbildungsmarktstatistik} of the Federal Employment Agency (BA) on apprenticeship applicants and positions reported between 2008 and 2018.

\Cref{tab:vocational_main} presents three sets of estimates: Panel A focuses on new apprentices and applicants with a high school degree, Panel B on total new apprenticeship contracts, applicants, and positions, and Panel C on the share of apprentices and applicants with a high school degree. Panel A shows a large and positive effect of state-level unemployment on the number of new apprentices with a high school diploma. A 1~pp increase in the state UR in t-1 is associated with a 5.1 percent increase in new apprentices with a high school degree. However, the results are not statistically significant for years after 2008. Consistent with this, a 1~pp increase in the state UR increases the number of applicants with \textit{Abitur} by around 6.4-7.3 percent. This aligns with the main findings on college enrollment, suggesting that a significant share of high school graduates make skill-specific investments rather than enroll in college.\footnote{As I do not explicitly observe the year of high school graduation of new apprentices, I focus on the year and state/CZ of the contract signing instead. As a result, the effects may include, but are not limited to, recent high school graduates. Effect sizes and baseline shares are thus not directly comparable to the effects on college enrollment studied earlier.}

\begin{table}[!htbp]
    \centering
    \begin{adjustbox}{max width=\textwidth}
    \begin{threeparttable}
		\caption{Effects on the apprenticeship market}
		\label{tab:vocational_main}
        \scriptsize{
        \begin{tabular} {l ccccc} 
            \toprule
            \toprule
            & \multicolumn{2}{c}{Contracts} & \multicolumn{2}{c}{Applicants} & \multicolumn{1}{c}{Positions} \\
             \cmidrule(lr){2-3}   \cmidrule(lr){4-5}
            & (1) & (2) & (3) & (4) & (5) \\ 
            \midrule
\textit{Panel A: ln(high school)} & & & & & \\
State UR (t-1)          &       0.051** &       0.018  &       0.073*** &       0.064***  &                   \\
                        &     (0.021)   &     (0.015)   &     (0.020)     &     (0.019)     &                   \\
                        &      [6,915]      &      [8,245]      &      [8,237]      &      [1,373]         &                   \\
                        &                   &                   &                   &                   &                   \\
\textit{Panel B: ln(total)} & & & & & \\
State UR (t-1)          &       0.023** &       0.025**   &       0.041**    &       0.022  &       0.021  \\
                        &     (0.009)   &     (0.010)    &     (0.018)    &     (0.017)   &     (0.015)      \\
                        &     [35,442]      &     [33,566]      &     [34,253]      &      [5,709]      &      [5,570]      \\
                        &                   &                   &                   &                   &                   \\
\textit{Panel C: Share high school} & & & & & \\
State UR (t-1)          &       0.358   &      -0.069  &       0.779*   &       0.795***   &                   \\
                        &     (0.268)   &     (0.321)      &     (0.357)    &     (0.202)      &                   \\
                        &        [20.5]     &        [25.2]     &        [23.3]     &        [21.9]     &                   \\
             &  &  &  &  &  \\
            No. Cells   &  384  &  176  &  176  &  1,056  &  1,056  \\
            Level of analysis & State & State & State & CZ & CZ \\
            Time frame & 1995-2018 & 2008-2018 & 2008-2018 & 2008-2018 & 2008-2018 \\
            Data source & BIBB & BIBB & BA & BA & BA \\
            \bottomrule
        \end{tabular}
        }
		\begin{tablenotes}[flushleft] \scriptsize{
				\item \scriptsize \textit{Notes:} This table presents estimates from different regressions mirroring \Cref{eq:specification}, on the state level and CZ level, on new apprenticeship contracts, applicants, and positions. Panel A shows effects on the number (in logs) of high school graduates starting or applying for an apprenticeship. Panel B shows effects on the total number (in logs) of new apprenticeship contracts, applicants, and positions reported. Panel C shows effects on the share of high school graduates among new apprenticeship contracts and applicants. Outcome means in levels in brackets. Standard errors in parentheses allow for clustering at the state and year level. * $p$\textless 0.1, ** $p$\textless 0.05, *** $p$\textless 0.01. \\ \textit{Source}: Federal Institute for Vocational Education and Training (BIBB), Federal Employment Agency (BA).
                } 
		\end{tablenotes}
    \end{threeparttable}
    \end{adjustbox}
\end{table}

Panel B presents results for the total number of new apprentices, regardless of qualification. Depending on the sample period, a 1~pp increase in the state UR increases the number of new contracts by 2.3--2.5 percent. For the years 2008 to 2018, BA data indicate that higher URs also increase the total number of registered applicants by approximately 2.2--4.1 percent, depending on the level of analysis, significant at the state level. Simultaneously, economic shocks do not significantly increase the number of registered new positions. My results again contrast with most international evidence, which indicates that the number of new apprentices is mildly procyclical (see, e.g., \citesupp{luthi2020apprenticeships} for Switzerland and \citesupp{brunello2009effect} for a review). For Germany, however, \citesupp{baldi2014effect} find no significant effect of the business cycle on apprenticeship contracts between 1999 and 2012.\footnote{Note that the BA data only contains positions and applications posted at the employment agency. A higher number of posted positions or applicants may signal greater labor market frictions rather than higher supply or demand.} 

Lastly, Panel C presents results for the share of new apprentices and applicants with a high school degree. In the post-Bologna years (2008-2018), the larger relative increase in high school graduates leads to a higher share of such graduates among applicants. A 1~pp increase in the state-level UR increases the share of high school graduates by around 0.78~pp from a baseline of around 22 percent. This may come at the expense of school leavers with lower academic qualifications, who must now compete with high school leavers. However, I find no significant effects on apprenticeship contracts, and the data do not permit direct testing of such displacement effects.

\Cref{tab:vocational_decomp} repeats the decomposition analysis for the apprenticeship market. For overall contracts, there seems to be a clear difference in the effects of national and state-specific labor market conditions. Consistent with \Cref{tab:vocational_main}, national conditions positively affect the number of new contracts with high school graduates. In particular, state-specific unemployment reduces both the number of new contracts and the number of contracts with high school graduates. After 2008, both state-specific effects are smaller and statistically insignificant, and the effect on total contracts turns positive. For the number of registered applicants, both national and state-specific UR increases are associated with a higher number of applicants in the following years. For the total number of registered positions, there is again no significant effect.

\begin{table}[H]
    \centering
    \begin{adjustbox}{max width=\textwidth}
    \begin{threeparttable}
		\caption{Decomposed effects on the apprenticeship market}
		\label{tab:vocational_decomp}
        \scriptsize{
        \begin{tabular} {l ccccc} 
            \toprule
            \toprule
            & \multicolumn{2}{c}{Contracts} & \multicolumn{2}{c}{Applicants} & \multicolumn{1}{c}{Positions} \\
             \cmidrule(lr){2-3}   \cmidrule(lr){4-5}
            & (1) & (2) & (3) & (4) & (5) \\ 
            \midrule
\textit{Panel A: ln(high school)} & & & & & \\
National UR (t-1)           &       0.084***&       0.042** &       0.069** &       0.063**   &                   \\
                            &     (0.028)   &     (0.019)   &     (0.029)   &     (0.024)      &                   \\
$\text{UR}^\Delta$ (t-1)	&      -0.085***&      -0.022  &       0.080*** &       0.067**    &                   \\
                            &     (0.029)   &     (0.013)   &     (0.025)  &     (0.025)      &                   \\
                            &      [6,915]      &      [8,245]      &      [8,237]      &      [1,373]         &                   \\
& & & & & \\            
\textit{Panel B: ln(total)} & & & & & \\
National UR (t-1)           &       0.033** &       0.029*  &       0.036*  &       0.012   &       0.021       \\
                            &     (0.012)   &     (0.014)   &     (0.018)  &     (0.014)   &     (0.021)      \\
$\text{UR}^\Delta$ (t-1)	&      -0.017** &       0.018  &       0.049*  &       0.046*  &       0.020     \\
                            &     (0.008)   &     (0.013)   &     (0.025)   &     (0.024)   &     (0.024)    \\
                            &     [35,442]    &    [33,566]   &    [34,253]   &    [5,709]    &    [5,570]      \\
& & & & & \\                                    
\textit{Panel C: Share high school} & & & & & \\
National UR (t-1)           &       0.653*  &       0.377  &       0.865  &       1.109***   &                   \\
                            &     (0.336)   &     (0.507)   &     (0.575)  &     (0.337)     &                   \\
$\text{UR}^\Delta$ (t-1)	&      -0.857***&      -0.820** &       0.634  &       0.090    &                   \\
                            &     (0.288)   &     (0.298)   &     (0.561)   &     (0.360)     &                   \\
                            &        [20.5]     &        [25.2]     &        [23.3]     &        [21.9]     &      \\
& & & & & \\            
             &  &  &  &  &  \\
            No. Cells   &  384  &  176  &  176  &  1,056  &  1,056  \\
            Level of analysis & State & State & State & CZ & CZ \\
            Time frame & 1995-2018 & 2008-2018 & 2008-2018 & 2008-2018 & 2008-2018 \\
            Data source & BIBB & BIBB & BA & BA & BA \\
            \bottomrule
        \end{tabular}
        }
		\begin{tablenotes}[flushleft] \scriptsize{
				\item \scriptsize \textit{Notes:} This table presents estimates based on \Cref{eq:mechanisms}, on the state and CZ level, on new apprenticeship contracts, applicants, and positions. Panel A shows effects on the number (in logs) of high school graduates starting or applying for an apprenticeship. Panel B shows effects on the total number (in logs) of new apprenticeship contracts, applicants, and positions reported. Panel C shows effects on the share of high school graduates among new apprenticeship contracts and applicants. Outcome means in levels in brackets. Standard errors in parentheses allow for clustering at the state and year level. * $p$\textless 0.1, ** $p$\textless 0.05, *** $p$\textless 0.01. \\ \textit{Source}: Federal Institute for Vocational Education and Training (BIBB), Federal Employment Agency (BA).
                } 
		\end{tablenotes}
    \end{threeparttable}
    \end{adjustbox}
\end{table}

Negative effects on overall contracts and positive effects on applicants registering at the BA may very well reflect increased frictions on the apprenticeship market. A zero effect on posted positions may still be consistent with this, e.g., when the true demand for apprentices is lower, but more of the remaining positions are posted due to frictions. Procyclical effects of state-level labor demand on the apprenticeship market align with the international evidence (e.g., \cite{luthi2020apprenticeships, brunello2009effect} as discussed above). Together with the decomposition results on college enrollment, these effects suggest that national labor demand shocks shift students at the margin between academic and applied universities towards applied universities, visible in attainment rather than contemporaneous enrollment, while more localized shocks operate at the margin of college going.

\FloatBarrier
\clearpage
\section{More evidence from the Panel Study of School Leavers}

\begin{table}[H]
    \begin{center}
	\begin{adjustbox}{max width=0.6\textwidth}
	\begin{threeparttable}
		\caption{Within-person expectation updating between the survey waves}
		\label{tab:updating}
		\scriptsize{
		        \begin{tabular} {l ccc}
	\toprule \toprule
    & \multicolumn{1}{c}{Vocational} & \multicolumn{1}{c}{Academic} & \multicolumn{1}{c}{Relative Returns} \\
 & (1) & (2) & (3)\\
\midrule
$\Delta$ State UR               &      -0.024   &      -0.012   &       0.011   \\
                                &     (0.032)   &     (0.039)   &     (0.040)   \\
\midrule
                No. students & 31,141 & 31,371 & 33,450 \\
                Cohort FE, state FE & yes & yes & yes \\
                Individual controls & yes & yes & yes \\
				\bottomrule
		\end{tabular}
		}
		\begin{tablenotes}[flushleft]
				\item \scriptsize{ \textit{Notes:} This table regresses the within-person change in each raw expectation measure between the first survey wave (December of the senior year) and the second wave (December of the graduation year) on the change in the state UR between the two interviews. The sample covers students with expectation reports in both waves. Individual controls include gender, SES, and the standardized GPA and its square. All specifications include cohort and state fixed effects. Standard errors in parentheses allow for clustering at the state level. * $p$\textless 0.1, ** $p$\textless 0.05, *** $p$\textless 0.01. \\ \textit{Source}: Federal Employment Agency (BA), and DZHW, cohorts 2008, 2012, 2015, and 2018.
                }
		\end{tablenotes}
	\end{threeparttable}
	\end{adjustbox}
    \end{center}
\end{table}

\begin{figure}[H]
	\centering
	\begin{minipage}{\textwidth}
	\caption{Marginal state UR effects on expectations by covariates (separate degrees)}
    \vspace{-1em}
 	\label{fig:margins_separate}
    \begin{center}
    \begin{subfigure}[b]{0.75\textwidth} 
        \centering
        \small{Panel A: Expected value of a college degree}
	\includegraphics[scale=0.7]{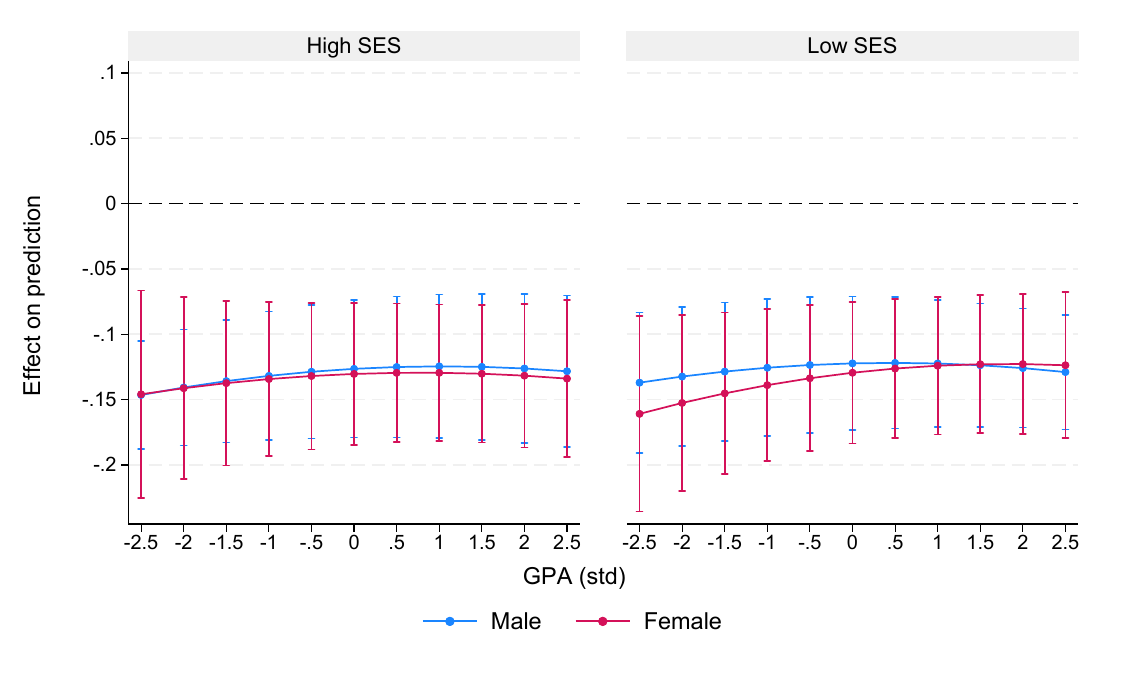} \\
    \end{subfigure}
    \vspace{-1.5em}    
    \vspace{1em} 
    \begin{subfigure}[b]{0.75\textwidth}
        \centering
        \small{Panel B: Expected value of a vocational degree}
	\includegraphics[scale=0.7]{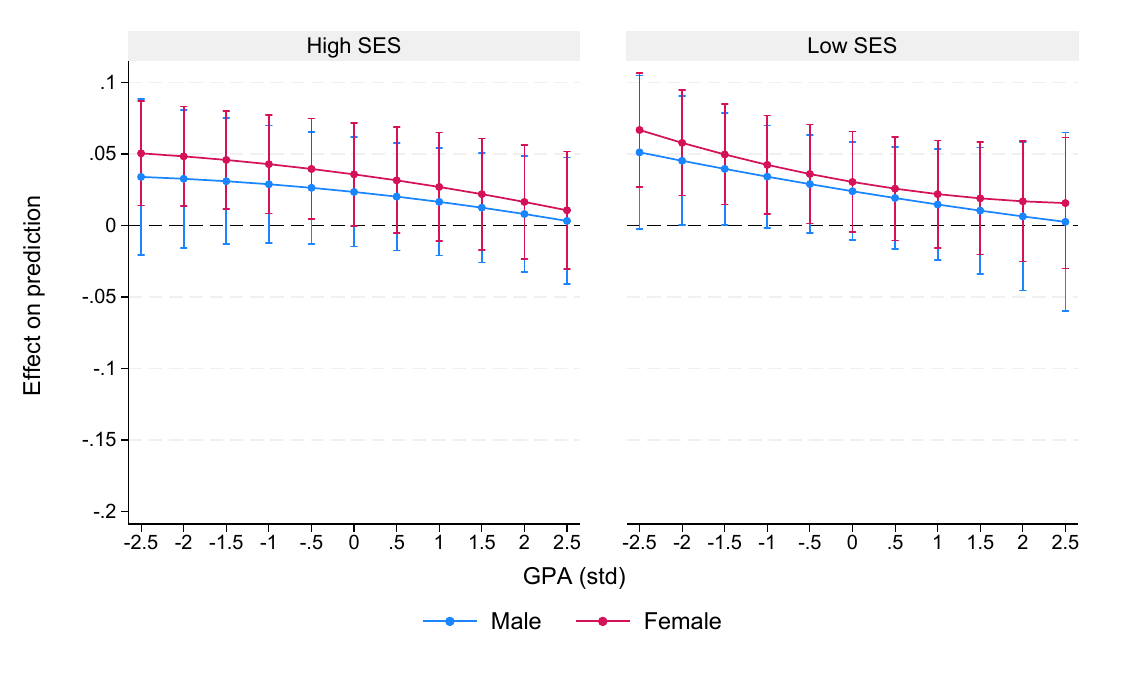} \\

    \end{subfigure}

    \end{center}

    \vspace{-1em}
    {\scriptsize \textit{Notes:} This figure plots the marginal effect of a positive 1~pp deviation of the state UR from its trend one year before high school graduation on the standardized explicitly expected general value of a college degree in (Panel A) and a vocational degree (Panel B). Estimates are based on \Cref{eq:expectations} with the UR interacted with the full set of controls (see \Cref{tab:expectations_zreturns}). \\ \textit{Source}: Federal Employment Agency (BA), and DZHW, years 2008, 2012, 2015, and 2018.
    \par}
    \end{minipage}

\end{figure}

\begin{figure}[H]
	\centering
	\begin{minipage}{\textwidth}
	\caption{Marginal state UR effects on enrollment intentions by covariates}
    \vspace{-1em}
 	\label{fig:margins_intentions}
	\begin{center}
	\includegraphics[scale=0.7]{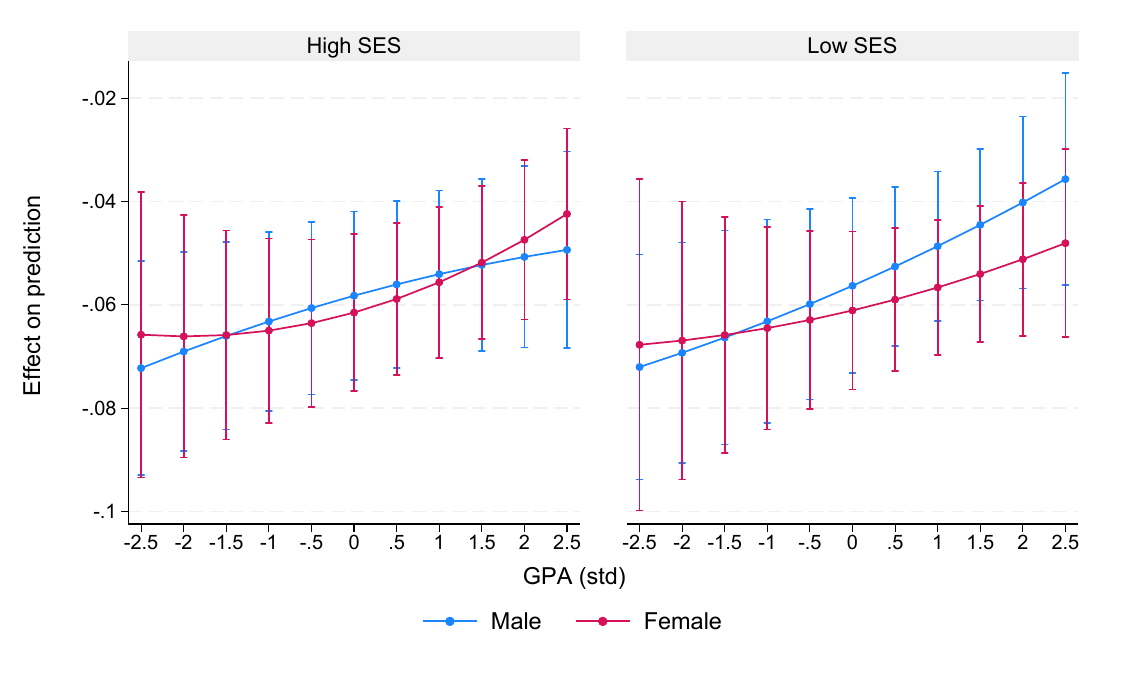} \\
	\end{center}
    \vspace{-1em}
    {\scriptsize \textit{Notes:} This figure plots the marginal effect of a positive 1~pp deviation of the state UR from its trend one year before high school graduation on the probability of intending to enroll, along the standardized GPA distribution and separately by gender and SES. Estimates are based on the fully interacted version of \Cref{eq:expectations} on the estimation sample of \Cref{tab:intentions}, Panel A. \\ \textit{Source}: Federal Employment Agency (BA), and DZHW, cohorts 2008, 2012, 2015, and 2018.
    \par}
    \end{minipage}
\end{figure}

\begin{table}[!htbp]
    \begin{center}
	\begin{adjustbox}{max width=0.75\textwidth}
	\begin{threeparttable}
		\caption{Intention to enroll: alternative time controls}
		\label{tab:intentions_specs}
		\scriptsize{
		        \begin{tabular} {l ccc}
	\toprule \toprule
    & \multicolumn{3}{c}{Intends to enroll (any)} \\
    \cmidrule(lr){2-4}
    & Quadratic trends & Linear trends & Cohort FE \\
 & (1) & (2) & (3)\\
\midrule
State UR (t-1)          &      -0.058***&      -0.052** &      -0.017** \\
                        &     (0.008)   &     (0.014)   &     (0.005)   \\
Wild cluster bootstrap $p$ &    0.010   &       0.013   &       0.008   \\
Outcome mean            &        0.65   &        0.65   &        0.65   \\
No. students            &     104,666   &     104,666   &     104,666   \\
				\bottomrule
		\end{tabular}
		}
		\begin{tablenotes}[flushleft]
				\item \scriptsize{ \textit{Notes:} This table presents estimates from \Cref{eq:expectations} with the intention to enroll as outcome, on the estimation sample of \Cref{tab:intentions}, Panel A. Column 1 uses state-specific quadratic trends as in the main specification, column 2 state-specific linear trends, and column 3 cohort fixed effects instead of trends. All columns include state fixed effects, policy controls, and individual controls (gender, SES, standardized GPA and its square). The wild cluster bootstrap row reports $p$-values from a bootstrap by state with 9,999 replications and Webb weights. Standard errors in parentheses allow for two-way clustering at the state and cohort level. * $p$\textless 0.1, ** $p$\textless 0.05, *** $p$\textless 0.01. \\ \textit{Source}: Federal Employment Agency (BA), and DZHW, cohorts 2008, 2012, 2015, and 2018.
                }
		\end{tablenotes}
	\end{threeparttable}
	\end{adjustbox}
    \end{center}
\end{table}

\begin{table}[H]
    \begin{center}
	\begin{adjustbox}{max width=0.75\textwidth}
	\begin{threeparttable}
		\caption{State UR effects on expected returns among students answering the personal-prospects question}
		\label{tab:expectations_peranswer}
		\scriptsize{
		        \begin{tabular} {l cccc}
	\toprule \toprule
    & \multicolumn{1}{c}{Vocational} & \multicolumn{1}{c}{Academic} & \multicolumn{1}{c}{Relative Returns} & \multicolumn{1}{c}{Personal} \\
 & (1) & (2) & (3) & (4)\\
\midrule
\textit{Panel A: State-level UR} & & & & \\
State UR (t-1)                  &       0.029   &      -0.125** &      -0.112** &      -0.220*** \\
                                &     (0.020)   &     (0.026)   &     (0.029)   &     (0.014)    \\
                                &      [0.280]  &      [0.019]  &      [0.015]  &      [0.002]   \\
&&&&\\
\textit{Panel B: Decomposed UR} & & & & \\
National UR (t-1)               &       0.025   &      -0.191***&      -0.156** &      -0.294*** \\
                                &     (0.033)   &     (0.028)   &     (0.029)   &     (0.016)    \\
$\text{UR}^{\Delta}$ (t-1)      &       0.038   &       0.029   &      -0.009   &      -0.047    \\
                                &     (0.066)   &     (0.054)   &     (0.048)   &     (0.031)    \\
&&&&\\
                No. students & 97,316 & 97,316 & 97,316 & 97,316 \\
                State FE, trends, policy controls & yes & yes & yes & yes \\
				\bottomrule
		\end{tabular}
		}
		\begin{tablenotes}[flushleft]
				\item \scriptsize{ \textit{Notes:} This table repeats the estimates of \Cref{tab:expectations_zreturns}, columns (1), (3), (5), and (7), on the subsample of students with a valid response to the personal-prospects question. The outcomes are the standardized expected returns to a vocational degree, to an academic degree, the implicitly expected relative returns to college, and the personally expected job prospects, one year before high school graduation. All specifications include individual controls (gender, SES, standardized GPA, and its square). Standard errors in parentheses allow for two-way clustering at the state and cohort level. Wild cluster bootstrap $p$-values in brackets impose the null and use 9,999 replications with Webb weights, clustering by state. * $p$\textless 0.1, ** $p$\textless 0.05, *** $p$\textless 0.01. \\ \textit{Source}: Federal Employment Agency (BA), and DZHW, cohorts 2008, 2012, 2015, and 2018.
                }
		\end{tablenotes}
	\end{threeparttable}
	\end{adjustbox}
    \end{center}
\end{table}

\begin{table}[H]
    \begin{center}
	\begin{adjustbox}{max width=0.75\textwidth}
	\begin{threeparttable}
		\caption{Expected returns and intentions among students answering the personal-prospects question}
		\label{tab:iv_peranswer}
		\scriptsize{
		        \begin{tabular} {l ccc}
	\toprule \toprule
    & \multicolumn{3}{c}{Intends to enroll} \\
    \cmidrule(lr){2-4}
    & \multicolumn{1}{c}{Any} & \multicolumn{1}{c}{Academic} & \multicolumn{1}{c}{Applied} \\
 & (1) & (2) & (3)\\
\midrule
\textit{Panel A: Sample of \Cref{tab:intentions}, Panel B} &&&\\
OLS                     &       0.101***&       0.061***&       0.035***\\
                        &     (0.012)   &     (0.008)   &     (0.005)   \\
2SLS                    &       0.464***&       0.316*  &       0.111   \\
                        &     (0.116)   &     (0.180)   &     (0.148)   \\
Anderson-Rubin 95\% conf.\ set & $[0.29, 1.32]$ & $[-0.05, 3.03]$ & $[-2.10, 0.47]$ \\
First-stage $F$         &        13.8   &        13.8   &        13.8   \\
No. students            &     103,791   &     100,679   &     100,679   \\
&&&\\
\textit{Panel B: Sample of \Cref{tab:expectations_zreturns}} &&&\\
OLS                     &       0.093***&       0.056***&       0.032***\\
                        &     (0.013)   &     (0.009)   &     (0.005)   \\
2SLS                    &       0.553***&       0.401*  &       0.121   \\
                        &     (0.158)   &     (0.232)   &     (0.180)   \\
Anderson-Rubin 95\% conf.\ set & $[0.34, 3.59]$ & $[-0.05, 7.21]$ & $[-4.77, 0.56]$ \\
First-stage $F$         &         9.9   &        10.1   &        10.1   \\
No. students            &      97,316   &      94,341   &      94,341   \\
				\bottomrule
		\end{tabular}
		}
		\begin{tablenotes}[flushleft]
				\item \scriptsize{ \textit{Notes:} This table repeats the estimation of \Cref{tab:intentions}, Panel B, on the subsample of students with a valid response to the personal-prospects question. Panel A restricts the estimation sample of \Cref{tab:intentions}, Panel B. Panel B restricts the estimation sample of \Cref{tab:expectations_zreturns}. Each column regresses the intention indicator on the standardized expected relative return to college, by OLS and by 2SLS with the state UR one year before graduation as the instrument, with individual controls. The Anderson-Rubin confidence sets impose the null and use the wild cluster bootstrap by state with 9,999 replications and Webb weights \citepsupp{davidson2014confidence}. First-stage $F$ statistics are cluster-robust. All specifications include state fixed effects, state-specific quadratic trends, and policy controls. Standard errors in parentheses allow for two-way clustering at the state and cohort level in the OLS rows and for clustering by state in the 2SLS rows. * $p$\textless 0.1, ** $p$\textless 0.05, *** $p$\textless 0.01. \\ \textit{Source}: Federal Employment Agency (BA), and DZHW, cohorts 2008, 2012, 2015, and 2018.
                }
		\end{tablenotes}
	\end{threeparttable}
	\end{adjustbox}
    \end{center}
\end{table}

\begin{table}[H]
    \begin{center}
	\begin{adjustbox}{max width=0.8\textwidth}
	\begin{threeparttable}
		\caption{Individual-level enrollment half a year after graduation}
		\label{tab:sbp_enrolled}
		\scriptsize{
		        \begin{tabular} {l cccc}
	\toprule \toprule
& All & Women & Men & Weighted \\
& (1) & (2) & (3) & (4) \\
\midrule
State UR (t-1)   &      -0.026   &      -0.049   &      -0.072   &      -0.032   \\
                 &     (0.041)   &     (0.030)   &     (0.042)   &     (0.034)   \\
\midrule
No. students     &      31,718   &      21,400   &      12,335   &      31,716   \\
State FE, trends, policy controls & Yes & Yes & Yes & Yes \\
Individual controls & Yes & Yes & Yes & Yes \\
				\bottomrule
		\end{tabular}
		}
		\begin{tablenotes}[flushleft]
				\item \scriptsize{ \textit{Notes:} This table presents linear probability estimates of being enrolled in higher education in December of the graduation year on the state unemployment rate in the senior high school year, following \Cref{eq:expectations}. Higher education excludes \textit{Berufsakademien}, mirroring the student register. Individual controls include gender (columns 1 and 4), SES, and the standardized GPA and its square. Columns 1 and 4 exclude conscription-affected men of the 2008 cohort, and column 3 includes them. Column 4 additionally uses the survey's longitudinal weights. Standard errors in parentheses allow for two-way clustering at the state and cohort level in columns 2--4, and for clustering at the state level in column 1. * $p$\textless 0.1, ** $p$\textless 0.05, *** $p$\textless 0.01. \\ \textit{Source}: Federal Employment Agency (BA), and DZHW, years 2008, 2012, 2015, and 2018.
                }
		\end{tablenotes}
	\end{threeparttable}
	\end{adjustbox}
    \end{center}
\end{table}

\begin{table}[H]
    \begin{center}
	\begin{adjustbox}{max width=0.6\textwidth}
	\begin{threeparttable}
		\caption{Expected returns and enrollment: OLS and 2SLS, women}
		\label{tab:iv_expectations}
		\scriptsize{
		        \begin{tabular} {l cc}
	\toprule \toprule
& OLS & 2SLS \\
& (1) & (2) \\
\midrule
Expected relative returns (W1)   &       0.056** &       0.439   \\
                                 &     (0.012)   &     (0.545)   \\
Anderson-Rubin 95\% confidence set &             & $[-0.622, 7.293]$ \\
\addlinespace
First stage: State UR (t-1)      &               &      -0.081** \\
                                 &               &     (0.022)   \\
First-stage $F$                  &               &         8.8   \\
Reduced form: State UR (t-1)     &               &      -0.035   \\
                                 &               &     (0.031)   \\
\midrule
No. students                     &      20,872   &      20,872   \\
State FE, trends, policy controls & Yes & Yes \\
Individual controls              & Yes & Yes \\
				\bottomrule
		\end{tabular}
		}
		\begin{tablenotes}[flushleft]
				\item \scriptsize{ \textit{Notes:} This table relates enrollment in higher education in December of the graduation year to the expected relative return to college measured in the senior year, for women. Column 1 reports OLS. Column 2 instruments the expected relative return with the state unemployment rate in the senior high school year and reports the corresponding first stage and reduced form. The Anderson-Rubin confidence set imposes the null and uses a wild cluster bootstrap with 9,999 replications and Webb weights, clustering by state \citepsupp{davidson2014confidence}. The first-stage $F$ statistic is cluster-robust. Individual controls include SES and the standardized GPA and its square. Standard errors in parentheses allow for clustering by state in column 2 and two-way clustering at the state and cohort level in column 1 and the first-stage and reduced-form rows. * $p$\textless 0.1, ** $p$\textless 0.05, *** $p$\textless 0.01. \\ \textit{Source}: Federal Employment Agency (BA), and DZHW, years 2008, 2012, 2015, and 2018.
                }
		\end{tablenotes}
	\end{threeparttable}
	\end{adjustbox}
    \end{center}
\end{table}

\begin{table}[H]
    \begin{center}
	\begin{adjustbox}{max width=0.85\textwidth}
	\begin{threeparttable}
		\caption{State UR, the reported influence of study costs, and perceived difficulties}
		\label{tab:costinfl}
		\scriptsize{
		        \begin{tabular} {l c ccc}
	\toprule \toprule
	& Influence of & \multicolumn{3}{c}{Perceived as difficulty} \\
	\cmidrule(lr){3-5}
	& study costs (std.) & Labor market & Financing & Admission \\
	& (1) & (2) & (3) & (4) \\
	\midrule
State UR (t-1)          &      -0.070** &      -0.077***&      -0.057***&      -0.041** \\
                        &     (0.021)   &     (0.007)   &     (0.009)   &     (0.011)   \\
                        &     [0.047]   &     [0.001]   &     [0.011]   &     [0.049]   \\
\midrule
Outcome mean            &        2.72   &        0.33   &        0.28   &        0.40   \\
No. students            &     112,491   &     113,299   &     113,299   &     113,299   \\
State FE, trends, policy controls &      yes      &      yes      &      yes      &      yes      \\
				\bottomrule
		\end{tabular}
		}
		\begin{tablenotes}[flushleft]
				\item \scriptsize{ \textit{Notes:} This table presents estimates from \Cref{eq:expectations} with elicited decision factors as outcomes, measured half a year before graduation. Column 1 uses the standardized influence of study costs on the education decision. Its outcome mean refers to the raw item, which ranges from 1 (no influence) to 5 (large influence). Columns 2--4 use indicators equal to one if the student perceives the development of the labor market, the financing of their education, or admission restrictions (\textit{numerus clausus}) as a difficulty for the education decision. Columns 2--4 share a common sample. Individual controls include gender, SES, and the standardized GPA and its square. The wild cluster bootstrap $p$-values in brackets impose the null and use 9,999 replications with Webb weights, clustering by state. Standard errors in parentheses allow for two-way clustering at the state and cohort level. * $p$\textless 0.1, ** $p$\textless 0.05, *** $p$\textless 0.01. \\ \textit{Source}: Federal Employment Agency (BA), and DZHW, cohorts 2008, 2012, 2015, and 2018.
                }
		\end{tablenotes}
	\end{threeparttable}
	\end{adjustbox}
    \end{center}
\end{table}

\FloatBarrier
\clearpage
\section{Evidence from the Socio-economic Panel}
\label{sec:soep}

To test credit constraints as a mechanism, I use survey data from the German Socio-Economic Panel (SOEP). The SOEP is a longitudinal survey of approximately 15,000 private households in Germany, tracking individuals’ educational trajectories, including their state and year of high school graduation and postsecondary education \citep{goebel2019german}. This comes with the advantage that I can now explicitly estimate within-person choices, including both college types and apprenticeships as outcomes. It also provides household and parental income and education as measures of SES. The sample covers school leavers from academic-track and comprehensive schools (\textit{Gymnasium}, \textit{Gesamtschule}) with any leaving degree, the closest survey analogue to the \textit{Abitur} sample of the administrative data. Restricting to \textit{Abitur} holders (1,758 of the 1,851 school leavers) leaves the estimates virtually unchanged, for example $-0.035$ instead of $-0.034$ for tertiary education per 1~pp UR. I start by re-estimating the main results and underlying heterogeneities before I test credit constraints directly.

\subsection{Heterogeneity by SES}
\label{sec:soep_heterogeneity}

As the student register contains no information on socioeconomic background, I use the German Socio-economic Panel (SOEP v39) to estimate effects across students with and without college-educated parents (high vs. low SES) and across gender. The SOEP data contain information on the first post-secondary enrollment spell of 1,851 high school graduates between 1995 and 2018 and information on the state of graduation. In contrast to the student register, it also allows for direct estimation of business cycle effects on the probability of starting an apprenticeship.

\Cref{tab:abi_postsec_soep} summarizes the main results for different enrollment choices (any college, vocational training, academic university, applied university, and neither or missing) in each panel and for the whole sample, as well as splits by gender and SES in each column. Overall, the main results follow a pattern similar to that observed in the administrative data. The overall enrollment effect is -3.4~pp and appears larger at academic universities (Panel D) than at applied universities (Panel E). However, these effects are not statistically significant, and the baseline probability of college enrollment (61 percent) is considerably lower than in the student register. 

Panel B shows that vocational training is positively and significantly affected by higher unemployment rates (+2.9~pp). The effect is largest and statistically significant among female high school graduates (+4.3~pp). A simple explanation for this is that a higher share of women obtain the \textit{Abitur} \citep{goldin2006homecoming, riphahn2015reversal}. Men are more likely to begin an apprenticeship after obtaining an intermediate secondary school degree and to forgo high school. Hence, the remaining men might be preselected and have stronger study intentions. In line with this, the adverse enrollment effects at academic universities are also stronger for women.\footnote{Gendered business cycle effects align with \cite{leibing2023gender}, who document stronger associations between the expected returns to college and actual enrollment for female high school graduates, and \cite{bietenbeck2023tuition}, who find larger effects of tuition fees on female college attainment in Germany.} Splitting by SES, the enrollment effects on overall college enrollment seem to be stronger and are statistically significant for low SES graduates, which seems intuitive. \Cref{sec:expectations} presents results for expectations as the main mechanism across gender, SES, and along the GPA distribution of graduates.

\begin{table}[!h]
    \begin{center}
	\begin{adjustbox}{max width=\textwidth}
	\begin{threeparttable}
		\caption{Enrollment effects in the Socio-economic Panel}
		\label{tab:abi_postsec_soep}
		\scriptsize{
		        \begin{tabular} {l c c cccc}
	\toprule \toprule
&  & \multicolumn{2}{c}{By gender} & \multicolumn{2}{c}{By SES} \\ 
 \cmidrule(lr){3-4} \cmidrule(lr){5-6} 
& \multicolumn{1}{c}{Main} & \multicolumn{1}{c}{Female} & \multicolumn{1}{c}{Male} & \multicolumn{1}{c}{High} & \multicolumn{1}{c}{Low}  \\
& (1) & (2) & (3) & (4) & (5) \\ 
\midrule
\textit{Panel A: Any college} & & & & & \\
State UR (t-1)       &      -0.034   &      -0.038   &      -0.041   &      -0.025   &      -0.044*  \\
            &     (0.021)   &     (0.026)   &     (0.026)   &     (0.054)   &     (0.023)   \\
\\ Outcome mean&        0.61   &        0.59   &        0.64   &        0.74   &        0.56   \\
No. graduates&       1,851   &       1,035   &         816   &         571   &       1,280   \\
& & & & & \\
\textit{Panel B: Vocational training} & & & & & \\
State UR (t-1)       &       0.029** &       0.043*  &       0.025   &       0.045   &       0.026   \\
            &     (0.013)   &     (0.022)   &     (0.022)   &     (0.033)   &     (0.020)   \\
\\ Outcome mean&        0.28   &        0.31   &        0.25   &        0.15   &        0.34   \\
No. graduates&       1,851   &       1,035   &         816   &         571   &       1,280   \\
& & & & & \\
\textit{Panel C: Neither or missing} & & & & & \\
State UR (t-1)       &       0.005   &      -0.005   &       0.016   &      -0.020   &       0.018   \\
            &     (0.013)   &     (0.010)   &     (0.021)   &     (0.021)   &     (0.013)   \\
\\ Outcome mean&        0.10   &        0.10   &        0.11   &        0.11   &        0.10   \\
No. graduates&       1,851   &       1,035   &         816   &         571   &       1,280   \\
& & & & & \\
\textit{Panel D: University}  & & & & & \\
State UR (t-1)       &      -0.024   &      -0.049*  &      -0.003   &      -0.029   &      -0.031   \\
            &     (0.021)   &     (0.026)   &     (0.035)   &     (0.037)   &     (0.026)   \\
\\ Outcome mean&        0.57   &        0.54   &        0.60   &        0.73   &        0.50   \\
No. graduates&       1,567   &         883   &         684   &         461   &       1,106   \\
& & & & & \\
\textit{Panel E: Applied university} & & & & & \\
State UR (t-1)       &      -0.006   &       0.006   &      -0.027   &      -0.017   &      -0.003   \\
            &     (0.010)   &     (0.012)   &     (0.025)   &     (0.014)   &     (0.011)   \\
\\ Outcome mean&        0.10   &        0.09   &        0.11   &        0.09   &        0.10   \\
No. graduates&       1,567   &         883   &         684   &         461   &       1,106   \\

				\bottomrule
		\end{tabular}
		}
		\begin{tablenotes}[flushleft] 
				\item \scriptsize{ \textit{Notes:} This table presents estimates from a linear probability model with state fixed effects, as well as state-specific trends including a 2005-trend break and policy controls, for high school graduates with any high school degree, spanning graduation years 1995--2018 in a repeated cross-section. Panels A-C focus on all high school graduates. Panels D-E focus on high school graduates for whom enrollment information by institution is available. Standard errors in parentheses allow for clustering at the state and cohort level. * $p$\textless 0.1, ** $p$\textless 0.05, *** $p$\textless 0.01. \\ \textit{Source}: SOEP v39, Federal Employment Agency (BA).
                } 
		\end{tablenotes}
	\end{threeparttable}
	\end{adjustbox}
    \end{center}
\end{table}

\subsection{Credit Constraints}
\label{sec:credit}

One alternative mechanism to expectations suggested in the conceptual framework is credit constraints. In Germany, however, higher education has low direct costs, as there are no general tuition fees.\footnote{Between 2007 and 2014, some states introduced minor tuition fees (500~EUR per semester) with little effect on overall attainment \citepsupp[][]{bietenbeck2023tuition}. Private institutions can still charge tuition fees.} A means-tested federal aid program (\textit{BAFöG}) supports around 25 percent of full-time students. However, only 28 percent of formally eligible students (i.e., those under 30) receive any \textit{BAFöG}, either because they failed the means test, or because they did not apply \citepsupp{fidan2022loanaversion}. Similarly, only five percent of students take on a private loan. Of the average monthly income of students under 21, 66 percent is parental allowances, 13 percent is own earnings, twelve percent stems from federal aid \textit{BAFöG}, and nine percent is from other sources \citepsupp{middendorff2017wirtschaftliche}. 

If downturns reduce parental income, the main financial resource of students in Germany, then credit-constrained students may forgo college. The evidence on credit constraints is mixed. \citesupp{hazarika2002role, christian2007liquidity} suggest that enrollment is more procyclical for low-income households. However, \citesupp{cameron2004estimation, hilger2016parental} find that most individuals with enrollment intentions are not liquidity constrained; \citesupp{bulman2015returns} detect only small effects of tax credits on college-going. \citesupp{bulman2021lottery} find that only large lottery wins have a small positive impact. However, \citesupp{lovenheim2011effect,pan2014parents,manoli2018cash} find significant effects of increases in housing wealth, parental layoff, and tax refunds on enrollment, respectively. For Germany, \citesupp{goehausen2024housing} finds negative effects of housing costs on enrollment and student mobility. 

\Cref{tab:abi_inc_enroll} Panel A shows that a 1,000~EUR increase in net household income is statistically significantly associated with a 2.9~pp increase in the probability of attending college. Yet, given the mean net household income (about 3,678~EUR; (sd=1,804)), this is a very small estimate. Panel B shows that a one percent increase in income is associated with an increase in the probability of attending college of 0.096 percentage points. The effects on vocational training are reversed (-0.070~pp), and the effects on institutional choice are insignificant.


\begin{table}[!h]
    \begin{center}
	\begin{adjustbox}{max width=\textwidth}
	\begin{threeparttable}
		\caption{Household income and postsecondary education}
		\label{tab:abi_inc_enroll}
		\scriptsize{
		        \begin{tabular} {l c c cccc}
	\toprule \toprule
&  & &  & \multicolumn{2}{c}{By college type} \\ 
 \cmidrule(lr){5-6} 
& \multirowcell{2}{Tertiary} & \multirowcell{2}{Vocational \\ training} & \multirowcell{2}{Neither \\ or missing} & \multirowcell{2}{Academic} & \multirowcell{2}{Applied}  \\
 &  &  &  &  &  \\ 
 & (1) & (2) & (3) & (4) & (5) \\ 
\midrule
\textit{Panel A: Net income at graduation} &&&&&\\
$w_t$ (in 1,000 EUR) &       0.029** &      -0.019*  &      -0.009*  &       0.019   &       0.002   \\
            &     (0.010)   &     (0.010)   &     (0.005)   &     (0.020)   &     (0.012)   \\
&&&&&\\
\textit{Panel B: ln(Income)} &&&&&\\
ln($w_t$) &       0.096***&      -0.070** &      -0.027** &       0.062   &       0.014   \\
          &     (0.027)   &     (0.031)   &     (0.010)   &     (0.054)   &     (0.033)   \\
&&&&&\\
\textit{Panel C: Immediate change} &&&&&\\
ln($w_t$)-ln($w_{t-1}$) &       0.046   &      -0.023   &      -0.024   &       0.126   &      -0.094   \\
            &     (0.072)   &     (0.081)   &     (0.056)   &     (0.123)   &     (0.091)   \\
&&&&&\\
\textit{Panel D: Deviation from trend} &&&&&\\
ln($w_t$)-ln($\hat{w}_{t}$) &       0.041   &      -0.043   &       0.002   &       0.028   &       0.016   \\
            &     (0.071)   &     (0.086)   &     (0.044)   &     (0.087)   &     (0.043)   \\
&&&&&\\
Outcome mean &        0.61   &        0.31   &        0.08   &        0.55   &        0.10   \\
No. graduates &       1,090   &       1,090   &       1,090   &         957   &         957   \\
				\bottomrule 
		\end{tabular}
		}
		\begin{tablenotes}[flushleft] 
				\item \scriptsize{ \textit{Notes:} This table presents estimates from a linear probability model with state fixed effects, as well as state-specific trends including a 2005-trend break and policy, gender, and SES controls for high school graduates with any high school degree, spanning graduation years 1995--2018 in a repeated cross-section. Standard errors in parentheses allow for clustering at the state and year level. * $p$\textless 0.1, ** $p$\textless 0.05, *** $p$\textless 0.01. \\ \textit{Source}: SOEP v39, Federal Employment Agency (BA).
                } 
		\end{tablenotes}
	\end{threeparttable}
	\end{adjustbox}
    \end{center}
\end{table}

\Cref{tab:abi_ue_inc} shows effects of the state UR increases on different income measures. A 1~pp increase in the senior-year state UR is associated with a 38~EUR (about 1 percent) lower monthly household income, not statistically distinguishable from zero. However, household income of high-SES graduates increases with the state UR (7.9 percent, Panel B), while that of low-SES graduates, the group for which credit constraints would bind, declines by an imprecisely estimated 3.4 percent (Panel C).

\begin{table}[H]
    \begin{center}
	\begin{adjustbox}{max width=\textwidth}
	\begin{threeparttable}
		\caption{State UR and household income ($w_t$)}
		\label{tab:abi_ue_inc}
		\scriptsize{
		        \begin{tabular} {l cccc}
	\toprule \toprule
& \multirowcell{2}{$w_t$} & \multirowcell{2}{ln($w_t$)} & \multirowcell{2}{ln($w_t$) - \\ ln($w_{t-1}$)} & \multirowcell{2}{ln($w_t$) - \\ ln($\hat{w}_{t}$)} \\
&       &           &                                     &                                     \\ 
& (1) & (2) & (3) & (4) \\ 
\midrule
\textit{Panel A. All graduates} &&&&\\
State UR (t-1)   &      -0.038   &      -0.005   &      -0.001   &      -0.011   \\
            &     (0.068)   &     (0.022)   &     (0.007)   &     (0.015)   \\
\\ Outcome mean&        3.68   &        8.09   &        0.02   &       -0.04   \\
No. graduates&       1,447   &       1,447   &       1,369   &       1,107   \\
&&&&\\
\textit{Panel B. High SES} &&&&\\
State UR (t-1)    &       0.222*  &       0.079** &       0.004   &       0.013   \\
            &     (0.113)   &     (0.032)   &     (0.027)   &     (0.039)   \\
\\ Outcome mean&        4.59   &        8.32   &       -0.00   &       -0.04   \\
No. graduates&         447   &         447   &         416   &         317   \\
&&&&\\
\textit{Panel C. Low SES} &&&&\\
State UR (t-1)    &      -0.114   &      -0.034   &      -0.004   &      -0.011   \\
            &     (0.067)   &     (0.023)   &     (0.007)   &     (0.020)   \\
\\ Outcome mean&        3.27   &        7.99   &        0.03   &       -0.05   \\
No. graduates&       1,000   &       1,000   &         953   &         790   \\
				\bottomrule
		\end{tabular}
		}
		\begin{tablenotes}[flushleft] 
				\item \scriptsize{ \textit{Notes:} This table presents estimates from a model with state fixed effects, as well as state-specific trends including a 2005-trend break and policy controls for the effects of the state UR on household income ($w_t$, in 1,000~EUR) and alternative (change) measures for high school graduates with any high school degree, spanning graduation years 1995--2018 in a repeated cross-section. Standard errors in parentheses allow for clustering at the state and year level. * $p$\textless 0.1, ** $p$\textless 0.05, *** $p$\textless 0.01. \\ \textit{Source}: SOEP v39, Federal Employment Agency (BA).
                } 
		\end{tablenotes}
	\end{threeparttable}
	\end{adjustbox}
    \end{center}
\end{table}


After conditioning on gender and SES, there can still be unobserved shocks on the household level that affect income and the decision to enroll in college. Hence, the estimates do not have a causal interpretation. However, as potential confounding factors, such as health shocks, similarly affect college enrollment and income, the estimates are more likely to be upward-biased and overstate the role of income. Panels C and D use within-household, across-time changes and deviations from each household's predicted income trend before the graduation year as income measures to account for potential shocks around graduation. A 1~pp higher change in income relative to the pre-graduation year is associated with an insignificant 0.046~pp increase in the probability of attending college (Panel C). Similarly, a 1~pp higher deviation in income from the pre-graduation linear prediction has no significant effect on enrollment.

\FloatBarrier
\clearpage
\section{Capacity Constraints}
\label{sec:capacity}

Another alternative mechanism for procyclical enrollment is supply-side adjustment, for example, a reduction in the number of students a university would admit to a given program. For the U.S., \citesupp{deming2017budget} show that a 10 percent increase in spending increases enrollment by 3 percent. To complement the spending data with a more direct measure of capacity, I draw on the \textit{Hochschulkompass} of the German Rectors' Conference (HRK), the official register of study programs at German higher-education institutions. For each state and winter semester from 2006 to 2018, I observe the number of undergraduate study programs on offer and, among these, the number subject to a local admission restriction (\textit{\"ortliche Zulassungsbeschr\"ankung}, a local \textit{numerus clausus}). The register pools all institution types and is not broken out by academic versus applied universities. A local restriction is imposed when applications to a program exceed the number of places a department supplies, so the count of restricted programs is a revealed measure of binding capacity. If universities reduce capacity when public budgets decline, the number of restricted programs would rise with higher UR. As restrictions depend on applications as well as places, the count is an equilibrium measure of excess demand. Since enrollment demand itself falls in downturns, a stable count of restricted programs is suggestive rather than conclusive evidence against capacity reductions.

\Cref{tab:spending} shows little sign of cyclical capacity adjustment. Real spending is essentially flat in the current year across all institution types (Panel A, columns 1--3). With a one-year lag (Panel B), higher unemployment is associated with somewhat lower spending overall and at academic universities. Still, the estimates are not statistically significant at conventional levels and are consistent with funding formulas that track student numbers with a lag. This modest budget sensitivity does not translate into a contraction of study-place supply: the number of study programs does not fall, and the number subject to a local admission restriction does not rise, when unemployment is high (columns 4--5). If anything, the total number of programs rises slightly when unemployment is high. 

\begin{table}[H]
    \begin{center}
	\begin{adjustbox}{max width=\textwidth}
	\begin{threeparttable}
		\caption{Cyclical adjustment of spending and study-program supply}
		\label{tab:spending}
		\scriptsize{
		        \begin{tabular} {l c c c c c }
	\toprule \toprule
& \multicolumn{3}{c}{Real spending (mn EUR)} & \multicolumn{2}{c}{Study programs (count)}  \\
\cmidrule(lr){2-4} \cmidrule(lr){5-6}
& \multicolumn{1}{c}{Overall} & \multicolumn{1}{c}{Academic} & \multicolumn{1}{c}{Applied} & \multicolumn{1}{c}{Total} & \multicolumn{1}{c}{Restricted}  \\
\cmidrule(lr){2-2} \cmidrule(lr){3-3} \cmidrule(lr){4-4} \cmidrule(lr){5-5} \cmidrule(lr){6-6}
& (1) & (2) & (3) & (4) & (5)  \\
\midrule
\textit{Panel A: Current unemployment} & &   &    \\
State UR (t)			&      -0.004   &      -0.005   &       0.001   &       0.016*  &       0.025   \\
                    	&     (0.011)   &     (0.011)   &     (0.013)   &     (0.008)   &     (0.029)   \\
& & & \\
\textit{Panel B: Lagged unemployment} & &   &    \\
State UR (t-1)			&      -0.016   &      -0.017   &      -0.010   &       0.017   &       0.040   \\
                    	&     (0.009)   &     (0.010)   &     (0.011)   &     (0.010)   &     (0.033)   \\
& & & \\
Outcome mean (in levels)  &        6,918   &        5,214   &        1,704   &         565   &         270   \\
No. state-year cells      &         207   &         207   &         207   &         207   &         207   \\
				\bottomrule 
		\end{tabular}
		}
		\begin{tablenotes}[flushleft] 
				\item \scriptsize{ \textit{Notes:} This table presents estimates from regressions of five outcomes on the state unemployment rate, years 2006--2018. Columns 1--3 use the log real annual spending of higher-education institutions per state (in million EUR). Columns 4--5 use the log number of undergraduate study programs offered per state, and among them the number subject to a local admission restriction (\textit{\"ortliche Zulassungsbeschr\"ankung}). The program-supply columns pool all institution types. All specifications include state fixed effects and state-specific quadratic trends with a 2005 break, mirroring \Cref{eq:specification}, and control for an indicator for the years in which tuition fees were charged and for double cohorts. All values are deflated to reflect 2020 real values. Spending excludes university hospitals. Standard errors in parentheses allow for two-way clustering at the state and year level. * $p$\textless 0.1, ** $p$\textless 0.05, *** $p$\textless 0.01. \\ \textit{Source}: Finance statistics of institutions of higher education, \textit{Regionaldatenbank} (columns 1--3), HRK \textit{Hochschulkompass} (columns 4--5).
                } 
		\end{tablenotes}
	\end{threeparttable}
	\end{adjustbox}
    \end{center}
\end{table}

Because the expenditure data do not separate teaching from research, the spending estimates have limited information value for the teaching margin specifically. Program-supply measures speak to it more directly and show no tightening. The spending measure excludes university hospitals as they rely on different funding schemes than their associated universities. As their medical degrees are always oversubscribed, with study places allocated via a central clearinghouse (\textit{Stiftung für Hochschulzulassung}, until 2010: \textit{Zentralstelle für die Vergabe von Studienplätzen}), my measure can not respond to state budgets by design. Overall, federal funding arrangements appear to stabilize aggregate capacity, and supply-side restrictions play a secondary role at best. This is consistent with \citesupp{marcus2019effect}, who find no effects of large double cohort shocks (see \Cref{sec:setup}) on the number of study programs using GPA-based admission cutoffs as a proxy for capacity constraints.

\FloatBarrier
\clearpage
\section{Robustness and Additional Results}
\label{sec:oa_robustness}

This section reports robustness checks and supplementary results. The overall and academic-university estimates are stable across the form of the state-specific trend (\Cref{fig:spec_curve}), the break year (\Cref{tab:rob_trendbreaks}), the timing of the unemployment measure relative to graduation (\Cref{tab:rob_leadlag}), and replacing the parametric trend with a Hodrick--Prescott filter (\Cref{tab:rob_hp}, \Cref{fig:hp_filter}). The small applied-university enrollment margin is more sensitive to these choices. The filter involves an arbitrary smoothing choice without institutional interpretation and tests robustness towards a smooth trend-cycle decomposition. Its fitted trends nearly coincide with the quadratic-break trend (\Cref{fig:hp_filter}). In contrast to the policy-motivated 2005 break, the filter is purely data driven and itself contested \citepsupp{hamilton2018why}. The section further reports the effects by period and for the graduation-year unemployment rate (\Cref{tab:period_split}, \Cref{tab:gradyear_ur}), and bootstrap inference for the 16 state clusters (\Cref{tab:wcb}). The main effects by gender are presented in \Cref{sec:heterogeneity}.


\begin{table}[H]
    \begin{center}
	\begin{adjustbox}{max width=0.75\textwidth}
	\begin{threeparttable}
		\caption{Robustness towards alternative trend breaks}
		\label{tab:rob_trendbreaks}
		\scriptsize{
		        \begin{tabular} {l cc cc cc}
	\toprule \toprule

& \multicolumn{2}{c}{Overall} & \multicolumn{2}{c}{Academic} & \multicolumn{2}{c}{Applied}  \\
\cmidrule(lr){2-3} \cmidrule(lr){4-5} \cmidrule(lr){6-7}
& (1) & (2) & (3) & (4) & (5) & (6) \\
\midrule
\textit{Panel A: trend break in 2004} & &   &  &   &   &   \\
State UR (t-1)			&      -2.127***&               &      -1.879***&               &      -0.248***&               \\
                    	&     (0.334)   &               &     (0.329)   &               &     (0.055)   &               \\
... $\times \,$ pre 2008 &    &      -2.217***&               &      -2.028***&               &      -0.189*  \\
                    			&   &     (0.400)   &               &     (0.354)   &               &     (0.103)   \\
... $\times \,$ post 2008&              &      -2.303***&               &      -2.172***&               &      -0.132   \\
                    	&             &     (0.563)   &               &     (0.495)   &               &     (0.165)   \\
p-value ($\beta_{\Delta}$)  &         &       [0.717]   &               &       [0.538]   &               &       [0.400]   \\
& &   &  &   &   &   \\
\textit{Panel B: trend break in 2006} & &   &  &   &   &   \\
State UR (t-1)			&      -1.873***&               &      -1.913***&               &       0.041   &               \\
                    	&     (0.438)   &               &     (0.465)   &               &     (0.106)   &               \\
... $\times \,$ pre 2008 &   &      -2.048***&               &      -2.188***&               &       0.140   \\
                    	&   &     (0.469)   &               &     (0.420)   &               &     (0.119)   \\
... $\times \,$ post 2008&     &      -2.281***&               &      -2.552***&               &       0.272   \\
                    	&    &     (0.602)   &               &     (0.517)   &               &     (0.184)   \\
p-value ($\beta_{\Delta}$)  &       &       [0.256]   &               &       [0.096]   &               &       [0.124]   \\
& &   &  &   &   &   \\
Outcome mean      &        71.2   &        71.2   &        54.2   &        54.2   &        17.0   &        17.0   \\
No. region-cohort cells       &        2,289   &        2,289   &        2,289   &        2,289   &        2,289   &        2,289   \\
No. high school graduates    &     6,296,662   &     6,296,662   &     6,296,662   &     6,296,662   &     6,296,662   &     6,296,662   \\
				\bottomrule 
		\end{tabular}
		}
		 \begin{tablenotes}[flushleft]
				\item \scriptsize{ \textit{Notes:} This table presents estimates from \Cref{eq:specification} for the share of high school graduates with \textit{Abitur} enrolling in different college types, spanning high school cohorts 1995--2018, but using different trend breaks in $g_s(t)$. Panel A uses a trend break in 2004, and Panel B in 2006, with the quadratic trend centered at the respective break year. Standard errors in parentheses allow for two-way clustering at the state and cohort level. * $p$\textless 0.1, ** $p$\textless 0.05, *** $p$\textless 0.01. \\ \textit{Source}: Student register, Federal Employment Agency, \textit{Regionaldatenbank}.
                }
		 \end{tablenotes}
	\end{threeparttable}
	\end{adjustbox}
    \end{center}
\end{table}

\begin{table}[H]
    \begin{center}
	\begin{adjustbox}{max width=0.8\textwidth}
	\begin{threeparttable}
		\caption{Robustness: business-cycle variation from a Hodrick--Prescott filter}
        \label{tab:rob_hp}
		\scriptsize{
		        \begin{tabular} {l ccc}
	\toprule \toprule
	& Overall & Academic & Applied \\
	& (1) & (2) & (3) \\
	\midrule
	\textit{Panel A: Main specification (state-specific quadratic trend, 2005 break)} & & & \\
	State UR (t-1)              &      -2.102*** &      -2.047*** &      -0.055   \\
	                            &     (0.378)   &     (0.373)   &     (0.063)   \\
    No. CZ-cohort cells            &      2,289  &       2,289   &      2,289   \\
	No. high school graduates     &  6,296,662   &   6,296,662   &   6,296,662   \\
	& & & \\
	\textit{Panel B: HP filter, $\lambda = 6.25$} & & & \\
	Cyclical state UR (t-1)     &      -1.314*** &      -1.131** &      -0.182   \\
	                            &     (0.325)   &     (0.386)   &     (0.120)   \\
    No. CZ-cohort cells         &     {2,193}   &       {2,193}   &   {2,193}   \\
    No. high school graduates  &   {6,071,511}   &  {6,071,511}   &   {6,071,511}   \\
	& & & \\
	\textit{Panel C: HP filter, $\lambda = 100$} & & & \\
	Cyclical state UR (t-1)     &      -1.401*** &      -1.043*  &      -0.358** \\
	                            &     (0.472)   &     (0.501)   &     (0.141)   \\
    No. CZ-cohort cells         &     {2,193}   &       {2,193}   &   {2,193}   \\
    No. high school graduates  &   {6,071,511}   &  {6,071,511}   &   {6,071,511}   \\
	& & & \\
	\midrule
	Outcome mean                &        71.2   &        54.2   &        17.0   \\
	State FE, policy controls   &      yes      &    yes      &      yes      \\
				\bottomrule
		\end{tabular}
		}
		 \begin{tablenotes}[flushleft]
				\item \scriptsize{\textit{Notes:} Panel A presents estimates from \Cref{eq:specification} with the state unemployment rate in the senior year ($t-1$), including state-specific quadratic trends with a 2005 break. In Panels B and C, the cyclical components of the state unemployment rate and the enrollment share are instead extracted with a Hodrick--Prescott filter \citepsupp{hodrick1997postwar}, applied within commuting zones, and cyclical enrollment is regressed on the lagged cyclical unemployment rate. As the state unemployment series is constant across the commuting zones of a state, filtering within commuting zones is equivalent to filtering the state series, except at interpolated cells. The smoothing parameter is $\lambda = 6.25$ \citepsupp[for annual data,][]{ravn2002adjusting} in Panel B and $\lambda = 100$ in Panel C. Panels B and C drop the 1995 cohort, which has no lagged cyclical unemployment rate. Empty cells (\textit{Nullkohorten}) are linearly interpolated before filtering. The outcome is the share of high school graduates with \textit{Abitur} enrolling at any college (column 1), academic universities (column 2), or applied universities (column 3). Policy controls include tuition fee introductions and abolitions, as well as double cohorts. Each cell is weighted by the number of high school graduates with \textit{Abitur}. Standard errors in parentheses allow for two-way clustering at the state and cohort level. * $p$\textless 0.1, ** $p$\textless 0.05, *** $p$\textless 0.01. \\ \textit{Source}: Student register, Federal Employment Agency, \textit{Regionaldatenbank}.}
		 \end{tablenotes}
	\end{threeparttable}
	\end{adjustbox}
    \end{center}
\end{table}

\begin{table}[H]
    \begin{center}
	\begin{adjustbox}{max width=0.8\textwidth}
	\begin{threeparttable}
		\caption{Robustness: attainment with a common minimum completion horizon}
		\label{tab:rob_horizon}
		\scriptsize{
		        \begin{tabular} {l ccc}
	\toprule \toprule
    & Overall & Academic & Applied \\
 & (1) & (2) & (3)\\
\midrule
\textit{Panel A: Baseline, cohorts 1995--2016} & & & \\
State UR (t-1)          &      -0.640   &      -1.039***&       0.399*  \\
                        &     (0.410)   &     (0.255)   &     (0.196)   \\
No. CZ-cohort cells       &       2,097   &       2,097   &       2,097   \\
No. high school graduates &   5,727,918   &   5,727,918   &   5,727,918   \\
\addlinespace
\textit{Panel B: Cohorts 1995--2013} & & & \\
State UR (t-1)          &      -0.393   &      -0.884***&       0.491***\\
                        &     (0.267)   &     (0.211)   &     (0.147)   \\
No. CZ-cohort cells       &       1,809   &       1,809   &       1,809   \\
No. high school graduates &   4,863,473   &   4,863,473   &   4,863,473   \\
\midrule
State FE, trends, policy controls &      yes      &      yes      &      yes      \\
				\bottomrule
		\end{tabular}
		}
		\begin{tablenotes}[flushleft]
				\item \scriptsize{ \textit{Notes:} This table presents estimates from \Cref{eq:specification} for the share of high school graduates with \textit{Abitur} graduating from any college (column 1), academic universities (column 2), or applied universities (column 3), with first-time graduations counted through 2019. Panel B restricts the sample to high school cohorts 1995--2013, for which every cohort has at least six years to complete a degree by 2019. Policy controls include tuition fee introductions and abolitions, as well as double cohorts. Each cell is weighted by the number of high school graduates with \textit{Abitur}. Standard errors in parentheses allow for two-way clustering at the state and cohort level. * $p$\textless 0.1, ** $p$\textless 0.05, *** $p$\textless 0.01. \\ \textit{Source}: Exam register, Federal Employment Agency, \textit{Regionaldatenbank}.
                }
		\end{tablenotes}
	\end{threeparttable}
	\end{adjustbox}
    \end{center}
\end{table}

\begin{figure}[H]
	\centering
	\begin{minipage}{\textwidth}
 	\caption{HP-filter fit comparison}
    \vspace{-1em}
    \label{fig:hp_filter}
    \begin{center}
    \includegraphics[width=\textwidth]{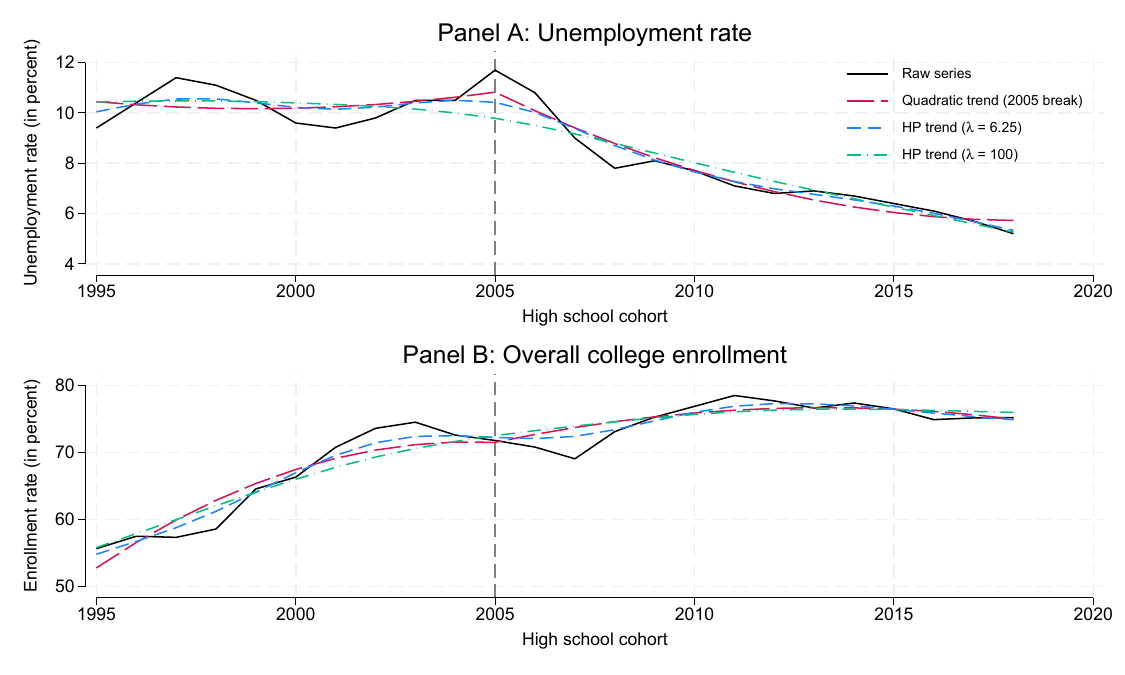} \\
    \end{center}
    \vspace{-1em}
    {\scriptsize \textit{Notes:} Each panel plots the population-weighted national series (solid black) together with three candidate trends: the state-specific quadratic trend with a continuous 2005 break (red long dash, as in \Cref{eq:trend}), and Hodrick--Prescott trends with smoothing parameter $\lambda = 6.25$ (blue dash) and $\lambda = 100$ (green dash-dot). The vertical line marks 2005. \\ \textit{Source}: Student register, Federal Employment Agency (BA), \textit{Regionaldatenbank}, years 1995--2018.
    \par}
    \end{minipage}
\end{figure}

\begin{table}[H]
    \begin{center}
	\begin{adjustbox}{max width = \textwidth}
	\begin{threeparttable}
		\caption{Robustness towards different leads, lags, and moving averages}
        \label{tab:rob_leadlag}
		\scriptsize{
		        \begin{tabular}{l c c c}
	\toprule \toprule
& \multicolumn{1}{c}{Overall} & \multicolumn{1}{c}{Academic} & \multicolumn{1}{c}{Applied}  \\
\cmidrule(lr){2-2} \cmidrule(lr){3-3} \cmidrule(lr){4-4}
& (1) & (2) & (3) \\ 
\midrule
\textit{Panel A: Two years to graduation} &    &   &   \\
State UR (t-2)		&      -1.601***             &      -1.739***             &       0.138            \\
                    	&     (0.262)           &     (0.218)             &     (0.154)              \\
& &   &    \\
\textit{Panel B: 2-year moving average} & &   &     \\
State UR (t, t-1)		&      -2.287***          &      -2.168***        &      -0.119            \\
                    	&     (0.458)        &     (0.530)          &     (0.099)            \\
& &    &   \\
\textit{Panel C: 3-year moving average} & &   &     \\
State UR (t, t-1, t-2) 			&      -3.354***         &      -3.378***      &       0.024       \\
                    &     (0.442)       &     (0.413)        &     (0.154)        \\
& &   &   \\
\textit{Panel D: Graduation year} & &   &   \\
{State UR (t)}		&      {-1.000}            &      {-0.886}         &    {-0.114}           \\
               &     {(0.593)}           &     {(0.662)}           &     {(0.124)}            \\
& &   &    \\
\textit{Panel E: One year after graduation} & &   &   \\
State UR (t+1)		&       1.004            &       1.083*         &      -0.080           \\
               &     (0.643)           &     (0.600)           &     (0.097)            \\
& &   &    \\
Outcome mean      &        71.2    &        54.2    &        17.0    \\
No. region-cohort cells       &        2,289   &        2,289   &        2,289     \\
No. high school graduates    &     6,296,662   &     6,296,662   &     6,296,662    \\
				\bottomrule 
		\end{tabular}
		}
		 \begin{tablenotes}[flushleft] 
				\item \scriptsize{ \textit{Notes:} This table presents estimates from \Cref{eq:specification} for the share of high school graduates with \textit{Abitur} enrolling in different college types, spanning high school cohorts 1995--2018, but using different UR measures in each Panel. Panel A uses the state UR two years before high school graduation, Panel B uses the two-year moving average in the respective graduation and senior year, Panel C additionally averages over the second year before graduation, Panel D uses the state UR in the graduation year, and Panel E uses the lead UR one year after high school graduation. Standard errors in parentheses allow for two-way clustering at the state and cohort level. * $p$\textless 0.1, ** $p$\textless 0.05, *** $p$\textless 0.01. \\ \textit{Source}: Student register, Federal Employment Agency, \textit{Regionaldatenbank}.
                } 
		 \end{tablenotes}
	\end{threeparttable}
	\end{adjustbox}
    \end{center}
\end{table}


\begin{table}[H]
    \begin{center}
	\begin{adjustbox}{max width=\textwidth}
	\begin{threeparttable}
		\caption{Upstream effects on the number of high school graduates}
    	\label{tab:hsgraduates}
		\scriptsize{
		        \begin{tabular} {l cc cc cc}
	\toprule \toprule
& \multicolumn{2}{c}{log(total graduates)} & \multicolumn{2}{c}{log(male graduates)} & \multicolumn{2}{c}{log(female graduates)}  \\
\cmidrule(lr){2-3} \cmidrule(lr){4-5} \cmidrule(lr){6-7}
& (1) & (2) & (3) & (4) & (5) & (6) \\ 
\midrule
\textit{Panel A: Graduation year assuming G9} & &   &  &   &   &   \\
State UR (t-3)                &       0.027** &               &       0.029** &               &       0.029** &               \\
                    &     (0.012)   &               &     (0.013)   &               &     (0.013)   &               \\
... $\times \,$ pre 2008 &               &       0.032** &               &       0.029*  &               &       0.022   \\
                    &               &     (0.014)   &               &     (0.014)   &               &     (0.021)   \\
... $\times \,$ post 2008 &               &       0.028** &               &       0.029** &               &       0.018   \\
                    &               &     (0.012)   &               &     (0.013)   &               &     (0.019)   \\
p-value ($\beta_{\Delta}$)               &               &       0.522   &               &       0.993   &               &       0.580   \\
& &   &  &   &   &   \\
Outcome mean (in levels)                 &      2,769   &      2,769   &      1,194   &      1,194   &      1,539   &      1,539   \\
No. region-cohort cells                  &        2,193   &        2,193   &        2,193   &        2,193   &        2,193   &        2,193   \\
& &   &  &   &   &   \\
\textit{Panel B: Graduation year assuming G8} & &   &  &   &   &   \\
State UR (t-2)                &       0.026** &               &       0.027***&               &       0.026** &               \\
                    &     (0.009)   &               &     (0.009)   &               &     (0.010)   &               \\
... $\times \,$ pre 2008 &               &       0.026** &               &       0.026** &               &       0.024*  \\
                    &               &     (0.010)   &               &     (0.010)   &               &     (0.013)   \\
... $\times \,$ post 2008 &               &       0.029** &               &       0.032** &               &       0.026*  \\
                    &               &     (0.011)   &               &     (0.011)   &               &     (0.015)   \\
p-value ($\beta_{\Delta}$)               &               &       0.549   &               &       0.156   &               &       0.532   \\
& &   &  &   &   &   \\
Outcome mean (in levels)                 &      2,751   &      2,751   &      1,185   &      1,185   &      1,528   &      1,528  \\
No. region-cohort cells                  &        2,289   &        2,289   &        2,289   &        2,289   &        2,289   &        2,289   \\
& &   &  &   &   &   \\
\textit{Panel C: Senior year} & &   &  &   &   &   \\
State UR (t-1)                &       0.012   &               &       0.012   &               &       0.009   &               \\
                    &     (0.008)   &               &     (0.008)   &               &     (0.009)   &               \\
... $\times \,$ pre 2008 &               &       0.016*  &               &       0.019** &               &       0.010   \\
                    &               &     (0.009)   &               &     (0.007)   &               &     (0.012)   \\
... $\times \,$ post 2008 &               &       0.022   &               &       0.030** &               &       0.014   \\
                    &               &     (0.014)   &               &     (0.011)   &               &     (0.020)   \\
p-value ($\beta_{\Delta}$)               &               &       0.514   &               &       0.214   &               &       0.699   \\
& &   &  &   &   &   \\
Outcome mean (in levels)                 &      2,751   &      2,751   &      1,185   &      1,185   &      1,528   &      1,528  \\
No. region-cohort cells                  &        2,289   &        2,289   &        2,289   &        2,289   &        2,289   &        2,289   \\
& &   &  &   &   &   \\
\textit{Panel D: Graduation year} & &   &  &   &   &   \\
State UR (t)  &      -0.010   &               &      -0.009   &               &      -0.014   &               \\
                    &     (0.009)   &               &     (0.011)   &               &     (0.009)   &               \\
... $\times \,$ pre 2008 &               &      -0.009   &               &      -0.005   &               &      -0.018*  \\
                    &               &     (0.009)   &               &     (0.010)   &               &     (0.010)   \\
... $\times \,$ post 2008 &               &      -0.008   &               &       0.001   &               &      -0.022   \\
                    &               &     (0.016)   &               &     (0.018)   &               &     (0.014)   \\
p-value ($\beta_{\Delta}$)               &               &       0.904   &               &       0.562   &               &       0.732   \\
& &   &  &   &   &   \\
Outcome mean (in levels)                 &      2,751   &      2,751   &      1,185   &      1,185   &      1,528   &      1,528  \\
No. region-cohort cells                  &        2,289   &        2,289   &        2,289   &        2,289   &        2,289   &        2,289   \\
				\bottomrule 
		\end{tabular}
		}
		\begin{tablenotes}[flushleft] 
				\item \tiny{ \textit{Notes:} This table presents estimates from \Cref{eq:specification} for the number of high school graduates with \textit{Abitur} by gender as outcome, spanning high school cohorts 1996--2018 in Panel A and 1995--2018 in Panels B-D. Policy controls include tuition fee introductions and abolitions, as well as double cohorts. ``G9'' assumes three years of upper-secondary education. In comparison ``G8'' assumes two years of upper-secondary education \citepsupp[see][on enrollment effects the so-called ``G8 reform.'']{marcus2019effect} Standard errors in parentheses allow for two-way clustering at the state and cohort level. * $p$\textless 0.1, ** $p$\textless 0.05, *** $p$\textless 0.01. \\ \textit{Source}: Student register, Federal Employment Agency, \textit{Regionaldatenbank}.
                } 
		\end{tablenotes}
	\end{threeparttable}
	\end{adjustbox}
    \end{center}
\end{table}

\begin{table}[H]
    \begin{center}
	\begin{adjustbox}{max width=0.8\textwidth}
	\begin{threeparttable}
		\caption{Robustness: controlling for pre-senior-year unemployment}
		\label{tab:rob_lagur}
		\scriptsize{
		        \begin{tabular} {l ccc}
	\toprule \toprule
    & Overall & Academic & Applied \\
 & (1) & (2) & (3)\\
\midrule
\textit{Panel A: Baseline, cohorts 1996--2018} & & & \\
State UR (t-1)          &      -1.946***&      -1.923***&      -0.023   \\
                        &     (0.434)   &     (0.440)   &     (0.085)   \\
\addlinespace
\textit{Panel B: Adding pre-senior-year URs} & & & \\
State UR (t-1)          &      -2.064***&      -1.910***&      -0.154   \\
                        &     (0.595)   &     (0.627)   &     (0.105)   \\
State UR (t-2)          &       0.112   &      -0.113   &       0.226   \\
                        &     (0.417)   &     (0.425)   &     (0.129)   \\
State UR (t-3)          &      -1.156***&      -1.135***&      -0.021   \\
                        &     (0.366)   &     (0.319)   &     (0.149)   \\
\midrule
No. CZ-cohort cells       &       2,193   &       2,193   &       2,193   \\
No. high school graduates &   6,071,511   &   6,071,511   &   6,071,511   \\
State FE, trends, policy controls &      yes      &      yes      &      yes      \\
				\bottomrule
		\end{tabular}
		}
		\begin{tablenotes}[flushleft]
				\item \scriptsize{ \textit{Notes:} This table presents estimates from \Cref{eq:specification} for the share of high school graduates with \textit{Abitur} enrolling at any college (column 1), academic universities (column 2), or applied universities (column 3). Panel B adds the state unemployment rates two and three years before graduation as controls. Both panels span high school cohorts 1996--2018, as the unemployment rate in $t-3$ is not available for the 1995 cohort. Policy controls include tuition fee introductions and abolitions, as well as double cohorts. Each cell is weighted by the number of high school graduates with \textit{Abitur}. Standard errors in parentheses allow for two-way clustering at the state and cohort level. * $p$\textless 0.1, ** $p$\textless 0.05, *** $p$\textless 0.01. \\ \textit{Source}: Student register, Federal Employment Agency, \textit{Regionaldatenbank}.
                }
		\end{tablenotes}
	\end{threeparttable}
	\end{adjustbox}
    \end{center}
\end{table}

\begin{table}[H]
    \begin{center}
	\begin{adjustbox}{max width=0.85\textwidth}
	\begin{threeparttable}
		\caption{Robustness: enrollment per 18--19-year-old (population denominator)}
		\label{tab:rob_popdenom}
		\scriptsize{
		        \begin{tabular} {l cccc}
	\toprule \toprule
    & \multicolumn{3}{c}{First-time enrollments per capita} & \multicolumn{1}{c}{Cohort share} \\
    \cmidrule(lr){2-4} \cmidrule(lr){5-5}
    & Overall & Academic & Applied & with \textit{Abitur} \\
 & (1) & (2) & (3) & (4)\\
\midrule
State UR (t-1)          &      -0.251*  &      -0.239** &      -0.011   &       0.037   \\
                        &     (0.118)   &     (0.095)   &     (0.025)   &     (0.095)   \\
\midrule
Outcome mean            &        9.86   &        7.48   &        2.38   &       14.08   \\
No. CZ-cohort cells       &       2,193   &       2,193   &       2,193   &       2,193   \\
No. high school graduates &   6,071,511   &   6,071,511   &   6,071,511   &   6,071,511   \\
State FE, trends, policy controls &      yes      &      yes      &      yes      &      yes      \\
				\bottomrule
		\end{tabular}
		}
		\begin{tablenotes}[flushleft]
				\item \scriptsize{ \textit{Notes:} This table presents estimates from \Cref{eq:specification} with outcomes per 100 of the population aged 18--19. Columns 1--3 use the number of high school graduates with \textit{Abitur} enrolling for the first time at any college, academic universities, or applied universities. Column 4 uses the number of high school graduates with \textit{Abitur}, so the outcome is the share of the cohort obtaining the qualification. The denominator is the county population aged 18 to under 20 on December 31 of the senior year ($t-1$), from the \textit{Regionaldatenbank} population statistics (12411 tables), aggregated to commuting zones. The sample spans high school cohorts 1996--2018, as the population data start in 1995. Each cell is weighted by the population aged 18--19. Dropping commuting-zone-cohort cells with incomplete county-to-commuting-zone population coverage leaves the estimates unchanged. Standard errors in parentheses allow for two-way clustering at the state and cohort level. * $p$\textless 0.1, ** $p$\textless 0.05, *** $p$\textless 0.01. \\ \textit{Source}: Student register, Federal Employment Agency, \textit{Regionaldatenbank}.
                }
		\end{tablenotes}
	\end{threeparttable}
	\end{adjustbox}
    \end{center}
\end{table}

\begin{table}[H]
    \begin{center}
	\begin{adjustbox}{max width=\textwidth}
	\begin{threeparttable}
		\caption{Enrollment effects of CZ-level unemployment variation}
		\label{tab:abi_enroll_CZ}
		\scriptsize{
		        \begin{tabular} {l cc cc cc}
	\toprule \toprule

& \multicolumn{2}{c}{Overall} & \multicolumn{2}{c}{Academic} & \multicolumn{2}{c}{Applied}  \\
\cmidrule(lr){2-3} \cmidrule(lr){4-5} \cmidrule(lr){6-7}
& (1) & (2) & (3) & (4) & (5) & (6) \\ 
\midrule
\textit{Panel A: CZ FEs, without year FEs} & &   &  &   &   &   \\
CZ-level UR (t-1)			&      -1.031** &               &      -0.734   &               &      -0.297** &               \\
                    	&     (0.444)   &               &     (0.481)   &               &     (0.138)   &               \\
... $\times \,$ pre 2008 &     &      -1.021*  &               &      -0.789   &               &      -0.232   \\
                    	&     &     (0.531)   &               &     (0.486)   &               &     (0.155)   \\
... $\times \,$ post 2008 &     &      -1.015   &               &      -0.822   &               &      -0.193   \\
                    		&     &     (0.702)   &               &     (0.633)   &               &     (0.200)   \\
p-value ($\beta_{\Delta}$) &     &       [0.982]   &               &       [0.897]   &               &       [0.611]   \\
& &   &  &   &   &   \\
\textit{Panel B: CZ FEs, including year FEs} & &   &  &   &   &   \\
CZ-level UR (t-1)			&       0.111   &               &       0.214   &               &      -0.103   &               \\
                    	&     (0.413)   &               &     (0.334)   &               &     (0.199)   &               \\
... $\times \,$ pre 2008 &        &       0.111   &               &       0.211   &               &      -0.100   \\
                    	&       &     (0.397)   &               &     (0.323)   &               &     (0.173)   \\
... $\times \,$ post 2008 &       &       0.101   &               &       0.335   &               &      -0.234   \\
                    		&   &     (0.526)   &               &     (0.381)   &               &     (0.192)   \\
p-value ($\beta_{\Delta}$)  &      &       [0.967]   &               &       [0.415]   &               &       [0.391]   \\
& &   &  &   &   &   \\
\textit{Panel C: State FEs, without year FEs} & &   &  &   &   &   \\
CZ-level UR (t-1)			&      -0.126   &               &      0.205   &               &      -0.332** &               \\
                    	&     (0.238)   &               &     (0.339)   &               &     (0.155)   &               \\
... $\times \,$ pre 2008 &     &      -0.166   &               &      0.146   &               &      -0.312*  \\
                    	&     &     (0.239)   &               &     (0.336)   &               &     (0.151)   \\
... $\times \,$ post 2008 &     &      -0.025   &               &      0.355   &               &      -0.381** \\
                    		&     &     (0.252)   &               &     (0.351)   &               &     (0.164)   \\
p-value ($\beta_{\Delta}$) &     &       [0.512]   &               &       [0.416]   &               &       [0.404]   \\
& &   &  &   &   &   \\
\textit{Panel D: State FEs, including year FEs} & &   &  &   &   &   \\
CZ-level UR (t-1)			&      -0.017   &               &      0.313   &               &      -0.329** &               \\
                    	&     (0.196)   &               &     (0.314)   &               &     (0.142)   &               \\
... $\times \,$ pre 2008 &     &      -0.036   &               &      0.080   &               &      -0.117   \\
                    	&     &     (0.248)   &               &     (0.381)   &               &     (0.153)   \\
... $\times \,$ post 2008 &     &      0.003   &               &      0.544**  &               &      -0.541*** \\
                    		&     &     (0.191)   &               &     (0.250)   &               &     (0.145)   \\
p-value ($\beta_{\Delta}$) &     &       [0.849]   &               &       [0.068]   &               &       [0.008]   \\
& &   &  &   &   &   \\
Outcome mean      &        73.7   &        73.7   &        55.4   &        55.4   &        18.3   &        18.3   \\
No. CZ-cohort cells     &        1,907   &        1,907   &        1,907   &        1,907   &        1,907   &        1,907   \\
No. high school graduates   &     5,366,608   &     5,366,608   &     5,366,608   &     5,366,608   &     5,366,608   &     5,366,608   \\
				\bottomrule 
		\end{tabular}
		}
		\begin{tablenotes}[flushleft] 
				\item \scriptsize{ \textit{Notes:} This table presents estimates of the effect of the CZ-level unemployment rate in the senior year ($t-1$) on the share of high school graduates with \textit{Abitur} enrolling in different college types, spanning high school cohorts 1999--2018. Panels A and B include CZ fixed effects and CZ-specific quadratic trends with a 2005 break. Panels C and D include the state fixed effects and state-specific quadratic trends of \Cref{eq:specification}. Panels B and D additionally include cohort fixed effects. Policy controls include tuition fee introductions and abolitions, as well as double cohorts. Each cell is weighted by the number of high school graduates with \textit{Abitur}. Standard errors in parentheses allow for two-way clustering at the state and cohort level in Panels A and C, and at the state level in Panels B and D. * $p$\textless 0.1, ** $p$\textless 0.05, *** $p$\textless 0.01. \\ \textit{Source}: Student register, Federal Employment Agency, \textit{Regionaldatenbank}.
                } 
		\end{tablenotes}
	\end{threeparttable}
	\end{adjustbox}
    \end{center}
\end{table}

\begin{table}[H]
    \begin{center}
	\begin{adjustbox}{max width=0.85\textwidth}
	\begin{threeparttable}
		\caption{Effects of the graduation-year state UR}
		\label{tab:gradyear_ur}
		\scriptsize{
		        \begin{tabular} {l ccc}
	\toprule \toprule
& \multicolumn{1}{c}{1995--2018} & \multicolumn{2}{c}{1995--2016}  \\
\cmidrule(lr){2-2} \cmidrule(lr){3-4}
& Enrollment & Enrollment & Attainment \\
& (1) & (2) & (3) \\
\midrule
\textit{Panel A: Overall (any college)} & & & \\
State UR (t)     &      -1.000   &      -1.206*  &       0.348   \\
                 &     (0.593)   &     (0.587)   &     (0.426)   \\
Outcome mean     &        71.2   &        71.0   &        59.9   \\
\addlinespace
\textit{Panel B: Academic university} & & & \\
State UR (t)     &      -0.886   &      -1.171*  &      -0.267   \\
                 &     (0.662)   &     (0.639)   &     (0.397)   \\
Outcome mean     &        54.2   &        54.6   &        40.0   \\
\addlinespace
\textit{Panel C: Applied university} & & & \\
State UR (t)     &      -0.114   &      -0.035   &       0.615***\\
                 &     (0.124)   &     (0.130)   &     (0.169)   \\
Outcome mean     &        17.0   &        16.4   &        19.9   \\
\midrule
No. CZ-cohort cells       &       2,289   &       2,097   &       2,097   \\
No. high school graduates &   6,296,662   &   5,727,918   &   5,727,918   \\
State FE, trends, policy controls &      yes      &      yes      &      yes      \\
				\bottomrule
		\end{tabular}
		}
		\begin{tablenotes}[flushleft]
				\item \scriptsize{ \textit{Notes:} This table presents estimates from \Cref{eq:specification} for the effect of the state unemployment rate in the graduation year ($t$). The outcome is the share of high school graduates with \textit{Abitur} enrolling at (columns 1--2) or graduating from (column 3) any college (Panel A), academic universities (Panel B), or applied universities (Panel C). Column 1 spans high school cohorts 1995--2018. Columns 2 and 3 span cohorts 1995--2016, with attainment measured by 2019. Policy controls include tuition fee introductions and abolitions, as well as double cohorts. Each cell is weighted by the number of high school graduates with \textit{Abitur}. Standard errors in parentheses allow for two-way clustering at the state and cohort level. * $p$\textless 0.1, ** $p$\textless 0.05, *** $p$\textless 0.01. \\ \textit{Source}: Student register, exam register, Federal Employment Agency, \textit{Regionaldatenbank}.
                }
		\end{tablenotes}
	\end{threeparttable}
	\end{adjustbox}
    \end{center}
\end{table}

\Cref{tab:wcb} reports three bootstrap procedures for the main enrollment effects: a wild cluster bootstrap-t with the null imposed, with Rademacher and with Webb weights \citepsupp{cameron2008bootstrap, roodman2019fast}, and a pairs cluster bootstrap that resamples states with replacement \citepsupp[following the guide in][]{mackinnon2023cluster}.

\begin{table}[H]
    \begin{center}
	\begin{adjustbox}{max width=0.75\textwidth}
	\begin{threeparttable}
		\caption{Small-cluster bootstrap inference for the main enrollment effects}
		\label{tab:wcb}
		\scriptsize{
		        \begin{tabular} {l ccc}
	\toprule \toprule
& \multicolumn{1}{c}{Overall} & \multicolumn{1}{c}{Academic} & \multicolumn{1}{c}{Applied}  \\
& (1) & (2) & (3) \\
\midrule
State UR (t-1)                   &      -2.102   &      -2.047   &      -0.055   \\
\addlinespace
\textit{Wild cluster bootstrap-t (Rademacher)} & & & \\
p-value                          &   $<$0.001    &   $<$0.001    &       0.521   \\
95\% CI                          & [-2.661, -1.423] & [-2.470, -1.505] & [-0.216, 0.126] \\
\addlinespace
\textit{Wild cluster bootstrap-t (Webb)} & & & \\
p-value                          &   $<$0.001    &   $<$0.001    &       0.530   \\
95\% CI                          & [-2.659, -1.449] & [-2.471, -1.515] & [-0.214, 0.129] \\
\addlinespace
\textit{Pairs cluster bootstrap}  & & & \\
p-value                          &   $<$0.001    &   $<$0.001    &       0.434   \\
95\% CI                          & [-2.512, -1.547] & [-2.383, -1.617] & [-0.175, 0.104] \\
\midrule
No. CZ-cohort cells       &        2,289   &        2,289   &        2,289   \\
No. state clusters        &          16    &          16    &          16    \\
				\bottomrule
		\end{tabular}
		}
		\begin{tablenotes}[flushleft]
				\item \scriptsize{ \textit{Notes:} This table reports bootstrap inference for the senior-year state UR coefficients of \Cref{eq:specification} on the share of high school graduates with \textit{Abitur} enrolling at each college type, high school cohorts 1995--2018. The wild cluster bootstrap-t rows use Rademacher and Webb weights drawn per state, impose the null, and are t-pivoted, with $B = 9{,}999$ replications (\texttt{fwildclusterboot}). Their confidence intervals are obtained by test inversion. The pairs cluster bootstrap resamples the 16 states with replacement, $B = 999$ replications, with percentile confidence intervals. All procedures cluster by state. The main estimates instead use two-way clustering by state and cohort. Policy controls, state-specific quadratic trends with a 2005 break, and weighting follow \Cref{eq:specification}. \\ \textit{Source}: Student register, Federal Employment Agency, \textit{Regionaldatenbank}.
                }
		\end{tablenotes}
	\end{threeparttable}
	\end{adjustbox}
    \end{center}
\end{table}

\newpage
\begin{spacing}{1.5}
\bibliographystylesupp{aer}
\bibliographysupp{library}
\end{spacing}

\end{document}